\documentclass[iop]{emulateapj}

\usepackage{natbib}
\usepackage{color}
\usepackage{graphicx}
\usepackage{amsmath}
\usepackage{amssymb}
\usepackage{savesym}
\usepackage{color}
\usepackage{tipa}
\usepackage{import}
\usepackage{tabularx}
\usepackage{textcomp}
\usepackage{verbatimbox}

\newcommand{\centi}{\text{c}}
\newcommand{\kilo}{\text{k}}

\newcommand{\meter}{\text{m}}
\newcommand{\parsec}{\text{pc}}
\newcommand{\second}{\text{s}}

\newcommand{\kev}{\text{keV}}

\newcommand{\days}{\text{d}}

\newcommand{\ergs}{\text{erg s$^{-1}$}}

\newcommand{\E}[1]{\times10^{#1}}
\newcommand{\Msol}{{\rm M}_{\odot}}

\newcommand{\U}[3]{$#1_{-#3}^{+#2}$}
\newcommand{\Pl}[2]{$#1\pm#2$}

\newcolumntype{Y}{>{\centering\arraybackslash}X}

\newcolumntype{M}[1]{>{\centering\arraybackslash}m{#1}}
\newcolumntype{N}{@{}m{0pt}@{}}

\newcommand\Tstrut{\rule{0pt}{2.6ex}}         
\newcommand\Bstrut{\rule[-0.9ex]{0pt}{0pt}}   

\shorttitle{Eclipsing ULXs in M\,51}
\shortauthors{Urquhart \& Soria}

\begin{document}

\label{firstpage}

\title{Two eclipsing ultraluminous X-ray sources in M\,51}

\author{R. Urquhart\altaffilmark{1} and R. Soria\altaffilmark{1,2}}
\affil{$^1$International Centre for Radio Astronomy Research, Curtin University, GPO Box U1987, Perth, WA 6845, Australia\\$^2$Sydney Institute for Astronomy, School of Physics A28, The University of Sydney, Sydney, NSW 2006, Australia}

\email{ryan.urquhart@icrar.org}
\email{roberto.soria@curtin.edu.au}

\begin{abstract}
We present the discovery, from archival {\it Chandra} and {\it XMM-Newton} data, of X-ray eclipses in two ultraluminous X-ray sources (ULXs), located in the same region of the galaxy M\,51: CXOM51 J132940.0$+$471237 (ULX-1, for simplicity) and CXOM51 J132939.5$+$471244 (ULX-2). Three eclipses were detected for ULX-1, two for ULX-2. The presence of eclipses puts strong constraints on the viewing angle, suggesting that both ULXs are seen almost edge-on and are certainly not beamed towards us. Despite the similar viewing angles and luminosities ($L_{\rm X} \approx 2 \times 10^{39}$ erg s$^{-1}$ in the 0.3--8 keV band for both sources), their X-ray properties are different. ULX-1 has a soft spectrum, well fitted by Comptonization emission from a medium with electron temperature $kT_e \approx 1$ keV. ULX-2 is harder, well fitted by a slim disk with $kT_{\rm in} \approx 1.5$--1.8 keV and normalization consistent with a $\sim$10\,$M_{\odot}$ black hole. ULX-1 has a significant contribution from multi-temperature thermal plasma emission ($L_{\rm X,mekal} \approx 2 \times 10^{38}$ erg s$^{-1}$); about 10\% of this emission remains visible during the eclipses, proving that the emitting gas comes from a region slightly more extended than the size of the donor star. From the sequence and duration of the {\it Chandra} observations in and out of eclipse, we constrain the binary period of ULX-1 to be either $\approx$6.3 days, or $\approx$12.5--13 days. If the donor star fills its Roche lobe (a plausible assumption for ULXs), both cases require an evolved donor; most likely a blue supergiant, given the young age of the stellar population in that galactic environment.

\end{abstract}


\section{Introduction} \label{intro}

Ultraluminous X-ray sources (ULXs) are the high-luminosity end of the X-ray binary population, with X-ray luminosities $> 10^{39}$ erg s$^{-1}$, which is the approximate peak luminosity of Galactic stellar-mass black holes (BHs). The most likely explanation for the vast majority of ULXs is that they are stellar-mass BHs \citep[or neutron stars;][]{2014Natur.514..202B} accreting well above the critical accretion rate and their luminosity is a few times the classical Eddington limit of spherical accretion. Another possibility is that ULXs are powered by accreting BHs up to $\approx 80 M_{\odot}$ \citep{2010ApJ...714.1217B}, several times more massive than typical Galactic stellar-mass BHs ($M \sim 5$--$15 M_{\odot}$: \citealt{2012ApJ...757...36K}). In addition, ULXs may appear more luminous because their X-ray emission is partly collimated along our line of sight. This may happen at super-critical accretion rates, because of the predicted formation of a dense radiatively-driven disk outflow and a lower-density polar funnel, along which more photons can escape \citep{2011ApJ...736....2O,2014ApJ...796..106J,2015MNRAS.453.3213S}. Finally, some of the brightest ULXs may contain a population of intermediate-mass BHs \citep[$10^2\,\Msol\leq M \leq10^4\,\Msol$;][]{2009Natur.460...73F,2016ApJ...817...88Z}.  It is difficult to determine the relative contribution of those three factors (mass, accretion rate and viewing angle), and therefore also determine the true isotropic luminosity and accretion rate of ULXs, without at least a direct constraint on their viewing angle.

There is already indirect evidence that ULXs are not strongly beamed. Modelling of the optical light curve from the irradiated donor star in NGC\,7793-P13 showed \citep{2014Natur.514..198M} that the source is viewed at an angle $>$20$^{\circ}$ and more likely much higher; thus, in that case, super-Eddington accretion is the reason for the high luminosity, not a heavier BH or a down-the-funnel view. Studies of large ($\ga$100 pc) photo-ionized and/or shock-ionized plasma bubbles around ULXs \citep{2002astro.ph..2488P,2008AIPC.1010..303P} provide other clues about viewing angles: if fast-accreting BHs appeared as ULXs only for a narrow range of face-on inclinations, we would see many more of those large ionized bubbles without a ULX inside; moreover, the true X-ray luminosity of a beamed source (much lower than the apparent luminosity) would not be enough to explain the strong He\,{\footnotesize{II}} emission observed from some of the photo-ionized ULX bubbles \citep{2006IAUS..230..293P}. Both the fact that most ULX bubbles do contain a bright, central X-ray source, and the fact that (in photo-ionized bubbles) the apparent X-ray photon flux from the central source is consistent with the He\,{\footnotesize{II}} photon flux from the bubble, suggest that, statistically, ULXs are seen over a broad range of viewing angles. X-ray spectroscopic studies can also be used to qualitatively constrain ULX viewing angles: it was suggested \citep{2013MNRAS.435.1758S} that ULXs seen at lower inclination (down the polar funnel) have harder X-ray spectra while those seen at higher inclination (through the disk wind) have softer spectra with a lower-energy downturn, due to a higher degree of Compton scattering in the wind. This interpretation is consistent with the presence of absorption and emission features (interpreted as signatures of the outflow) in the X-ray spectra of ULXs with softer spectra \citep{2014MNRAS.438L..51M, 2015MNRAS.454.3134M}. It is also in agreement with a higher degree of short-term variability (interpreted as the imprint of a clumpy wind) in sources with softer spectra \citep{2015MNRAS.447.3243M}.  

\begin{figure}
    \centering
    \includegraphics[width=0.49\textwidth]{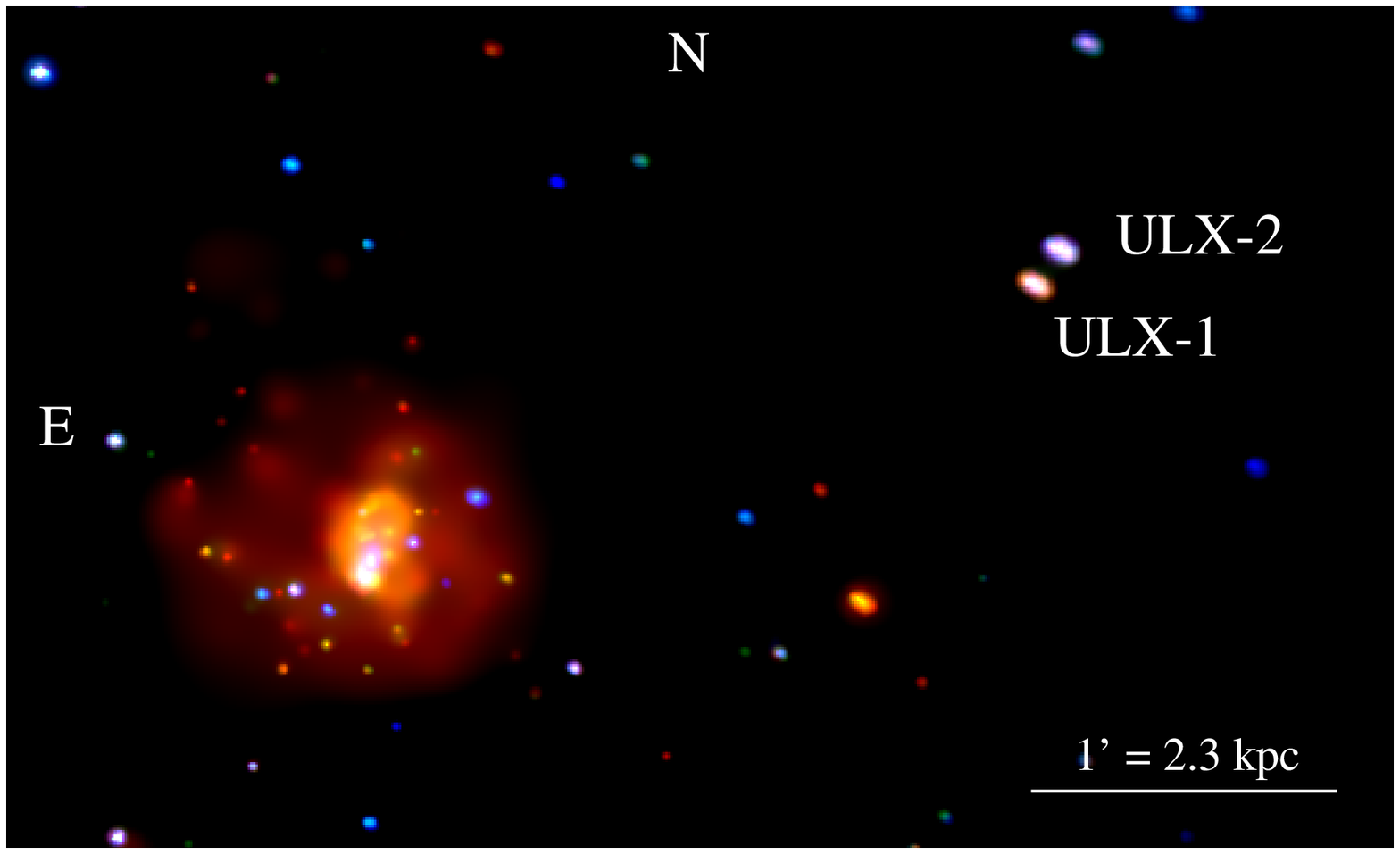}\\[-7pt]
    \includegraphics[width=0.49\textwidth]{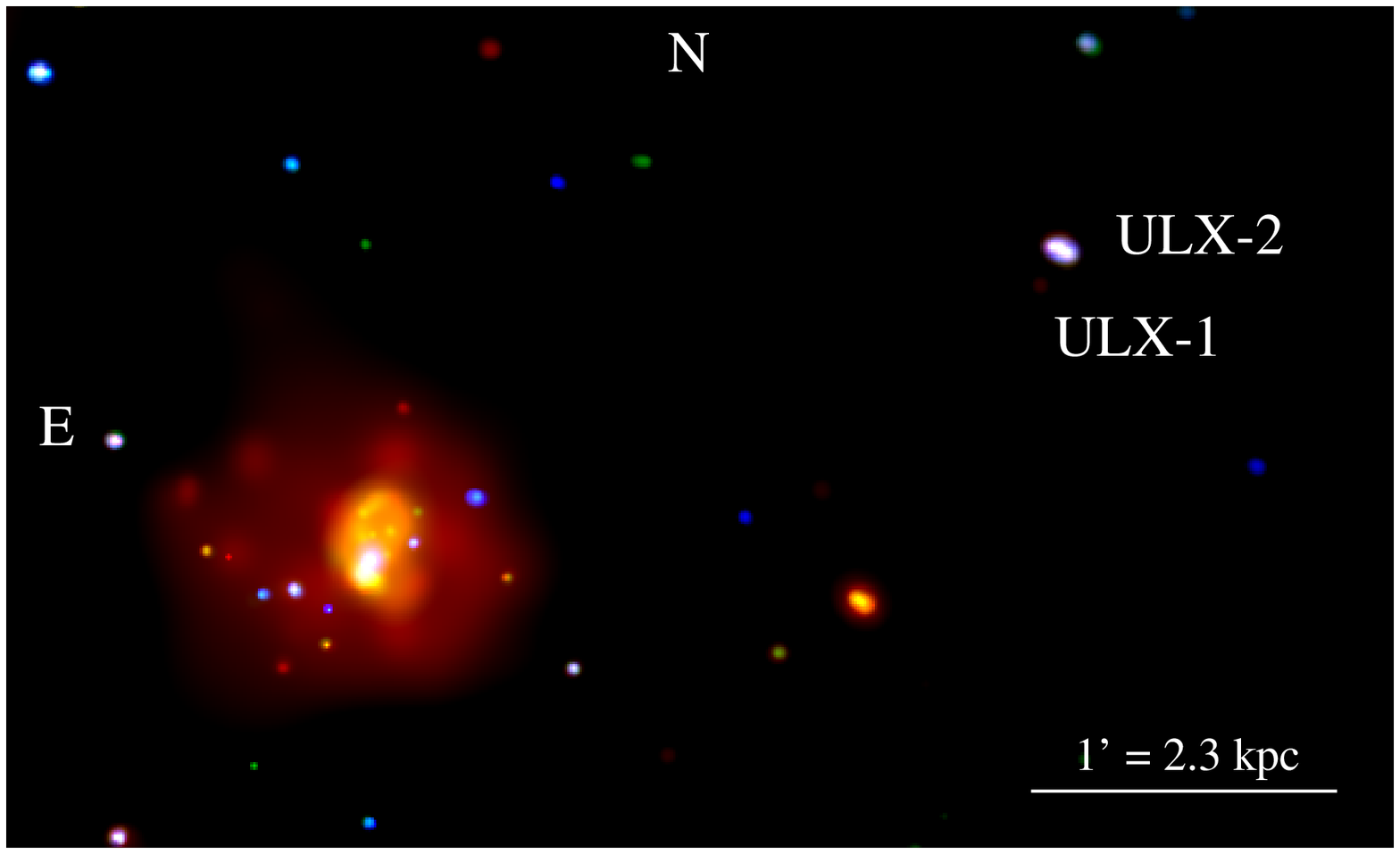}
    \caption{Top panel: {\it Chandra}/ACIS-S adaptively smoothed 190-ks image of M\,51 during ObsID 13814, showing the location of ULX-1 and ULX-2 with respect to the nuclear region. Red represents the 0.3--1 keV band, green the 1--2 keV band, and blue the 2--7 keV band. Bottom panel: as in the top panel, but only for the portion of ObsID 13814 during which ULX-1 is in eclipse (70 ks). }
    \label{chand_img}
    \vspace{0.3cm}
\end{figure}

Apart from those indirect or statistical arguments, until recently there was no bright extragalactic stellar-mass BH for which the viewing angle could be directly pinned down. We have now discovered two such sources, both located in the same spiral arm of the spiral galaxy M\,51; in fact, surprisingly, they appear projected in the sky within only $\approx$ $350\,\parsec$ of each other (see Figure \ref{chand_img}). Both sources have X-ray luminosities $\ga$10$^{39}$ erg s$^{-1}$, and crucially, they both show sharp X-ray drops and rebrightenings, which we interpret as eclipses by their donor stars, occulting the inner region of the disk. The presence of eclipses places a lower limit on the inclination angle ($i \gtrsim 75^{\circ}$) as we must be viewing the X-ray sources near edge-on. In this paper, we present the eclipse discovery and the main X-ray timing and spectral properties of the two sources. We will also briefly discuss more general implications and opportunities provided by the detection of eclipses, for our modelling of these systems. In a companion paper (Soria et al., in prep.) we will present a study of the optical counterparts and other interesting, newly-discovered properties of those same two ULXs, which show optical and radio evidence of jets and outflows.

\section{Targets of our study} \label{obs_sec}

M\,51, also known as the Whirlpool Galaxy, is an interacting face-on spiral at a distance of $8.0 \pm 0.6$ Mpc \citep{M51dist}. The two eclipsing sources discussed in this paper are those catalogued as CXOM51 J132940.0$+$471237 (henceforth, ULX-1) and CXOM51 J132939.5$+$471244 (henceforth, ULX-2) in \cite{2004ApJ...601..735T}. We re-estimated their positions using all the {\it Chandra} data available to-date, and obtained RA (J2000) $=13^{\rm{h}} 29^{\rm{m}} 39^{\rm s}.94$, Dec.~(J2000) $= +47^{\circ} 12' 36''.6$ for ULX-1, and RA (J2000) $= 13^{\rm{h}} 29^{\rm{m}} 39^{\rm s}44$, Dec.~(J2000) $= +47^{\circ} 12' 43''.3$ for ULX-2. Both positions are subject to the standard uncertainty in the absolute astrometry of {\it Chandra} pointings, $\approx$0''.6 at the 90\% confidence level\footnote{See http://cxc.harvard.edu/cal/ASPECT/celmon/}. A more precise determination of their positions is left to a follow-up study (Soria et al., in prep.) of their optical and radio counterparts.

ULX-1 and ULX-2 were first discovered as a single unresolved source by the {\it Einstein Observatory} \citep{1985ApJ...298..259P}. This was followed up with observations with {\it ROSAT} (source C in \citealt{1995A&A...295..289E}; source R7 in   \citealt{1995ApJ...438..663M}). The higher spatial resolution of {\it Chandra}'s Advanced CCD Imaging Spectrometer (ACIS) finally led to the two sources being resolved \citep[source 6 and source 5 in][]{2004ApJ...601..735T}. ULX-1 was found to be a relatively soft source, with very few counts above $2\,\kev$. ULX-2 was found to be variable, decreasing in luminosity by a factor of $\approx2.5$ between observations \citep{2004ApJ...601..735T}. Further spectral studies of the two sources, based on a 2003 {\it XMM-Newton} observation, were carried out by \cite{2005ApJ...635..198D}. With a much larger database of {\it Chandra} and {\it XMM-Newton} observation available since then, we have now studied the two sources in more detail, and found more intriguing properties. 

\begin{table}
\caption{Log of the {\it Chandra} and {\it XMM-Newton} observations used in this study.}
\begin{scriptsize}
\begin{center}
{\begin{tabular}{lccr}
\hline\hline
\multicolumn{1}{c}{Observatory} & \multicolumn{1}{c}{ObsID} & \multicolumn{1}{c}{Exp Time} & \multicolumn{1}{c}{Date}\Tstrut\\[2pt]
&&\multicolumn{1}{c}{(ks)}&\Bstrut\\[2pt]
\hline\hline
{\it Chandra} & \ \ \,354 & \ \,14.9 & 2000-03-21\Tstrut\\[2pt]
& \ \,1622 & \ \,26.8 & 2001-06-23\\[2pt]
& \ \,3932 & \ \,48.0 & 2003-08-07\\[2pt]
& 13813 & 179.2 & 2012-09-09\\[2pt]
& 13812 & 157.5 & 2012-09-12\\[2pt]
& 15496 & \ \,41.0 & 2012-09-19\\[2pt]
& 13814 & 189.9 & 2012-09-20\\[2pt]
& 13815 & \ \,67.2 & 2012-09-23\\[2pt]
& 13816 & \ \,73.1 & 2012-09-26\\[2pt]
& 15553 & \ \,37.6 & 2012-10-10\Bstrut\\[2pt]
\hline
{\it XMM-Newton} & 0112840201 & 20.9 & 2003-01-15\Tstrut\\[2pt]
& 0212480801 & 49.2 & 2005-07-01\\[2pt]
& 0212480901 & \ \,closed & 	2005-07-01\\[2pt]
& 0303420101 & 54.1 & 2006-05-20\\[2pt]
& 0303420301 & \ \,closed & 	2006-05-20\\[2pt]
& 0303420201 & 36.8 & 2006-05-24\\[2pt]
& 0303420401 & \ \,closed & 2006-05-24\\[2pt]
& 0677980701 & 13.3 & 2011-06-07\\[2pt]
& 0677980801 & 13.3\footnote{Due to background flaring, only $\approx$2.5 ks of epoch 0677980801 can be used.} & 2011-06-11\Bstrut\\[2pt]
\hline
\end{tabular}}
\end{center}
\end{scriptsize}
\label{obs_table}
\end{table}

\begin{table*}
\caption{ULX-1 eclipse times and net count rates for the {\it Chandra} observations. Square brackets signify that the eclipse continues beyond the start/end of the observation for some time.}
\begin{scriptsize}
\begin{center}
{\begin{tabular}{ccccccc}
\hline\hline
\multicolumn{1}{c}{ObsID} & \multicolumn{1}{c}{In Eclipse} & \multicolumn{1}{c}{Out of Eclipse} & \multicolumn{1}{c}{MJD Observation} &\multicolumn{1}{c}{MJD Eclipse} & \multicolumn{1}{c}{Count Rate in Eclipse} & \multicolumn{1}{c}{Count Rate out of Eclipse}\Tstrut\\[2pt]
&\multicolumn{1}{c}{(ks)}&\multicolumn{1}{c}{(ks)}&&&\multicolumn{1}{c}{($10^{-3}$\,ct\,s$^{-1}$)}&\multicolumn{1}{c}{($10^{-3}$\,ct\,s$^{-1}$)}\Bstrut\\[2pt]
\hline\hline
\ \ \,354 & \ \,0 & \ \,15 &51715.34--51715.51&&& $18.1\pm1.1$\Tstrut\\[2pt]
\ \,1622 & 25 & \ \ \,2& 52083.78--52084.09& 52083.81--[52084.09] & $1.2\pm0.2$ & $10.7\pm1.9$\\[2pt]
\ \,3932 & \ \,0 & \ \,48 &52858.61--52859.16 &&& $12.0\pm0.5$\\[2pt]
13813 & 40 & 139 & 56179.74--56181.82&[56179.74]--56180.20 & $0.4\pm0.1$ & $14.7\pm0.3$ \\[2pt]
13812 & \ \,0 & 157 & 56182.77--56184.59&&& $12.2\pm0.3$ \\[2pt]
15496 & \ \,0 & \ \,41 &56189.39--56189.86&&& $14.8\pm0.6$\\[2pt]
13814 & 70 & 120 & 56190.31--56192.50& 56191.64--[56192.50] & $0.3\pm0.1$ & $15.6\pm0.4$\\[2pt]
13815 & \ \,0 & \ \,67 &56193.34--56194.12 &&& $14.0\pm0.5$\\[2pt]
13816 & \ \,0 & \ \,73 &56196.22--56197.06 &&& $12.0\pm0.4$\\[2pt]
15553 & \ \,0 & \ \,38 &56210.03--56210.47&&& $10.2\pm0.5$\Bstrut\\[2pt]
\hline
\end{tabular}}
\end{center}
\end{scriptsize}
\label{ec1_table}
\end{table*}

\begin{table*}
\caption{ULX-2 eclipse times and net count rates for the {\it Chandra} observations. Square brackets signify that the eclipse continues beyond the start/end of the observation for some time.}
\begin{scriptsize}
\begin{center}
{\begin{tabular}{ccccccc}
\hline\hline
\multicolumn{1}{c}{ObsID} & \multicolumn{1}{c}{In Eclipse} & \multicolumn{1}{c}{Out of Eclipse} & \multicolumn{1}{c}{MJD Observation} & \multicolumn{1}{c}{MJD eclipse} & \multicolumn{1}{c}{Count Rate in Eclipse} & \multicolumn{1}{c}{Count Rate out of Eclipse}\Tstrut\\[2pt]
&\multicolumn{1}{c}{(ks)}&\multicolumn{1}{c}{(ks)}&&\multicolumn{1}{c}{($10^{-3}$\,ct\,s$^{-1}$)}&\multicolumn{1}{c}{($10^{-3}$\,ct\,s$^{-1}$)}\Bstrut\\[2pt]
\hline\hline
\ \ \,354 & \ \,0 & \ \,15 &51715.34--51715.51&&& $18.0\pm1.1$\Tstrut\\[2pt]
\ \,1622 & \ \,0 & \ \,27 & 52083.78--52084.09& && $7.6\pm0.5$\\[2pt]
\ \,3932 & \ \,0& \ \,48 &52858.61--52859.16 &&& $9.9\pm0.5$\\[2pt]
13813 & 48 & 131 & 56179.74--56181.82& [56179.74]--56180.30 &$0.6\pm0.1$& $10.5\pm0.3$\\[2pt]
13812 & \ \,0 & 157 & 56182.77--56184.59&&& $10.7\pm0.3$ \\[2pt]
15496 & \ \,0 & \ \,41 &56189.39--56189.86&&& $13.7\pm0.6$\\[2pt]
13814 & \ \,0& 190 & 56190.31--56192.50& && $11.5\pm0.2$\\[2pt]
13815 & \ \,0 & \ \,67 &56193.34--56194.12&&& $3.3\pm0.2$\\[2pt]
13816 & \ \,0 & \ \,73 &56196.22--56197.06&&& $14.6\pm0.5$\\[2pt]
15553 & \ \,0 & \ \,38 &56210.03--56210.47&&& $7.9\pm0.5$\Bstrut\\[2pt]
\hline
\end{tabular}}
\end{center}
\end{scriptsize}
\label{ec2_table}
\end{table*}

\section{Data Analysis} \label{data_sec}

M\,51 was observed by {\it Chandra}/ACIS-S fourteen times between 2000 and 2012: two of those observations were too short ($\le$2 ks) to be useful, and another two did not include our sources in the field of view; the other 10 observations are listed in Table \ref{obs_table}. (See \citealt{2016..Kuntz} for a full catalog and discussion of all the {\it Chandra} sources in M\,51.) We downloaded the {\it Chandra} data from the public archives and re-processed them using standard tasks within the Chandra Interactive Analysis of Observations ({\small{CIAO}}) Version 4.7 software package \citep{2006SPIE.6270E..1VF}. Any intervals with high particle backgrounds were filtered out. We extracted spectra and light-curves for ULX-1 and ULX-2 using circular regions of $\approx$4\arcsec~radii and local background regions three times as large as the source regions. For each observation, background-subtracted light curves were created with the {\small{CIAO}} task {\it dmextract}. Spectra were extracted with {\it specextract}, and were then grouped to a minimum of 15 counts per bin, for $\chi^2$ fitting.


M\,51 was also observed by {\it XMM-Newton} nine times between 2003 and 2011, although no data were recorded on three occasions due to strong background flaring (Table \ref{obs_table}). We downloaded the {\it XMM-Newton} data from NASA's High Energy Astrophysics Science Archive Research Centre (HEASARC) archive. We used the European Photon Imaging Camera (EPIC) observations and re-processed them using standard tasks in the Science Analysis System ({\small SAS}) version 14.0.0 software package; we filtered out high particle background exposure intervals. Due to the lower spatial resolution of {\it XMM-Newton}/EPIC, the ULXs cannot be entirely visually resolved, although the elongated appearance of the EPIC source is consistent with the two separate {\it Chandra} sources (as discussed in Section \ref{eclipses_sec}). For each observation we extracted a single background-subtracted light-curve and spectrum for both sources combined, using a circular extraction region of 20\arcsec\ radius, and a local background region that is at least three times larger, does not fall onto any chip gap and is of similar distance to the readout nodes as the source region. Standard flagging criteria \verb|#XMMEA_EP| and \verb|#XMMEA_EM| were used for pn and MOS respectively, along with \verb|FLAG=0|. We also selected patterns 0--4 for pn and 0--12 for MOS. For our timing study, we extracted light-curves with the {\small SAS} tasks {\it evselect} and {\it epiclccorr}. For our spectral study, we extracted individual pn, MOS1 and MOS2 spectra with standard {\it xmmselect} tasks; whenever possible, we combined the pn, MOS1 and MOS2 spectra of each observation with {\it epicspeccombine}, to create a weighted-average EPIC spectrum. In some observations, the pn data were not usable because the source falls onto a chip gap; in those cases, we used only the MOS1 and MOS2 data in {\it epicspeccombine}. Finally, we grouped the spectra to a minimum of 20 counts per bin so that we could use Gaussian statistics. 

For both {\it Chandra} and {\it XMM-Newton} data, spectral fitting was performed with {\small XSPEC} version 12.8.2 \citep{1996ASPC..101...17A}. Timing analysis was conducted with standard {\small FTOOLS} tasks \citep{1995ASPC...77..367B}, such as {\it lcurve}, {\it efsearch} and {\it statistics}. Imaging analysis was done with HEASARC's {\small {DS9}} visualization package, and adaptive image smoothing with {\small{CIAO}}'s {\it csmooth} routine.

\begin{figure}
\centering
\includegraphics[width=0.48\textwidth]{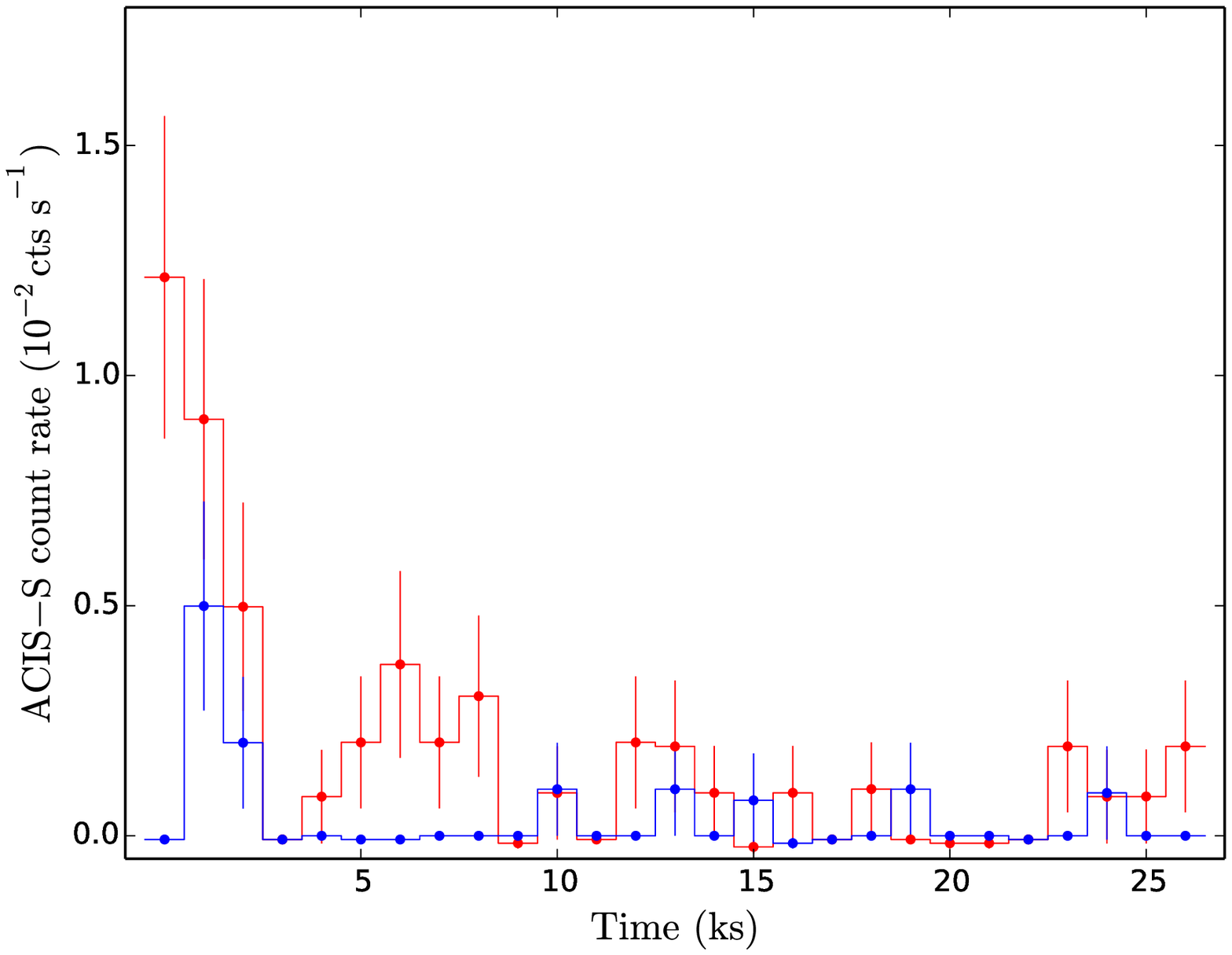}
\includegraphics[width=0.48\textwidth]{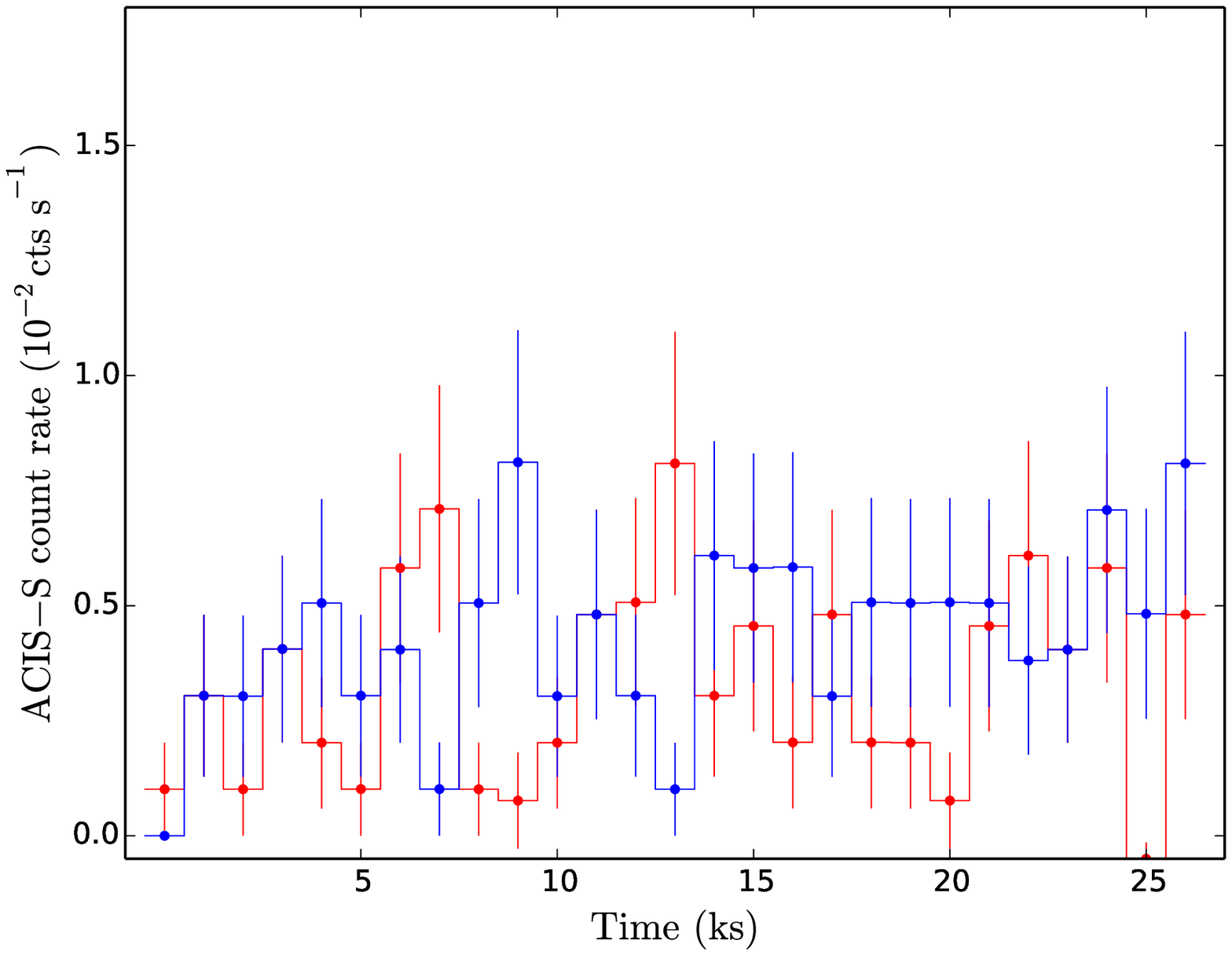}
 \caption{Top panel: {\it Chandra}/ACIS-S background-subtracted light-curve of ULX-1 from observation 1622, split into a soft band (0.3--1.2 keV: red datapoints) and a hard band (1.2--7.0 keV; blue datapoints). It shows a sharp drop in flux about 2 ks into the observation. The data are binned into 1000-s intervals. Bottom panel: as in the top panel, for ULX-2 in the same observation.}
  \label{1622_lc}
  \vspace{0.3cm}
\end{figure}

\begin{figure*}
\centering
\includegraphics[width=0.48\textwidth]{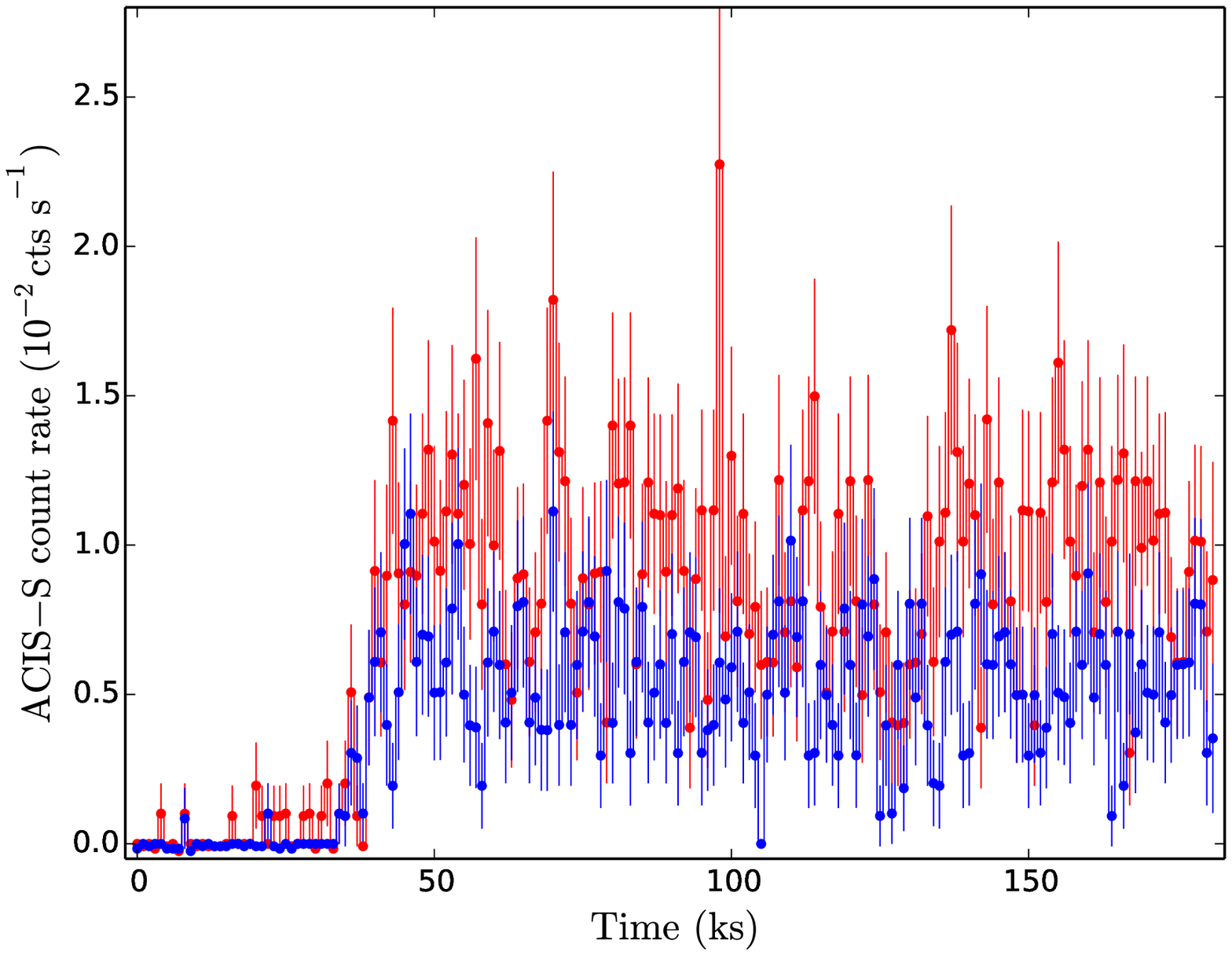}
\includegraphics[width=0.48\textwidth]{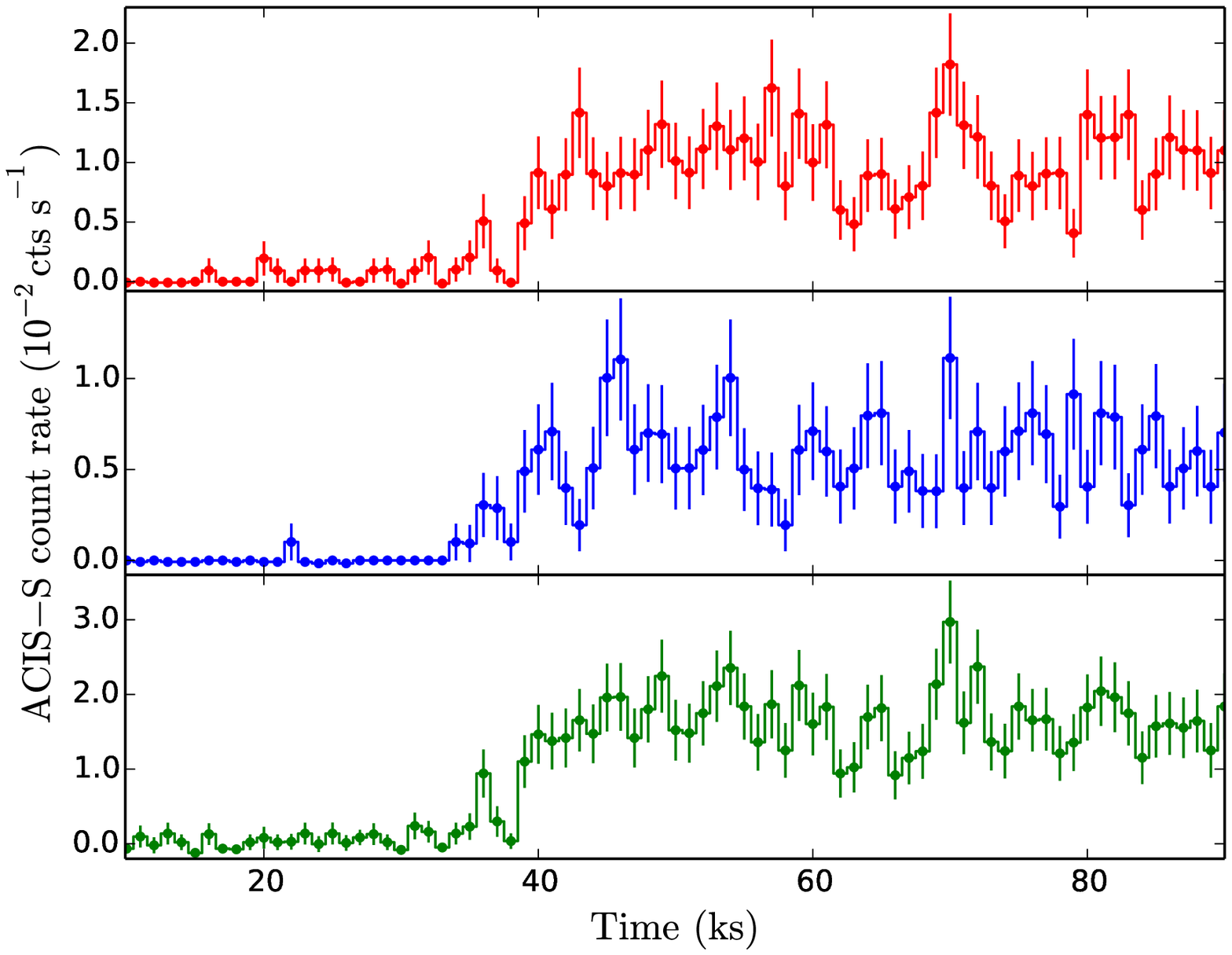}\\
\includegraphics[width=0.48\textwidth]{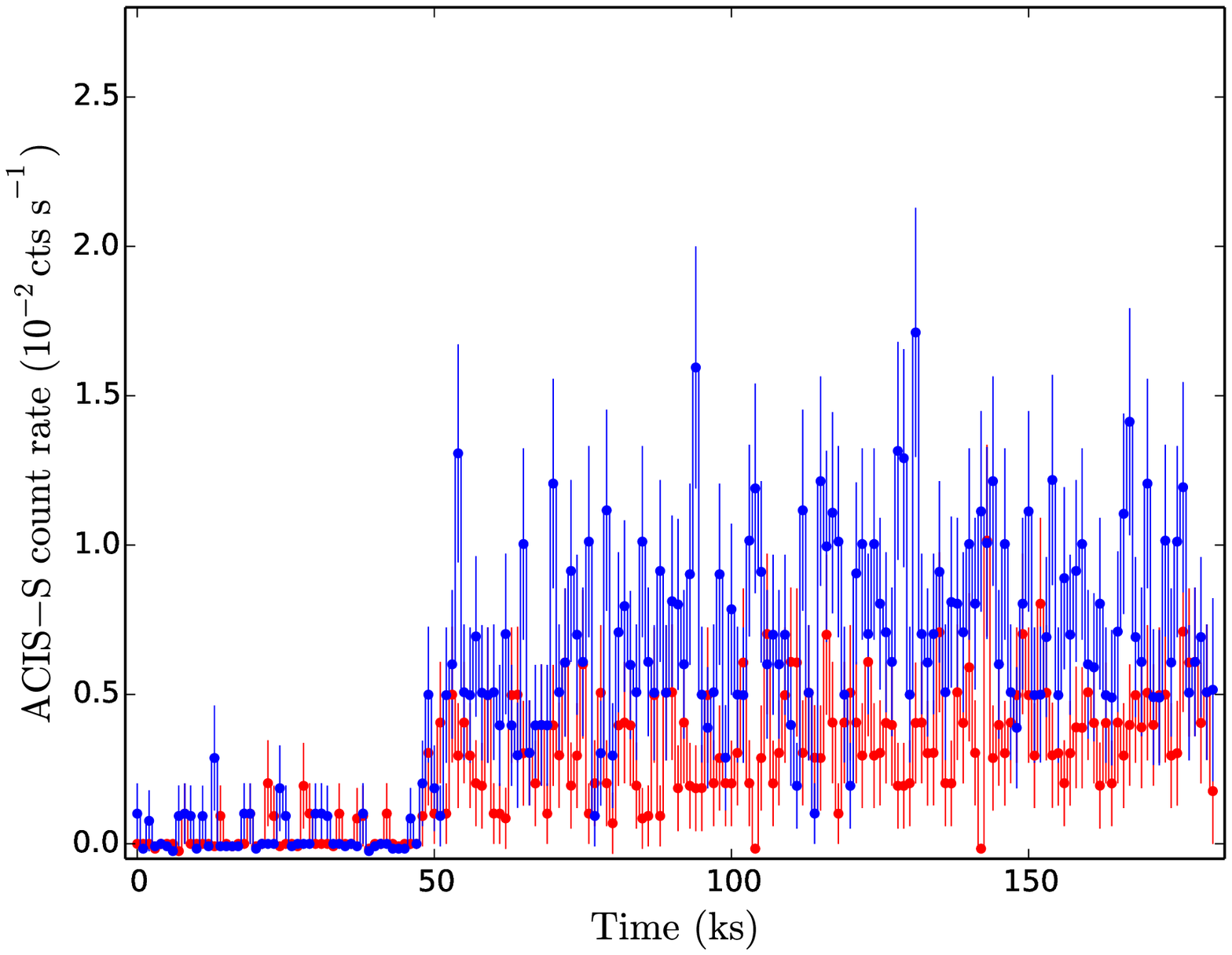}
\includegraphics[width=0.48\textwidth]{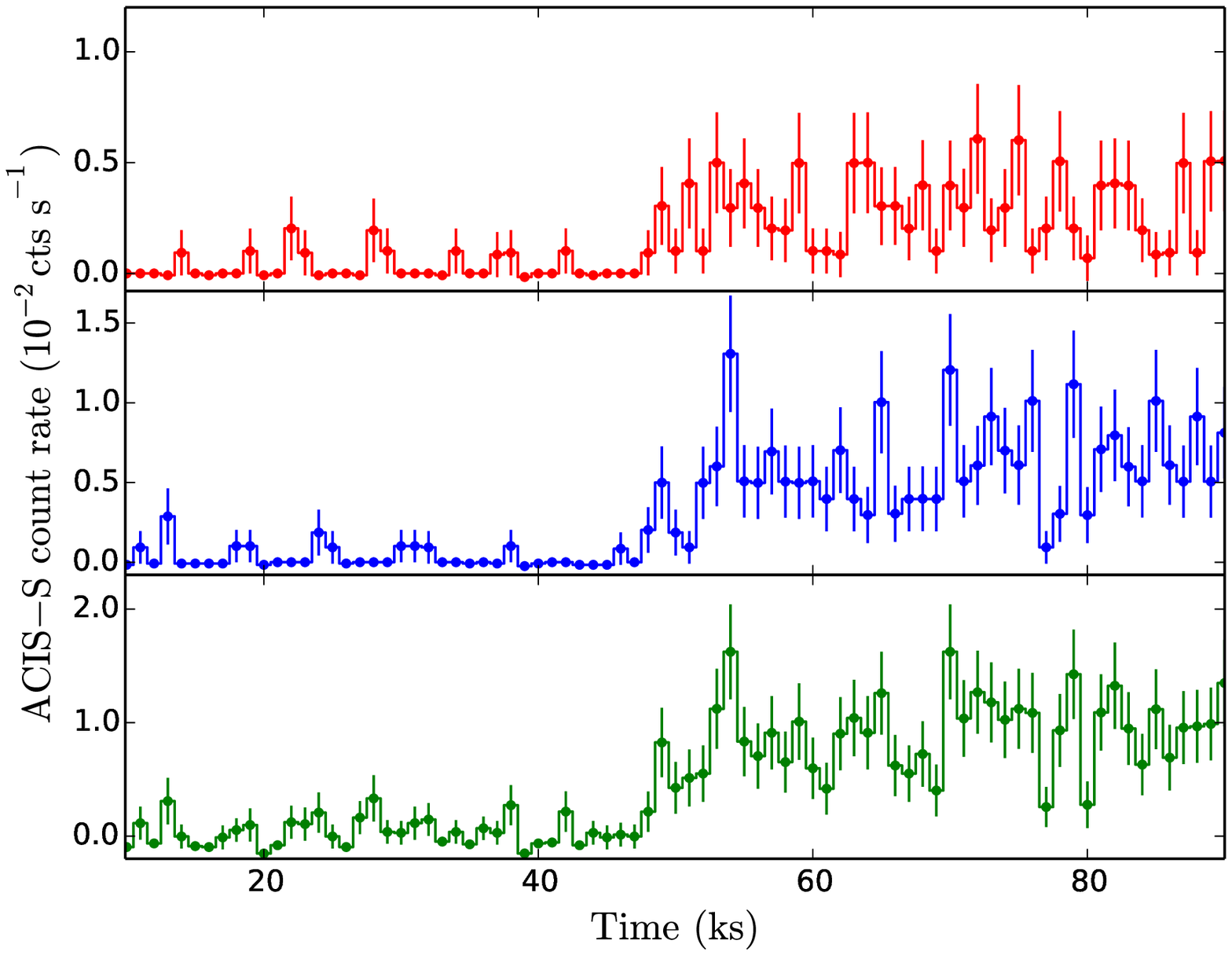}\\
 \caption{Top left panel: {\it Chandra}/ACIS-S background-subtracted light-curves of ULX-1 during ObsID 13813 (red for the 0.3--1.2 keV band, blue for the 1.2--7.0 keV band), showing the end of an eclipse about 40 ks into the observation. The data are binned into 1000-s intervals. Top right panel: soft (red curve, 0.3--1.2 keV), hard (blue curve, 1.2--7.0 keV) and total (green curve, 0.3--7.0 keV) {\it Chandra}/ACIS-S background-subtracted light-curves of ULX-1 during ObsID 13813, zoomed in around the time of eclipse egress. The data are binned into 1000-s intervals. Bottom left panel: as in the top left panel, for ULX-2 during the same {\it Chandra} observation showing the end of an eclipse about 50 ks into the observation. Bottom right panel: as in the top right panel, for ULX-2 during ObsID 13813, zoomed-in around the time of eclipse egress.}
  \label{13813_lc}
  \vspace{0.3cm}
\end{figure*}

\begin{figure}
\centering
\includegraphics[width=0.48\textwidth]{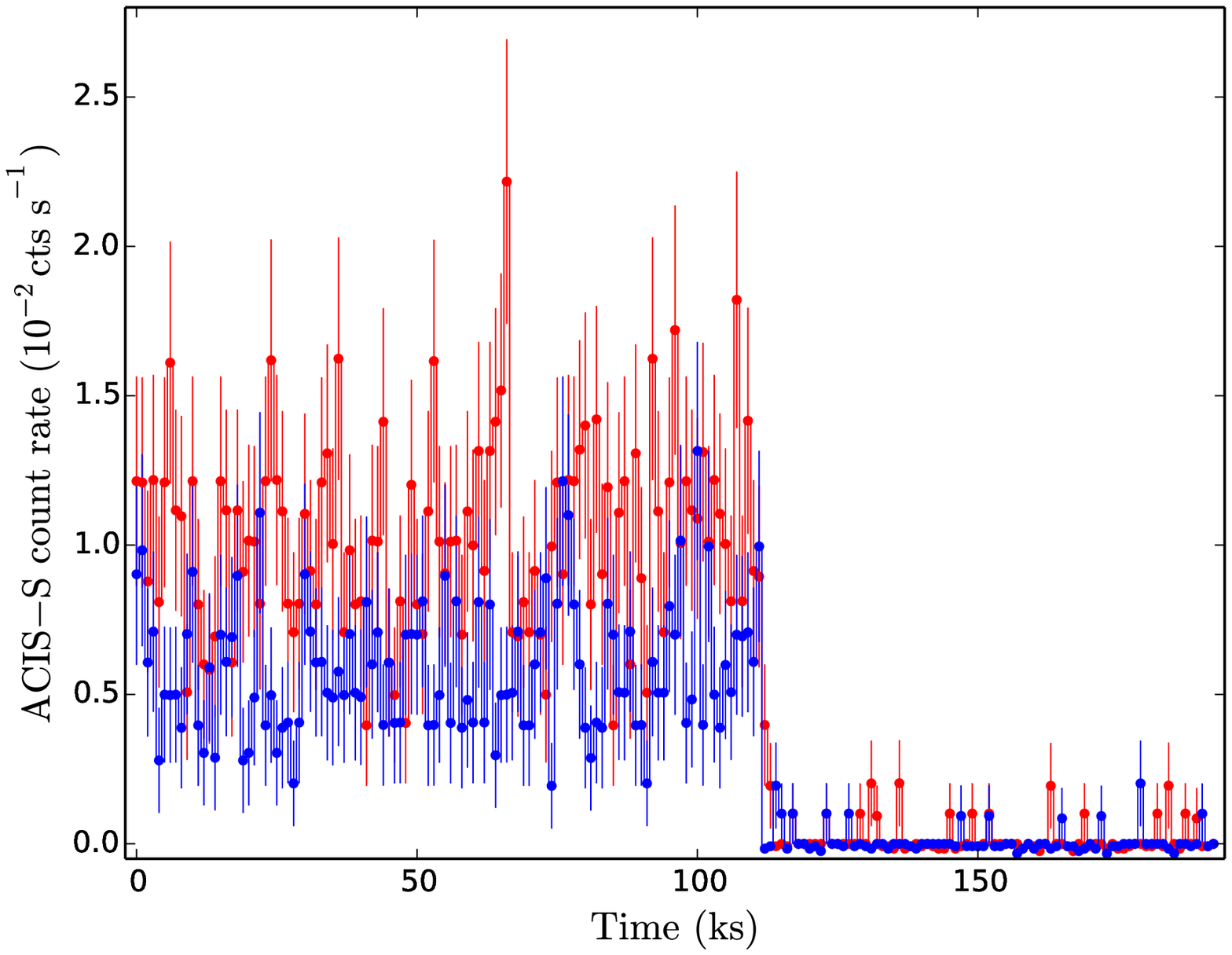}
\includegraphics[width=0.48\textwidth]{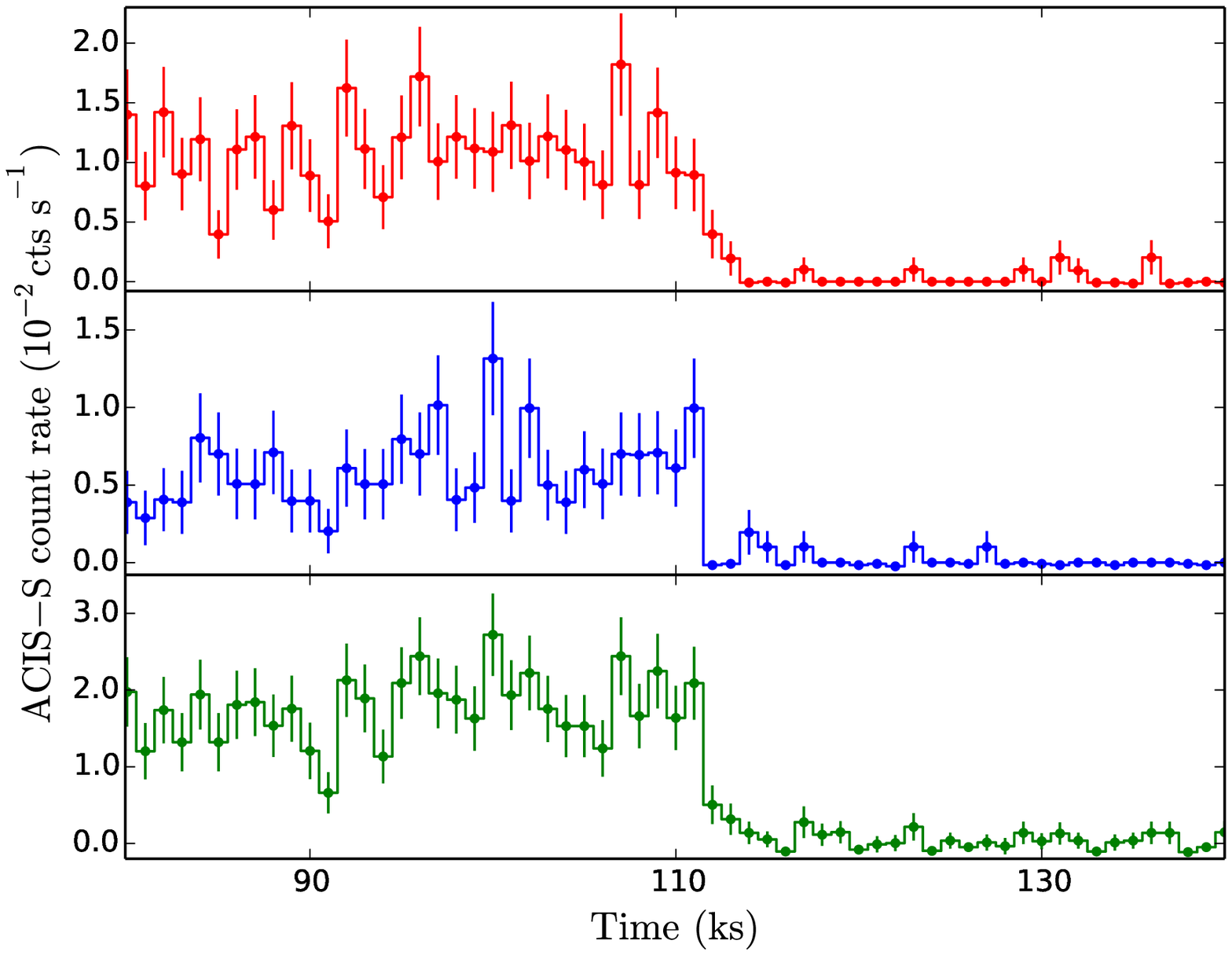}
\includegraphics[width=0.48\textwidth]{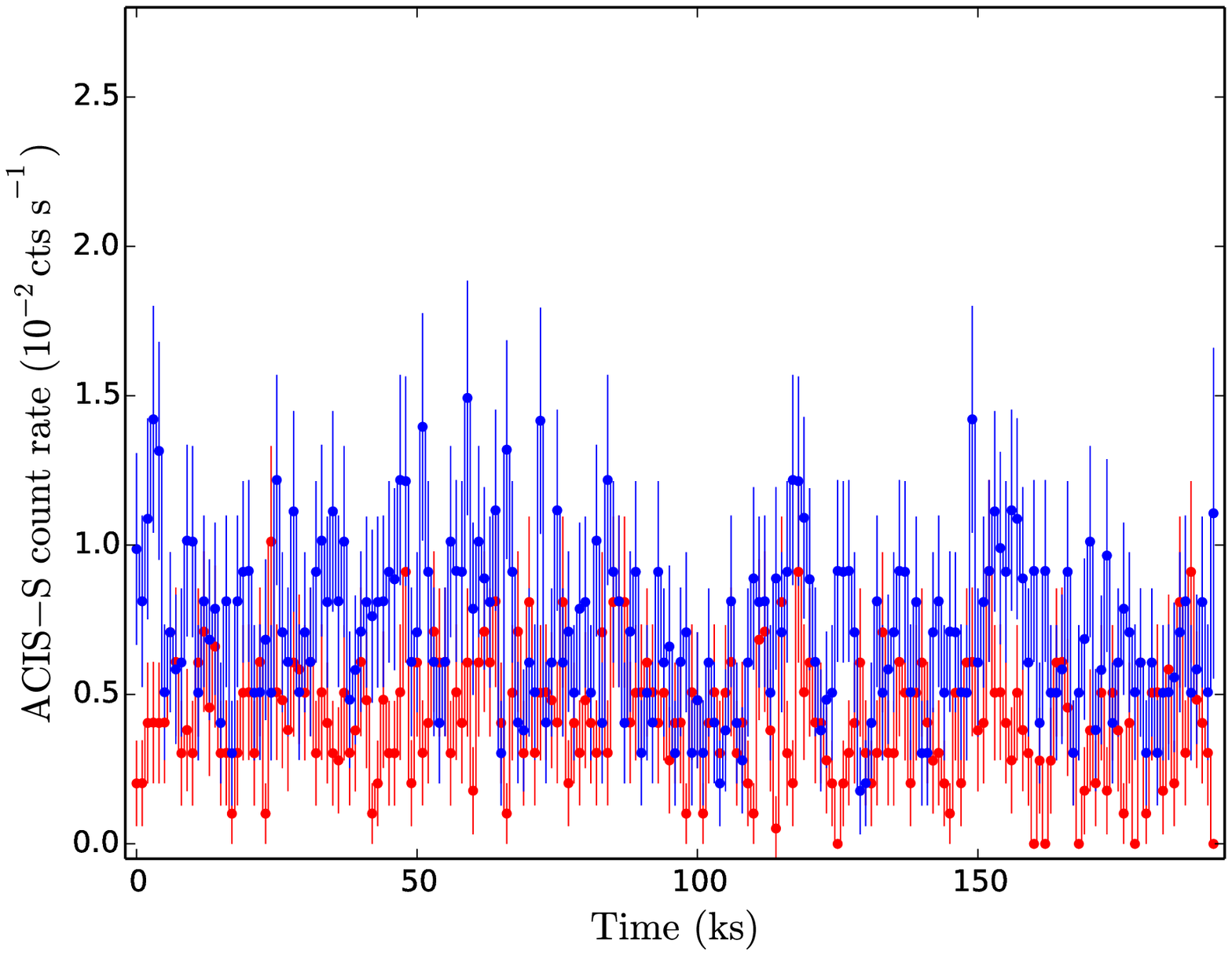}
 \caption{Top panel: {\it Chandra}/ACIS-S background-subtracted light-curves of ULX-1 during ObsID 13814 (red for the 0.3--1.2 keV band, blue for the 1.2--7.0 keV band), showing the beginning of an eclipse about 110 ks into the observation. The data are binned into 1000-s intervals. Middle panel: soft (red curve, 0.3--1.2 keV), hard (blue curve, 1.2--7.0 keV) and total (green curve, 0.3--7.0 keV) {\it Chandra}/ACIS-S background-subtracted light-curves of ULX-1 during ObsID 13814, zoomed in around the time of eclipse ingress. The data are binned into 1000-s intervals. Bottom panel: as in the top panel, for ULX-2 during the same {\it Chandra} observation.}
  \label{13814_lc}
  \vspace{0.3cm}
\end{figure}

\section{Results}

\subsection{Eclipses} \label{eclipses_sec}

\subsubsection{ULX-1 eclipses and dips in the {\it Chandra} data} \label{ulx1_ec_sec}

From our inspection of the {\it Chandra} light-curves, we have discovered 3 epochs (ObsIDs 1622, 13813 and 13814) in which the flux of ULX-1 is strongly reduced for at least part of the observation (Table \ref{ec1_table} and Figures \ref{1622_lc}, \ref{13813_lc}, \ref{13814_lc}). The transition between the long-term-average flux level and the lower level occurs too quickly ($\Delta t \sim 10^3$ s) to be explained by a state transition in the inflow, or a change in the mass accretion rate. Our identification of the low state in ObsID 1622 as a true stellar eclipse rather than a dip may be debatable, given that the flux drop happens right at the start of the observation; however, the presence of eclipses is very clear in ObsIDs 13813 (2012 September 9) and 13814 (2012 September 20), which show a low-to-high and a high-to-low transition, respectively. We also checked that ULX-1 is not at the edge of the chip, there are no instrumental glitches, and no other source in the field has a count-rate step change at the same time. We conclude that the simplest and most logical explanation is an eclipse of the X-ray emitting region by the donor star.  The flux during the eclipse is not exactly zero: by stacking the time intervals during eclipses, we can find a faint but statistically significant residual emission, softer than the emission outside eclipses. We will discuss the spectrum of the residual emission in Section \ref{spectral_sec}.

The way ULX-1 enters the eclipse in ObsID 13814 (Figure \ref{13814_lc}) is also interesting. 
The transition to eclipse in the soft band (0.3--1.2 keV) appears less sharp than the transition in the hard band (1.2--7.0 keV): the soft-band count rate drops to effectively zero in $\approx$4 ks, while the same transition happens in $\la$1 ks for the hard band. This can be explained if the softer X-ray photons come from a more extended region that takes longer to be completely occulted than the effectively point-like central region responsible for the harder X-ray photons \citep{2001A&A...365L.282B}; for example, the softer emission may have contributions from the outer, cooler parts of an outflow. However, we cannot rule out that the discrepancy is simply due to small-number statistics.

\begin{figure}
\centering
\includegraphics[width=0.48\textwidth]{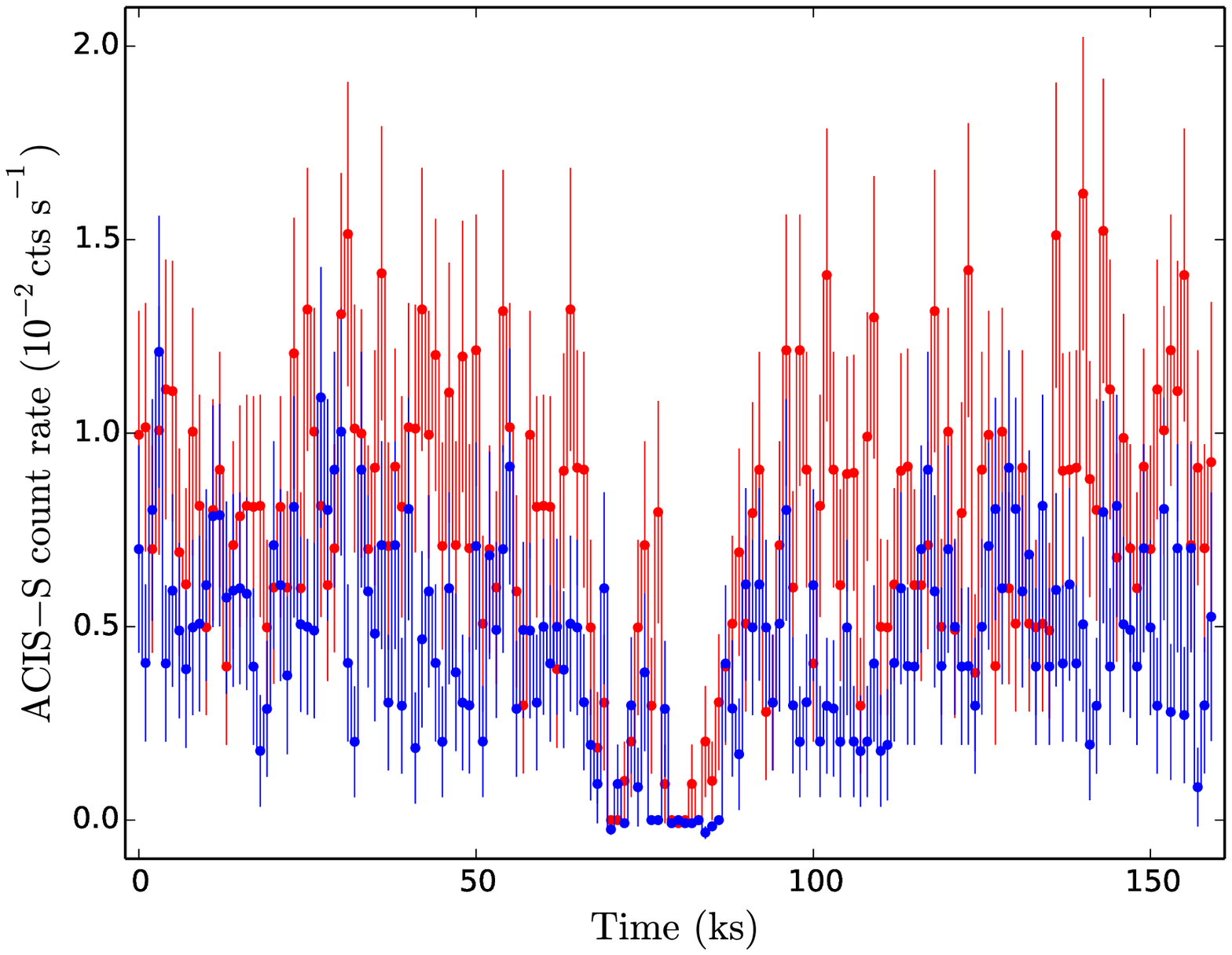}
\includegraphics[width=0.48\textwidth]{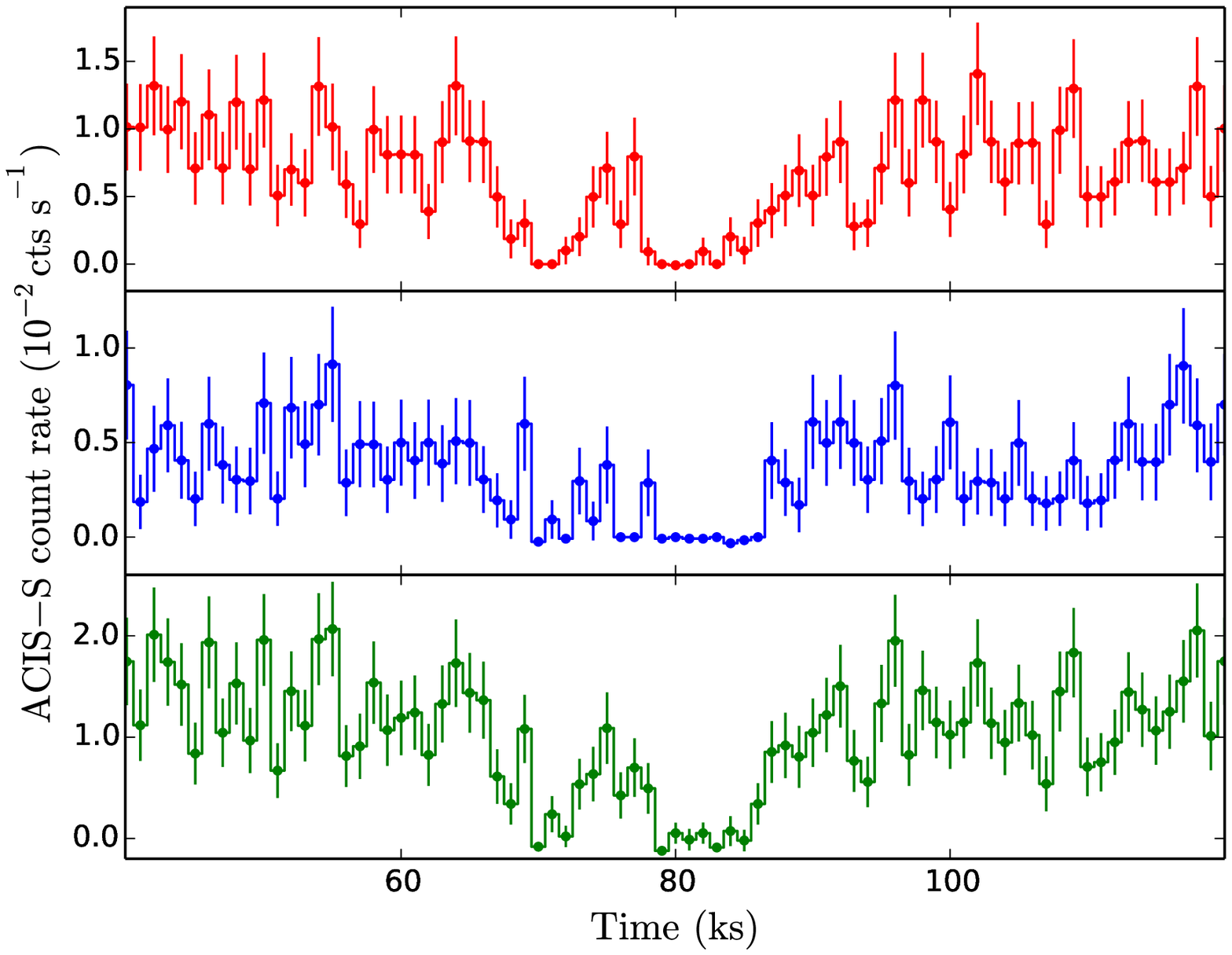}
\includegraphics[width=0.48\textwidth]{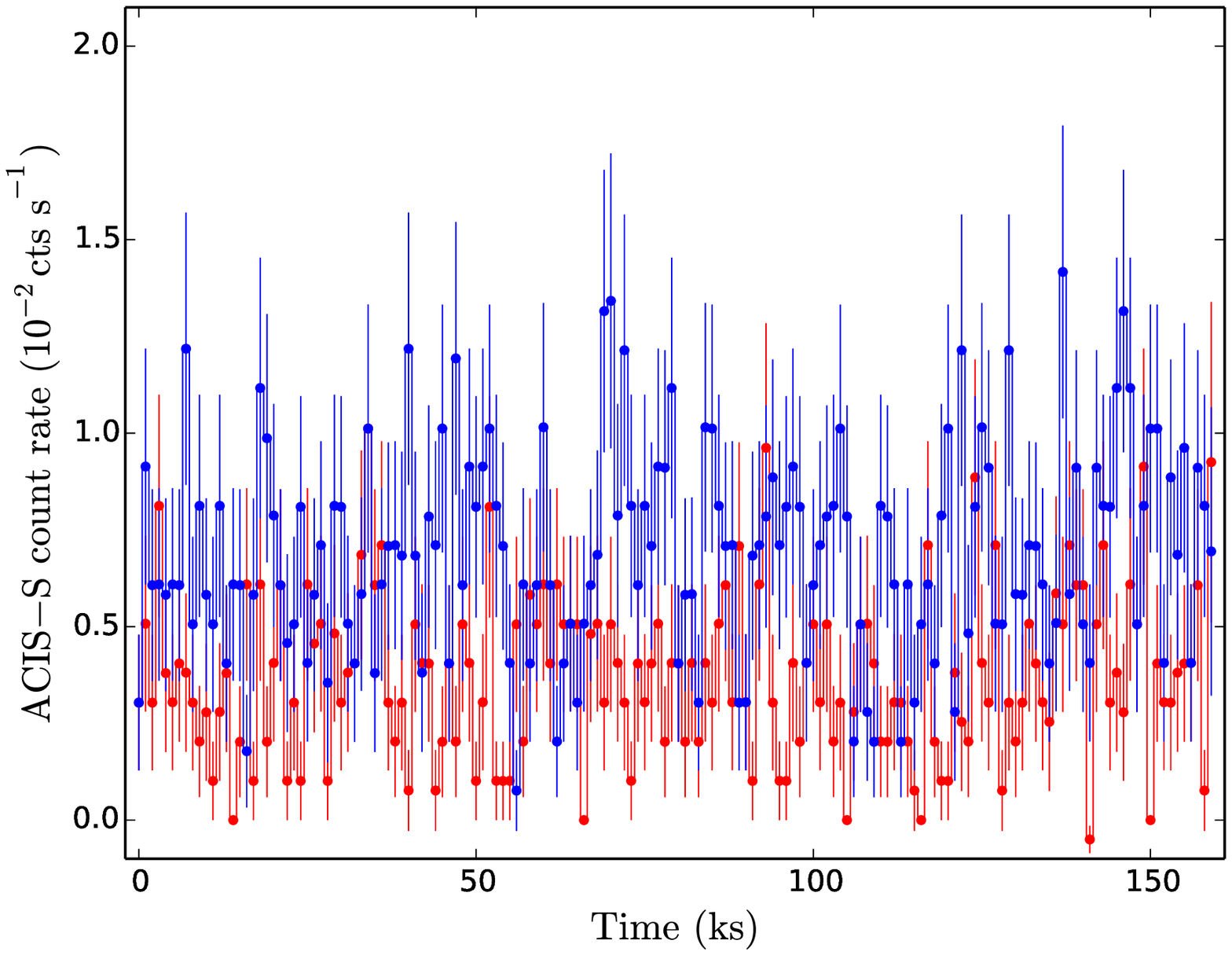}
 \caption{Top panel: as in Figure \ref{13814_lc}, for the {\it Chandra}/ACIS-S observation 13812, showing a dip around 70--90 ks into the observation. All data in this panel and in those below are binned to $1000\,\second$. Middle panel: zoomed-in view of the dip in the soft band (red datapoints), hard band (blue datapoints) and total band (green datapoints). Botttom panel:  as in Figure \ref{13814_lc}, for observation 13812.}
  \label{13812_lc}
  \vspace{0.3cm}
\end{figure}

Finally, we find a deep dip in the {\it Chandra} light curve of ULX-1 during ObsID 13812 (Figure \ref{13812_lc}). The count rate drops to zero and then recovers to the pre-dip level, just like during an eclipse. However, the short duration ($\approx$20 ks) and double-dipping substructure of this phase suggest that this occultation is not due to the companion star; we suggest that it is more likely the result of lumps or other  inhomogeneities in the thick outer rim of the disk, or is caused by the accretion stream overshooting the point of impact in the outer disk and covering our view of the inner regions \citep{1987A&A...178..137F,1996ApJ...470.1024A}.
Analogous X-ray dips are seen in several Galactic X-ray binaries \citep[e.g.,][]{1982ApJ...253L..61W,2001A&A...365L.282B,2003A&A...412..799H,2006A&A...445..179D} and are interpreted as evidence of a high viewing angle.
Assuming that the occultation is produced by a geometrically thick structure in Keplerian rotation, we can estimate the angular extent of this feature by scaling the duration of the dipping phase to the binary period of ULX-1. If the period is $\approx$6\,d (see Section \ref{ulx1_bp_sec}), the occulting structure spans $\Delta \phi \approx 14^{\circ}$; for a $\approx$13\,d period, $\Delta \phi \approx 6^{\circ}$.

\subsubsection{ULX-2 eclipse in the {\it Chandra} data}

In the same set of {\it Chandra} observations, we also discovered one eclipse in ULX-2, in observation 13813 (Figure \ref{13813_lc}, bottom panel). The abrupt nature of the transition from low to high count rates once again suggests that we are looking at an occultation by the companion star. Remarkably, the egress from the eclipse of ULX-2 happens only $\approx$8ks later than the egress from the ULX-1 eclipse, at MJD 56180.30 and 56180.20, respectively (cf.~bottom and top panels of Figure \ref{13813_lc}). The small but significant time difference guarantees that the two count-rate jumps seen in the two ULXs are not instrumental anomalies but real physical events. Moreover, we did extensive checks on other bright X-ray sources in the same ACIS-S3 chip, and found that none of them shows similar jumps around that time; this also rules out instrumental problems. We also examined the light-curves of ULX-2 in all other {\it Chandra} observations, including those where eclipses or dips were found in the light-curve of ULX-1 (bottom panels of Figures \ref{1622_lc}, \ref{13814_lc}, and \ref{13812_lc}). We found no other unambiguous eclipses or deep dips.

\begin{figure}
\centering
\includegraphics[width=0.48\textwidth]{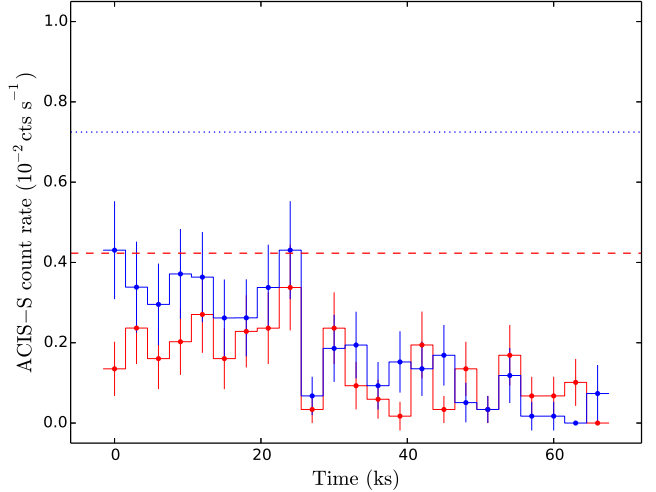}
 \caption{{\it Chandra}/ACIS-S background-subtracted light-curves for ULX-2 during observation 13815; red datapoints are for the 0.3--1.2 keV band, blue datapoints for the 1.2--7.0 keV band. Data are binned to 300 s. As a comparison, the dashed and dotted lines represent the average count rates for the soft and hard band, respectively, during the previous {\it Chandra} observation, ObsID 13814, taken 3 days earlier.}
  \label{13815_lc}
  \vspace{0.3cm}
\end{figure}

ULX-2 does show significant intra-observational variability in ObsID 13815. Throughout the 67-ks observation, the source displays a much lower count rate than its average out of eclipse count rate, in both the soft and the hard band (Table 3 and Figure \ref{13815_lc}). 
The count rate further decreases during that {\it Chandra} epoch, until it becomes consistent with a non-detection at the end of the observation. The decrease is slow enough (compared with the eclipse in ObsID 13813, Figure \ref{13813_lc}) to rule out a stellar occultation.
We do not have enough evidence or enough counts to test whether this flux decrease is due to intrinsic variability of ULX-2, or to an increased absorption by colder material in the outer disk. 
As usual, we checked the behaviour of ULX-1 and other bright sources in ObsID 13815 to ascertain that the lower count rate seen from ULX-2 is not an instrumental problem.

\subsubsection{ULX-2 eclipse in the {\it XMM-Newton} data}

\begin{figure}
\centering
\includegraphics[width=0.47\textwidth]{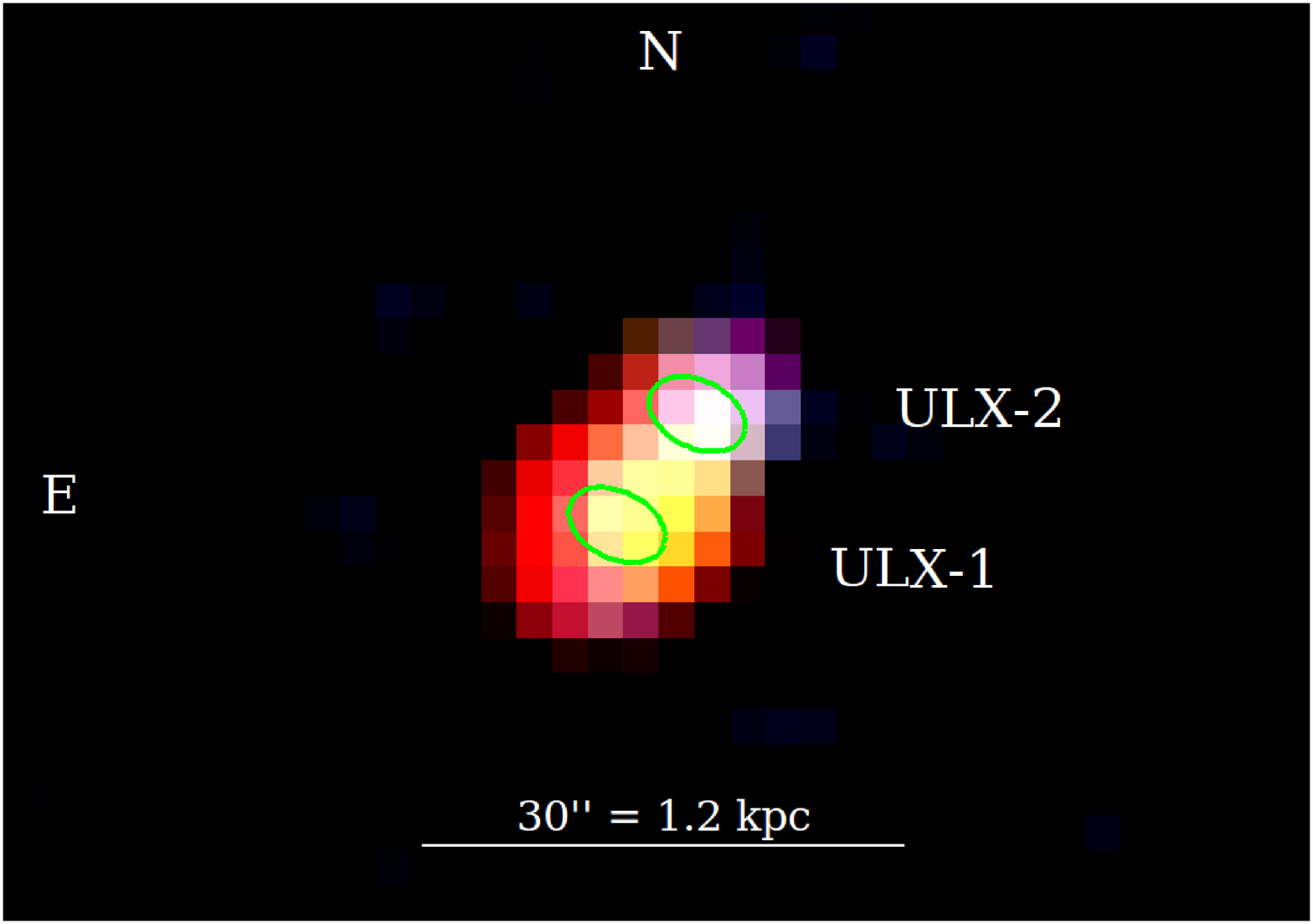}\\[1pt]
\includegraphics[width=0.47\textwidth]{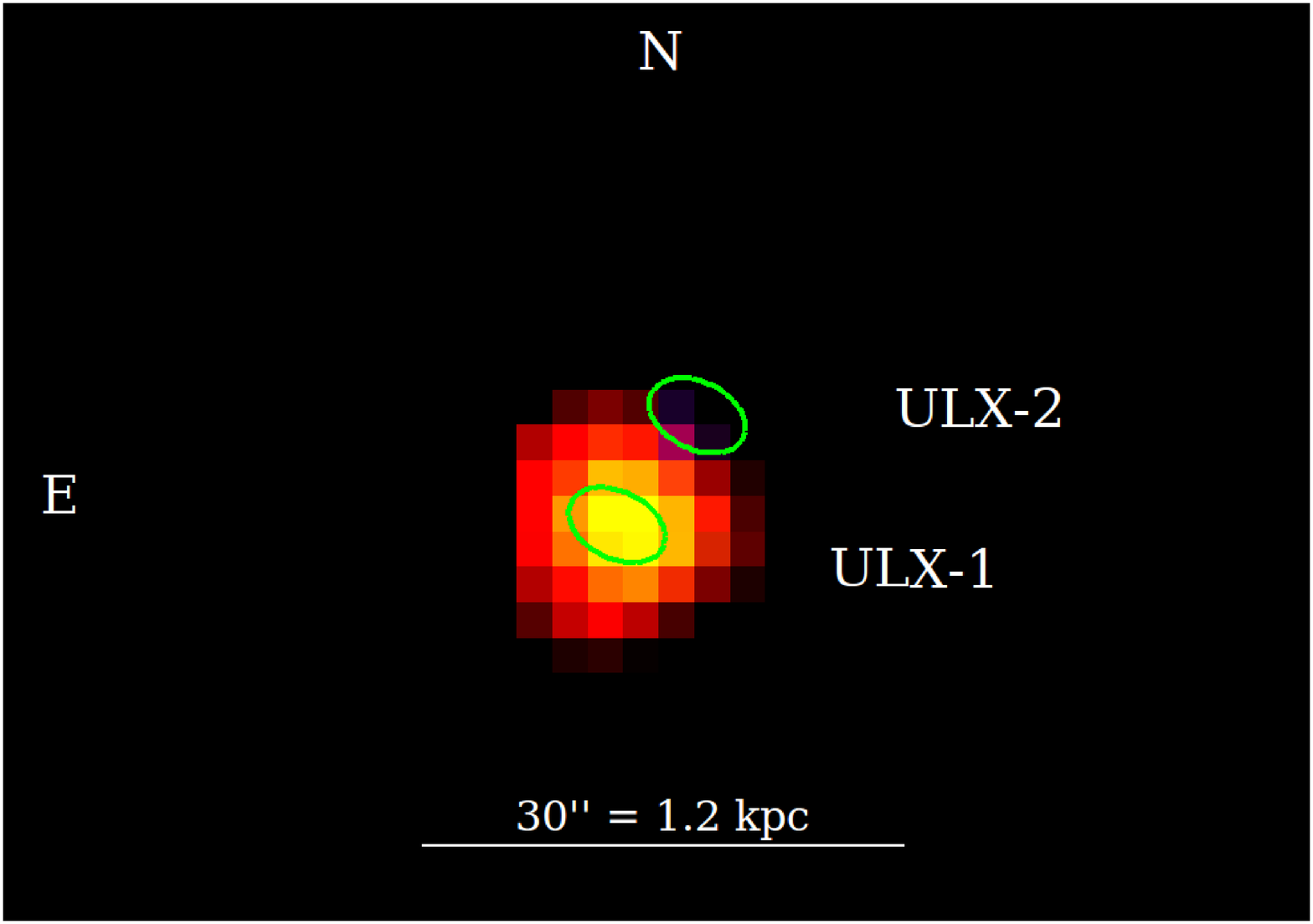}
 \caption{Top panel: stacked {\it XMM-Newton} MOS1+MOS2 image for non-eclipse interval of observation 0303420101. 
 Red represents photons in the $0.3$--1 keV band, green is for 1--2 keV and blue is for 2--7 keV. The green ellipses indicate the location of ULX-1 and ULX-2 as determined from the {\it Chandra}/ACIS-S images; their point spread functions appear elongated because the ULXs were observed a few arcmin away from the ACIS-S3 aimpoint. The two sources are not clearly resolved by {\it XMM-Newton}, but the color difference between the two ends of the peanut-shaped EPIC-MOS source is consistent with the color and spectral differences seen by {\it Chandra}. Bottom panel: as in the top panel, but for the ULX-2 eclipse interval of observation 0303420101.}
  \label{XMM_image}
  \vspace{0.3cm}
\end{figure}

\begin{figure}
\centering
\includegraphics[width=0.48\textwidth]{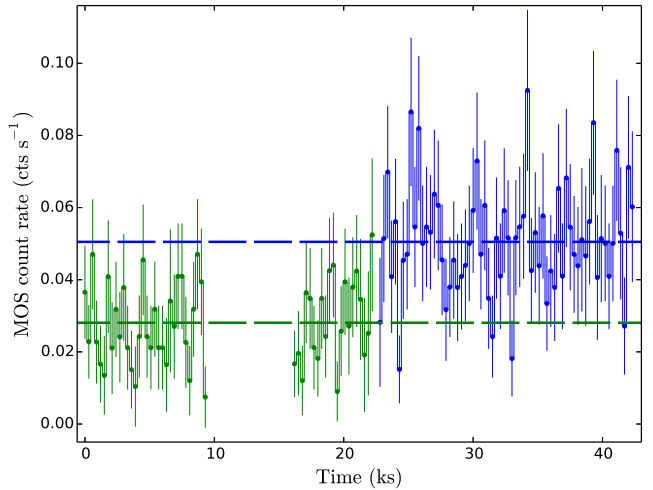}
 \caption{Background-subtracted {\it XMM-Newton}/EPIC MOS1+MOS2 light-curve for the unresolved ULX source in observation 0303420101. Time intervals affected by background flaring have been removed. The light-curve was extracted in the $0.2$--8 keV band and the datapoints have been binned to 300 s for display purposes.  
 The light-curve is broken into two sections: the first $\approx$22 ks (green datapoints) have a lower count rate and correspond to an eclipse of ULX-2; in the remaining $\approx$20 ks, both ULXs are out of eclipse (blue datapoints). Dotted lines indicate the average count rates for the two sub-intervals.} 
  \label{XMM_lc}
  \vspace{0.3cm}
\end{figure}

We then searched for possible eclipses of either ULX-1 or ULX-2 during the {\it XMM-Newton} observations. 
Due to the poorer spatial resolution of EPIC relative to ACIS-S, ULX-1 and ULX-2 are not completely resolved by {\it XMM-Newton}; however, the point spread function in the combined EPIC MOS1+MOS2 images is clearly peanut-shaped, consistent with the position and relative intensity of the two {\it Chandra} sources, and the upper source (ULX-2) has significantly harder colors (Figure \ref{XMM_image}, top panel). Firstly, we extracted and examined background-subtracted EPIC light-curves for the combined emission of the two unresolved sources, for each {\it XMM-Newton} observation. Because the two sources have comparable count rates (Tables 2 and 3), an eclipse in either source would cause the observed count rate to drop by a factor of $\approx$2. This is the scenario we find in observation 0303420101: there is an apparent increase in the observed EPIC-MOS count rate by a factor of $\approx$2, some 22 ks from the start of the observation, which we tentatively interpret as the egress from an eclipse (Figure \ref{XMM_lc}, top panel), superposed on short-term intrinsic variability. Unfortunately we cannot use EPIC-pn data for this crucial epoch, because the source falls onto a chip gap. To quantify the step change in the count rate between the first and second part of the observation (green and blue datapoints in Figure \ref{XMM_lc}), we performed a Kolmogorov-Smirnov (KS) test on the two distributions of datapoints, to determine whether they are drawn from different populations. We find a KS statistic of $0.65$ and p-value of $3.8\times10^{-11}$, suggesting that the two sections of the light-curve are indeed statistically different.
The average MOS1+MOS2 net count rate in the ``eclipse" part of the light-curve is $\approx 0.026 \pm 0.003$ ct s$^{-1}$ (90\% confidence limit), while in the ``non-eclipse" part it is $\approx 0.046 \pm 0.003$ ct s$^{-1}$.  Having ascertained from the X-ray light-curve that ObsID 0303420101 probably includes an eclipse, we extracted MOS1+MOS2 images from the low-rate and high-rate sections of that observation, and confirmed (Figure \ref{XMM_image}) that in the low-rate interval, the emission from ULX-2 is missing. 

We extracted and inspected the light-curves of every other {\it XMM-Newton} observation. No eclipses of ULX-1 and no further eclipses of ULX-2 were detected; however, several of those observations are much shorter than the typical {\it Chandra} observations, and the background count rate is much higher in the EPIC cameras. Thus, ruling out the presence of an eclipse as opposed to intrinsic variability is no easy task in some of the {\it XMM-Newton} observations.




\subsection{Constraints on the binary period of ULX-1} \label{ulx1_bp_sec}

We noted (Table \ref{ec1_table} and Section \ref{ulx1_ec_sec}) that for ULX-1, two fractions of eclipses are seen $\approx$12 days apart, in ObsID 13813 and ObsID 13814. The egress from the eclipse in ObsID 13813 occurs at MJD 56180.21; the ingress into the eclipse in ObsID 13814 occurs at MJD 56191.64. 
This enables us to place some constraints on binary period, which must be,
\begin{gather}
P \approx \frac{(11.43 + {\rm eclipse~duration})}{n} \ \, {\rm days}, \label{bp_eq}  
\end{gather}
with $n \ge 1$. To refine this constraint, we take into account that the minimum duration of an eclipse is $\approx$90 ks ($\approx$1.0 days), as observed in ObsID 13814. We also know that the maximum duration of an eclipse is $\approx$150 ks ($\approx$1.7 days) as this is the time between the start of the eclipse in ObsID 13814 and the start of the next observation, ObsID 13815, which has no eclipse. Assuming the shortest possible duration of the eclipse implies a binary period of $\approx$12.5$/n$ days. If we use the maximum eclipse time, $\approx$1.7 days, the binary period is $\approx$13.1$/n$ days.

We tested a range of eclipse durations and binary periods, to determine which combination of parameters is consistent with the observed sequence of eclipses/non-eclipses in our {\it Chandra} observations. Based solely on the minimum duration of an uninterrupted non-eclipse phase ($\approx$160 ks) and the minimum duration of the eclipse ($\approx$90 ks), the minimum acceptable binary period, from Equations \eqref{bp_eq}, is $P\approx$230 ks $\approx$2.7 d (that is, $n=4$). However, if the binary period were $\approx$3 days, the eclipse found during ObsID 13813 implies that another eclipse should be detected in ObsID 13812. The start of ObsID 13812 is only 2.56 days after the end of the eclipse in ObsID 13813. We do not find an eclipse in ObsID 13812, and this rules out a period of $\approx$3 days. Moreover, an eclipse time $\ga$90 ks over a period of about 3 days would imply that ULX-1 should be in eclipse $\ga$30\% of the time. A Roche-lobe filling donor star can eclipse a point-like X-ray source for such a long fraction of the orbit only for mass ratios $q \equiv M_2/M_1 \ga$ a few 100  \citep[Fig.~2 in][]{1976ApJ...208..512C}, which is impossible for any combination of compact objects and normal donor stars. Next, we consider the possibility that $n=3$ in Equation \eqref{bp_eq}, which corresponds to a period range between $\approx$4.08 and $\approx$4.38 days. In this case, too, we would have seen at least part and more likely all of an eclipse in ObsID 13812, which is not the case: this rules out the $n=3$ case, too. Therefore, the only two acceptable options for the binary period are $n=2$ ($P \sim 6$--6.5 days) or $n=1$ ($P \sim 12$--13 days).

\begin{figure}
\centering
\includegraphics[width=0.48\textwidth]{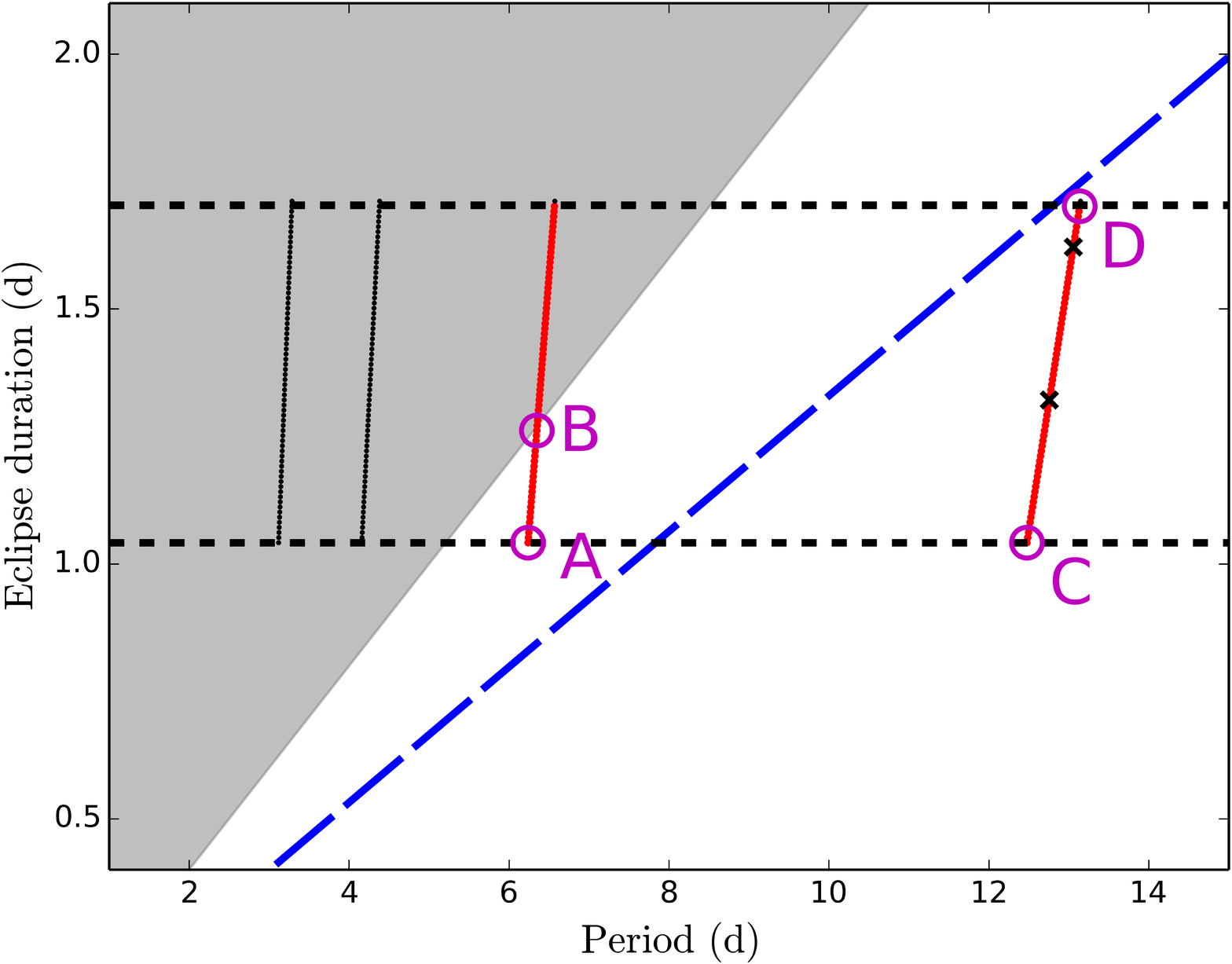}
 \caption{Test of potential binary periods for ULX-1, based on the spacing between observed eclipses in the {\it Chandra} series of observations. We know that $P \approx$(11.43 + eclipse duration)$/n$ days, with $n \ge 1$ and the eclipse duration is between 1.0 days and 1.7 days (horizontal dashed black lines). Each line segment represents a choice of $n$ (from left to right: $n = 4,3,2,1$), and is plotted between the minimum (1.04 days) and maximum (1.70 days) permitted value of the eclipse duration. On each segment, black intervals indicate a combination of period and eclipse duration that is not consistent with the sequence of {\it Chandra} observations; instead, red intervals do fit the observed data. The dashed blue line is the region of the parameter space where the eclipse duration is 13.3\% of the period, which is the observed eclipsing fraction from all {\it Chandra} and {\it XMM-Newton} observations. The grey shaded region marks the region of the parameter space where the eclipse duration is greater than 20\% of the period, which we consider less likely for empirical reasons (too far from the observed value). For each value of the ratio between eclipse duration and binary period, there is a unique value of the mass ratio $q(\theta)$ (see Section 5.3 for details). For $\theta = 90^{\circ}$, points A, B, C, D correspond to $q=3.6,9.7,0.25,1.2$, respectively; the two points marked with crosses correspond to $q(90^{\circ})=0.5$ and $1.0$. 
 Acceptable solutions in the shaded region require mass ratios $q(\theta) \geq q(90^{\circ}) \ga 10$ (Section \ref{ulx1_bp_sec}); such high values are ruled out in the case of a BH accretor, but are still possible if ULX-1 is powered by a neutron star.} 
  \label{ec_test}
  \vspace{0.3cm}
\end{figure}

We summarize the acceptable region of the period versus eclipse duration parameter space in Figure \ref{ec_test}. We iterated over all possible eclipse durations (1.04--1.70 days, in iteration steps of 0.01 days) and for values of $n = 1,2,3,4$, and compared the predicted occurrences of eclipses with what is detected in the seven {\it Chandra} observations between 2012 September 9 to 2012 October 10. Along the line corresponding to each value of $n$, some periods are consistent with the {\it Chandra} data (red intervals), others are ruled out (black intervals). In addition, for Roche-lobe-filling donors, eclipse durations $> 20\%$ of the binary period (dark shaded area in Figure \ref{ec_test}) require a mass ratio $q \ga 8$ at an inclination angle of 90$^{\circ}$, or $q \ga 10$ at an inclination of 80$^{\circ}$ \citep{1976ApJ...208..512C}. This is very implausible if the accretor is a BH, but it is acceptable for a neutron star accreting from an OB star. In the assumption that ULX-1 has a BH primary, the mass-ratio constraint further restricts the viable $n=2$ case to the narrow range $P = 6.23$--6.35 days, with an eclipse duration range of $1.04$--$1.26$ days. If we allow for a neutron star primary, the period can be as long as 6.55 days, corresponding to an eclipse fraction of 26\%. Finally, for the $n=1$ case, the predicted fractional time in eclipse goes from $\approx$8\% ($P = 12.48$\,d, eclipse duration $\approx$1.0\,d) to $\approx$13\% ($P = 13.14$\,d, eclipse duration $\approx$1.7\,d), with mass ratios $q \sim 0.3$--1, more typical of a BH primary orbiting an OB star.

Based on the previous analysis, we compared the predicted eclipse fractions with the total fraction of time ULX-1 was observed in eclipse. Over all {\it Chandra} epochs, the system is seen in eclipse for a total of $\approx$135 ks out of $\approx$835 ks, equating to a total eclipsing fraction of $\approx 16\%$. No eclipses of ULX-1 are significantly detected in 177 ks of {\it XMM-Newton}/EPIC observations; therefore, the combined eclipse fraction observed by {\it Chandra} plus {\it XMM-Newton} becomes $\approx$13.3\%. 
This is slightly lower than the predicted time in eclipse in the case of $n=2$ (fractional eclipse duration $\ga 16.7\%$: Figure \ref{ec_test}). Conversely, for the case of $n=1$, the observed time in eclipse is slightly larger than expected (between $\approx$8\% and $\approx$13\%). We do not regard such discrepancies as particularly significant, because of the limited and uneven sampling of the system; we may have been slightly lucky or slightly unlucky in catching ULX-1 during its eclipses.

\begin{figure}
\centering
\includegraphics[width=0.48\textwidth]{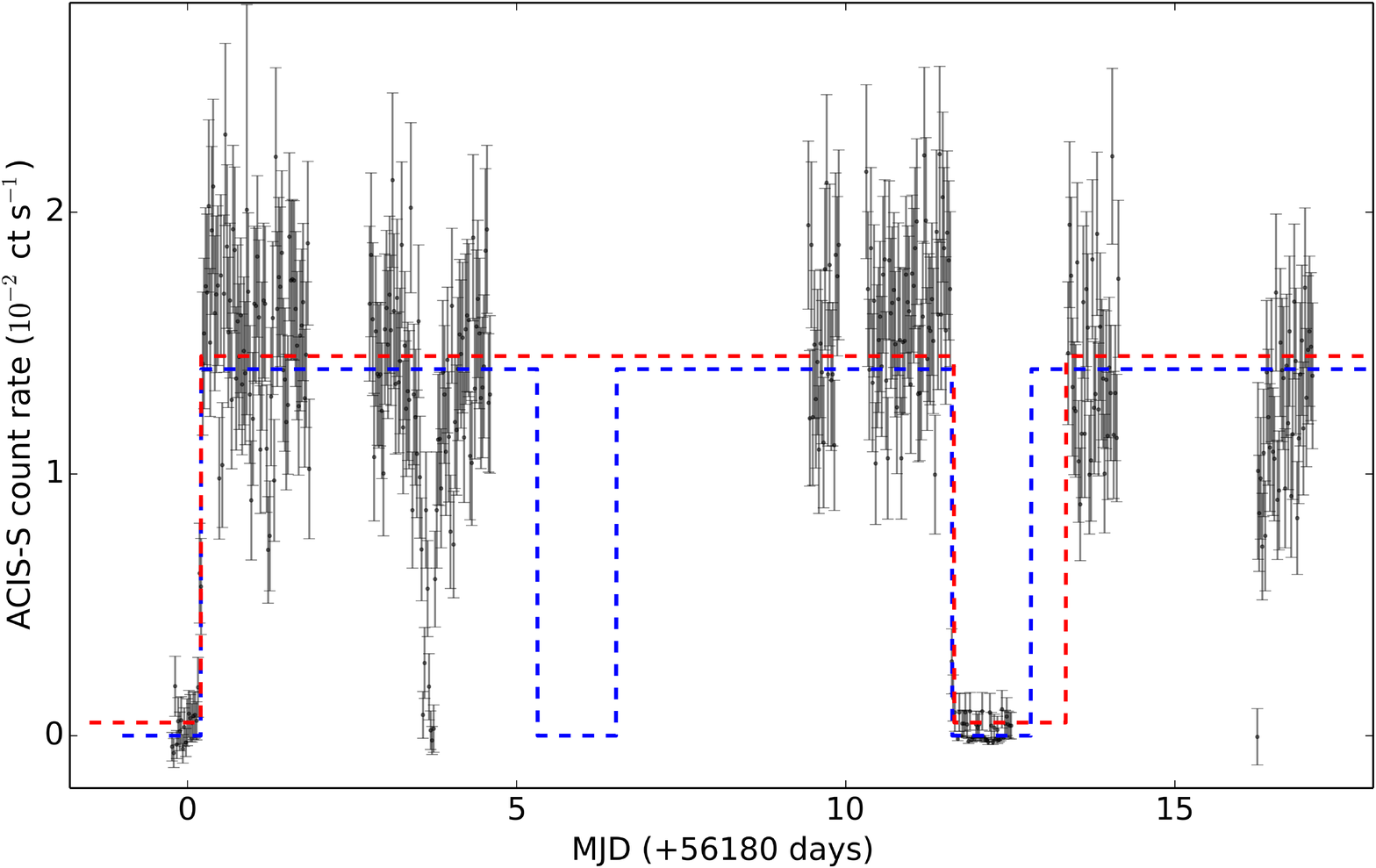}
 \caption{{\it Chandra}/ACIS-S background subtracted light-curve for ULX-1 for all epochs in September 2012 (in chronological order: ObsIDs 13813, 13812, 15496, 13814, 13815, 13816). We overlaid two schematic lightcurves corresponding to two alternative periods consistent with the observations: a $6.3$-day period with a $1.3$-day eclipse (dashed blue line), and a $13.1$-day period with a $1.7$-day eclipse (dashed red line, slightly shifted upwards for clarity). 
 In addition to the two eclipses, two shorter dips are also seen. The second dip appears only in the first data point from the final epoch, but is at approximately the same orbital phase as the first dip with respect to their preceding eclipses. The first 2 datapoints of the final epoch are plotted as 1000-s bins (to highlight the short dip), while all other datapoints are binned to 2000 s.}
  \label{full_lc}
  \vspace{0.3cm}
\end{figure}

We also note that dips in the X-ray flux can sometime provide phasing information in binary systems, if they are caused by bulging, denser material located where the accretion stream splashes onto the disk. For example, regular dips at phases $\sim$0.6--0.7 are sometimes seen in low-mass X-ray binaries \citep[{\it e.g. }EXO 0748-676; ][]{1989ApJ...340.1064L, 2003A&A...412..799H}, and other Roche-lobe overflow systems. We have already mentioned (Section \ref{ulx1_ec_sec}) that ULX-1 shows a dip in ObsID 13812. A second possible dip can also be seen in the full light-curve (Figure \ref{full_lc}) at the start of the final observation, ObsID 13816. Although only detected in a single 1000-s bin (the first 1000 s of the observation), this drop in flux appears to be intrinsic to ULX-1, as other nearby sources do not show this feature and there are no instrumental problems in those first 1000 s. Intriguingly, both dips appear to be at the same phase with respect to the preceding eclipses ({\it i.e.}, $\approx$3.5 days after the eclipse), which strengthens our confidence that the second dip is also real. For a binary period $\approx$6 days, the dips would be at phase $\approx$0.6; for the alternative period range $\approx$12.5--13 days, the dips would be at phase $\approx$0.25--0.30.


Finally, two eclipses were found for ULX-2. Unfortunately, the large time interval between the two eclipses seen by {\it Chandra} in 2012 September and by {\it XMM-Newton} in 2006 May precludes any attempt to constrain the binary period. All we can say is that the total fraction of time spent in eclipse by ULX-2 in our $\approx 1$Ms {\it Chandra} plus {\it XMM-Newton} dataset is $\approx$7\%. Since the minimum eclipse duration is 48 ks ($\approx$0.55 d), we expect the binary period to be $\sim$10 d.



\subsection{Hardness ratios in and out of eclipses} \label{col_sec}

\begin{figure}
\centering
\includegraphics[width=0.47\textwidth]{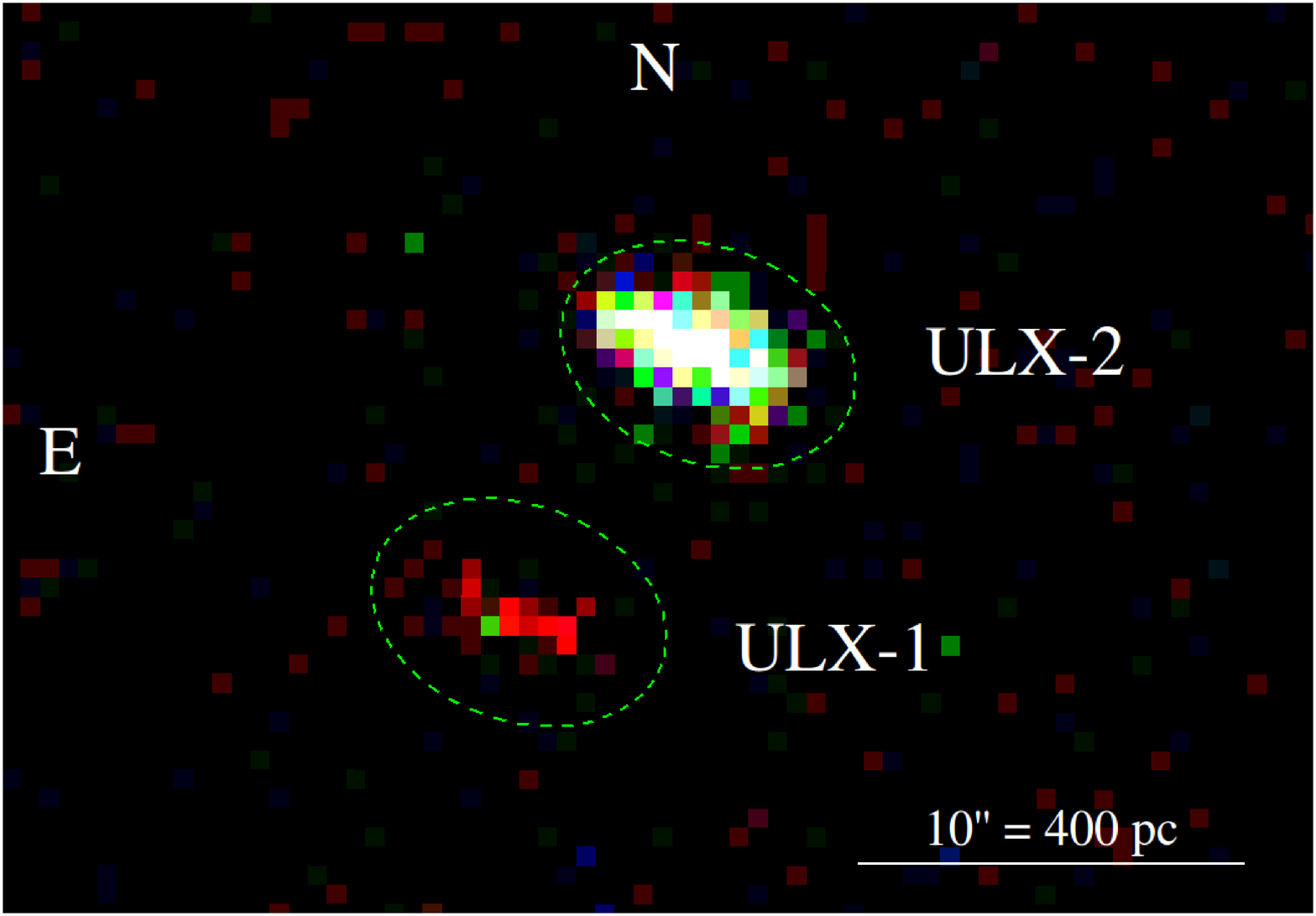}\\[4pt]
\includegraphics[width=0.47\textwidth]{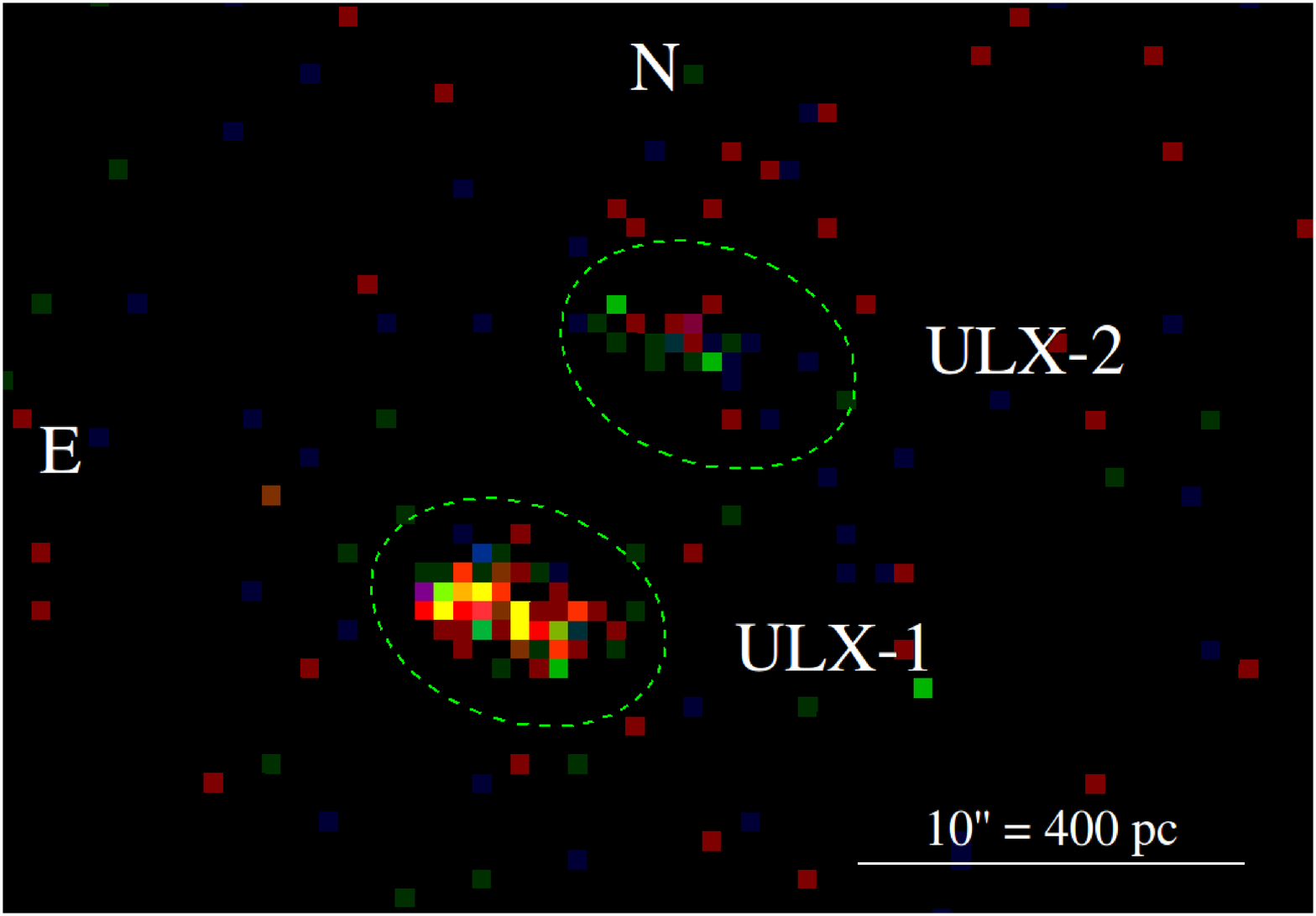}
 \caption{Top panel: stacked {\it Chandra}/ACIS-S image during the ULX-1 eclipse intervals from ObsIDs 1622, 13813 and 13814. Colors are red for 0.3--1 keV, green for 1--2 keV, blue for 2--7 keV. The dashed green ellipses represent the {\it Chandra} extraction regions for ULX-1 and ULX-2. Bottom panel: same as the top panel, for the ULX-2 eclipse interval during {\it Chandra} ObsID 13813.}
  \label{chan_res_im}
  \vspace{0.3cm}
\end{figure}

Residual emission is detected at the position of ULX-1 during eclipses (Table 2). This is particularly evident in ObsID 1622, with a residual eclipse count rate $\approx$10\% of the average out of eclipse count rate. It is also marginally significant in ObsIDs 13813 and 13814. The reason why the residual emission appears less significant in ObsID 13813 and 13814 than in ObsID 1622 is likely because of the decreased sensitivity of ACIS-S in the soft band between 2001 and 2012 \citep{2004SPIE.5488..251P}. We stacked the eclipse intervals from all three ULX-1 eclipses and display the resulting 135-ks ACIS-S X-ray-color image in Figure \ref{chan_res_im} (top panel). The residual emission of ULX-1 is centred at the same coordinates as the out of eclipse emission, and is unresolved, but appears softer (most photons below 1 keV). We also show (Figure \ref{chan_res_im}, bottom panel) the 48-ks ACIS-S image corresponding to the only {\it Chandra} eclipse of ULX-2; the signal-to-noise ratio is lower, but there is significant residual emission for ULX-2 in eclipse, as well.

\begin{figure}
    \centering
    \includegraphics[width=0.47\textwidth]{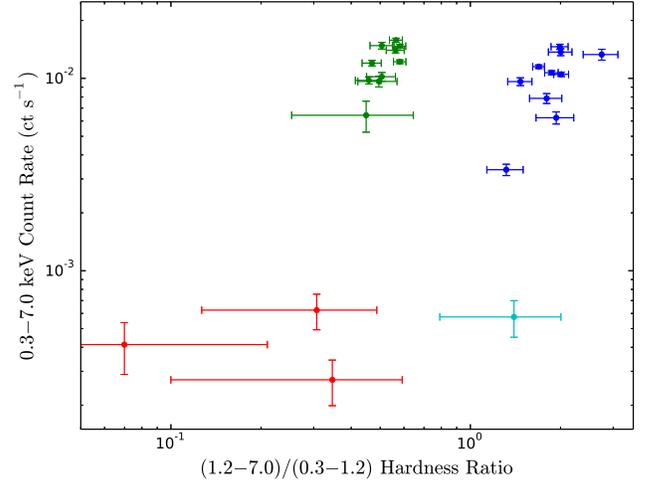}
    \caption{(1.2--7.0)/(0.3--1.2) hardness ratio versus 0.3--7.0 keV count rate for ULX-1 and ULX-2 in eclipse and non-eclipse intervals during the {\it Chandra} observations. Red datapoints correspond to ULX-1 in eclipse, green datapoints to ULX-1 out of eclipse, the single cyan datapoint to ULX-2 in eclipse and blue datapoints to ULX-2 out of eclipse. Colors have been corrected for the change in sensitivity of the ACIS-S detector over the years (Section \ref{col_sec}, Table \ref{HR_tab})}
    \label{hard_ratio_fig}
    \vspace{0.3cm}
\end{figure}

To quantify the colors and the color differences in and out of eclipse, we determined the hardness ratio between the net count rates in the 1.2--7 keV band and in the 0.3--1.2 keV band ({\it i.e.}, the same bands used in our light-curve plots). It appears (particularly in ObsID 13813) that ULX-1 is softer in eclipse than out of eclipse (Figure \ref{hard_ratio_fig} and Table \ref{HR_tab}).
The difference becomes more significant when we compare the hardness ratio of the stacked eclipse data (Table \ref{HR_tab}) with that of the stacked out of eclipse ones. For ULX-2, we cannot identify significant color differences in and out of eclipse, because of the short duration of the lone detected {\it Chandra} eclipse. Our hardness ratio study also clearly shows (Table \ref{HR_tab} and Figure \ref{hard_ratio_fig}) that ULX-1 is always softer than ULX-2, both in and out of eclipse. 

Another difference between the two ULXs is their degree of hardness ratio variability from epoch to epoch in the {\it Chandra} series. For ULX-1, all the 2012 observations are consistent with the same hardness ratio (Table \ref{HR_tab}). The source appears softer in ObsIDs 354, 1622 and 3932; however, this is misleading because such observations were taken in Cycle 1, Cycle 2 and Cycle 4, respectively, when ACIS-S was more sensitive to soft photons \citep{2004SPIE.5488..251P}. A rough way to account for this effect is to assume simple power-law models and use the {\it Chandra} X-Ray Center online installation of {\small{PIMMS}} (Version 4.8) to convert the observed count rates into ``equivalent" count rates that would have been observed in Cycle 13 (year 2012) when all the other observations took place. A more accurate conversion from observed count rates to Cycle 13-equivalent count rates requires proper spectral modelling in the various epochs. The corrected count rates listed in Table \ref{HR_tab} and plotted in Figure \ref{hard_ratio_fig}, for both ULX-1 and ULX-2, were obtained with the latter method, after we carried out the spectral analysis discussed in Section \ref{spectral_sec}; the best-fitting spectral models were convolved with response and auxiliary response functions of the detector at different epochs, to determine the predicted count rates. Inspection of the corrected count rates confirms that the hardness of ULX-1 is approximately constant; instead, that of ULX-2 is intrinsically variable from epoch to epoch.

Based on the observed hardness of the residual eclipse emission of ULX-1, we more plausibly interpret it as thermal-plasma emission, for the purpose of converting count rates into fluxes and luminosities. Assuming a temperature $\sim$0.5 keV, and using again the online {\small{PIMMS}} tool, we estimate a residual ULX-1 luminosity $L_{\rm X} \sim 10^{37}$ erg s$^{-1}$ in the 0.3--8 keV band. Again, this is only a simple, preliminary estimate. We will present a more accurate estimate of the residual emission based on spectral fitting, and we will discuss its physical origin, after we carry out a full spectral modelling of ULX-1 (Section \ref{spectral_mod_sec}). We do not have any constraints on plausible models for the residual eclipse emission of ULX-2; however, a selection of thermal-plasma and power-law models also give typical luminosities $L_{\rm X} \sim 10^{37}$ erg s$^{-1}$. What is clear is that it is harder than the residual emission of ULX-1.

\begin{table}[]
    \centering
    \caption{Hardness ratios of ULX-1 and ULX-2 for eclipsing and non-eclipsing intervals of the {\it Chandra} observations. Values in brackets are rescaled to their {\it Chandra} Cycle 13-equivalents.}
    \def\arraystretch{1.5}
    \begin{scriptsize}
    {\begin{tabular}{lcccc}
        \hline\hline
        Epoch& \multicolumn{2}{c}{ULX-1 hardness ratios} & \multicolumn{2}{c}{ULX-2 hardness ratios}\\
        & \multicolumn{1}{c}{Non-eclipsing} & \multicolumn{1}{c}{Eclipsing} & \multicolumn{1}{c}{Non-eclipsing} & \multicolumn{1}{c}{Eclipsing}\\[2pt]
       \hline\hline
       354 & \Pl{0.27}{0.04} && \Pl{1.71}{0.23} &\\[1pt]
        & [\Pl{0.49}{0.07}] && [\Pl{2.75}{0.36}] &\\[1pt]
       1622 & \Pl{0.28}{0.12} & \Pl{0.17}{0.10} & \Pl{1.31}{0.19} &\\[1pt]
        & [\Pl{0.40}{0.17}] & [\Pl{0.31}{0.18}] & [\Pl{1.90}{0.27}] &\\[1pt]       
       3932 & \Pl{0.33}{0.03} && \Pl{1.11}{0.10} &\\[1pt]
        & [\Pl{0.42}{0.04}] && [\Pl{1.43}{0.13}] &\\[1pt]
       13812 & \Pl{0.58}{0.03} && \Pl{1.87}{0.10} &\\[1pt]
       13813 & \Pl{0.58}{0.03} & \U{0.07}{0.14}{0.07} & \Pl{2.01}{0.11} & \Pl{1.4}{0.6} \\[1pt]
       13814 & \Pl{0.56}{0.03} & \Pl{0.35}{0.25} & \Pl{1.69}{0.08} &\\[1pt]
       13815 & \Pl{0.56}{0.04} && \Pl{1.32}{0.18} & \\[1pt]
       13816 & \Pl{0.47}{0.03} && \Pl{1.99}{0.13} &\\[1pt]
       15496 & \Pl{0.51}{0.04} && \Pl{2.00}{0.18} &\\[1pt]
       15553 & \Pl{0.51}{0.06} && \Pl{1.80}{0.22} &\\[1pt]
       stacked & \Pl{0.55}{0.01} & \Pl{0.21}{0.09} & \Pl{1.78}{0.04} & \Pl{1.4}{0.6}\\[1pt]
       & [\Pl{0.56}{0.01}] & [\Pl{0.23}{0.10}] & [\Pl{1.86}{0.06}] & [\Pl{1.4}{0.6}]\\[2pt] 
       \hline
    \end{tabular}}
    \end{scriptsize}
    \label{HR_tab}
\end{table}

\subsection{Spectral properties} \label{spectral_sec}

The presence of eclipses implies that both system are viewed at high inclination. Therefore, these two ULXs, although not exceptionally luminous, can help us investigate the relationship between the spectral appearance of ULXs and their viewing angles.
During out of eclipse intervals, both ULXs have sufficiently high count rates for multi-component spectral fitting. Here, we present the results of spectral fitting to the three longest {\it Chandra} observations: ObsIDs 13812 (158 ks), 13813 (179 ks) and 13814 (190 ks), taken between 2012 September 9--20. For each source, we fitted the three spectra simultaneously, keeping the intrinsic absorption column density and the parameters of any possible thermal-plasma components locked between them but leaving all other model parameters free. The reason for this choice is that we are assuming for simplicity that cold absorption and thermal-plasma emission vary on timescales longer than a few days, while the emission from the inner disk and corona may change rapidly. In addition, we assumed a line-of-sight absorption column $N_{\rm H,0} = 2 \times 10^{20}$ cm$^{-2}$ \citep{1990ARA&A..28..215D, 2005A&A...440..775K}.

\begin{table}[]
    \centering
    \caption{Goodness-of-fit $\chi^2_{\nu}$ fits for several alternative models simultaneously fitted to the spectra of ULX-1 and ULX-2 in {\it Chandra} epochs 13812, 13813 and 13814. Each model was multiplied by both a fixed line-of-sight and a free intrinsic {\it TBabs} component. {\it mk$_1$} and {\it mk$_2$} are two {\it mekal} components.}
    \def\arraystretch{1.4}
    \begin{scriptsize}
	{\begin{tabular}{lrr}
    \hline\hline
    \multicolumn{1}{c}{Model} & \multicolumn{2}{c}{$\chi_{\nu}^2$}\Tstrut\\[1pt]
    & \multicolumn{1}{c}{ULX-1} & \multicolumn{1}{c}{ULX-2}\Bstrut\\[1pt]
    \hline\hline
    {\it powerlaw} & $1.94\, (478.4/246)$&$ 0.99\, (259.3/261) $\Tstrut\\[1pt]
    {\it diskbb} & $1.61\, (395.8/246)$& $1.01\, (264.3/261)$\\[1pt]
    {\it diskir} & $1.25\, (297.2/237)$ & $0.96\, (241.8/252)$\\[1pt]
    {\it diskpbb} & $1.50\, (371.1/243)$ & $0.93\, (238.9/258)$\\[1pt]
    {\it cutoffpl} & $1.23\, (298.5/243)$& $0.93\, (239.5/258)$\\[1pt]
    {\it diskbb+powerlaw} & $1.39\, (334.6/240)$ & $0.96\, (245.6/255)$\\[1pt]
    {\it diskbb+cutoffpl} & $1.11\, (264.1/237)$& $0.93\, (234.8/252)$\\[1pt]
    {\it diskbb+comptt} & $1.17\, (278.2/237)$& $0.93\, (234.1/252)$\\[1pt]
    {\it bb+comptt} & $1.17\, (278.0/237)$& $0.93\, (234.1/252)$\\[1pt]
    {\it diskbb+powerlaw+mk$_1$+mk$_2$} & $1.02\, (241.6/236)$& $0.95\, (239.5/251)$\\[1pt]   
    {\it diskbb+comptt+mk$_1$+mk$_2$} & $1.01\, (234.4/233)$& $0.93\, (230.2/248)$\\[1pt]
    {\it bb+comptt+mk$_1$+mk$_2$} & $1.01\, (236.3/233)$& $0.94\, (233.0/248)$\\[1pt]
    {\it diskir+mk$_1$+mk$_2$} & $1.01\, (235.6/233)$ & $0.96\, (237.2/248)$\\[1pt]
    {\it diskpbb+mk$_1$+mk$_2$} & $1.00\, (238.3/239)$ & $0.92\, (234.3/254)$\Bstrut\\[1pt]
    \hline
    \end{tabular}}
    \end{scriptsize}
    \label{tab:my_label}
\end{table}

\subsubsection{Spectral models for ULX-1} \label{spectral_mod_sec}

The first obvious result of our modelling (Table \ref{tab:my_label}) is that the spectrum of ULX-1 is intrinsically curved, not well fitted by a simple power-law ($\chi^2_{\nu} \approx 1.9$) regardless of the value of $N_{\rm H,int}$. Therefore, we tried several other models, roughly belonging to two typical classes: disk-dominated models, in which the disk is responsible for most of the emission above 1 keV and for the high-energy spectral curvature; and Comptonization-dominated models, in which the disk (or other thermal component) provides the seed photon emission below 1 keV, and a cut-off power-law or Comptonization component provides the bulk of the emission above 1 keV. Much of the debate in the literature about the spectral classification of ULXs reduces to the choice between these two interpretations \citep[{\it e.g.,} ][]{2009MNRAS.397.1836G,2011NewAR..55..166F,2013MNRAS.435.1758S}. Finally, we tested whether the addition of a thermal-plasma emission component improves the fit: the justification for this component is that some ULXs (especially those seen at high viewing angles) may show emission features in the $\sim$1 keV region \citep{2014MNRAS.438L..51M,2015MNRAS.454.3134M}.

\begin{figure*}
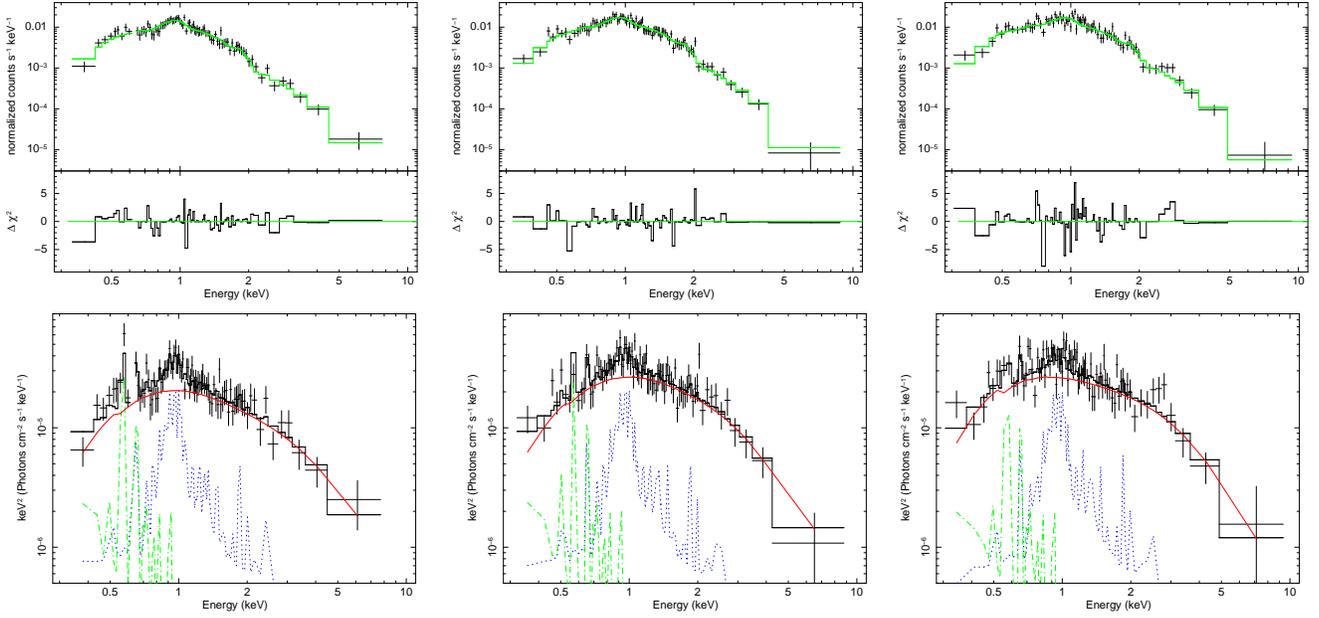

\centering
\includegraphics[width=0.22\textwidth,angle=270]{dat_ULX1_12.eps}
\includegraphics[width=0.22\textwidth,angle=270]{dat_ULX1_13.eps}
\includegraphics[width=0.22\textwidth,angle=270]{dat_ULX1_14.eps}\\[4pt]
\hspace{-0.07cm}
\includegraphics[width=0.22\textwidth,angle=270]{mod_ULX1_12.eps}
\hspace{0.13cm}
\includegraphics[width=0.22\textwidth,angle=270]{mod_ULX1_13.eps}
\includegraphics[width=0.22\textwidth,angle=270]{mod_ULX1_14.eps}
 \caption{Upper panels: {\it Chandra}/ACIS-S spectra of ULX-1 in ObsIDs 13812, 13813 and 13814, with model fits and $\chi^2$ residuals. The model is {\it TBabs} $\times$ {\it TBabs} $\times$ ({\it mekal} $+$ {\it mekal} $+$ {\it diskbb} $+$ {\it comptt}); see Table \ref{tab_ulx1_comptt} for the best-fitting parameters.
 Lower panels: unfolded spectra with model components. The dot-dashed green line represents the cooler {\it mekal} component, the dotted blue line represents the hotter {\it mekal} component and the solid red curve is the {\it comptt} component. The seed {\it diskbb} component does not appear in the plots because the direct contribution from the disk is negligible.}
  \label{b1_model}
  
  \vspace{0.3cm}
\end{figure*}

We started by fitting single-component disk models: {\it TBabs} $\times$ {\it TBabs} $\times$ {\it diskbb} for a standard disk \citep{1984PASJ...36..741M, 1986ApJ...308..635M}, and {\it TBabs} $\times$ {\it TBabs} $\times$ {\it diskpbb} for a slim disk \citep{2005ApJ...631.1062K}. They fare relatively better ($\chi^2_{\nu} \approx 1.6$ and $\chi^2_{\nu} \approx 1.5$, respectively) than a power-law model, but they are still not good fits. They also require a surprisingly low peak color temperature, $kT_{\rm in} \approx 0.6$--0.7 keV; this is inconsistent with the disk temperatures expected near or just above the Eddington limit \citep[$\approx1.0$--1.3 keV: {\it e.g.}, ][]{2004ApJ...601..428K,2006ARA&A..44...49R}, and would require a heavy stellar-mass BH (as we shall discuss later). 

Next, we tried adding a power-law component to the disk model: {\it TBabs} $\times$ {\it TBabs} $\times$ ({\it diskpbb} $+$ {\it powerlaw}). This is probably the most commonly used model in the literature for the classification of accretion states in stellar mass BHs (despite the interpretation problems caused by the unphysically high contribution of the power-law component at low energies). The quality of the fit improves slightly ($\chi^2_{\nu} \approx 1.4$) but there are still significant systematic residuals. One source of fit residuals is the high-energy downturn. By using instead a {\it TBabs} $\times$ {\it TBabs} $\times$ ({\it diskbb} $+$ {\it cutoffpl}) model, we obtain a substantially better fit ($\chi^2_{\nu} \approx 1.1$), with an F-test statistical significance $\approx (1-10^{-12})$. The presence of a high-energy downturn is of course one of the main spectral features of ULXs \citep{2006MNRAS.368..397S}, compared with stellar-mass BHs in sub-Eddington states. However, in this case the best-fitting cut-off energy $E \approx 1$ keV, much lower than the typical $\sim$5-keV high-energy cutoff seen in other ULXs \citep{2009MNRAS.397.1836G}; this is quantitative evidence that the spectrum of ULX-1 is extremely soft compared with average ULX spectra. Fitting a cutoff power-law alone (without the disk component) gives a $\chi^2_{\nu} \approx 1.2$; from this, we verify that an additional soft thermal component is significant to $>99.99\%$. There is still a third source of fit residuals, at energies around 1 keV, which we will discuss later.

The successful models discussed so far are phenomenological approximations of physical models; therefore, we fitted several alternative Comptonization models that produce a soft excess and a high-energy downturn: {\it TBabs} $\times$ {\it TBabs} $\times$ ({\it diskbb} $+$ {\it comptt}) \citep{1994ApJ...434..570T}, {\it TBabs} $\times$ {\it TBabs} $\times$ ({\it bb} $+$ {\it comptt}), and {\it TBabs} $\times$ {\it TBabs} $\times$ {\it diskir} \citep{2009MNRAS.392.1106G}. They provide moderately good fits, with $\chi^2_{\nu} \approx 1.2$ (Table \ref{tab:my_label}). In this class of Comptonization models, the disk component is used as the source of seed photons, and the electron temperature sets the location of the high-energy cutoff. For ULX-1, typical seed photon temperatures are $kT_0 \la 0.3$ keV, and the range of electron temperatures in the Comptonizing region is $kT_e \approx 0.8$--1.2 keV, with optical depths $\approx 7$--9. The electron temperature of the Comptonization region is substantially cooler than in most other two-component ULXs (where $kT_e \sim 1.5$--3 keV: \citealt{2009MNRAS.397.1836G}). This is the physical reason why ULX-1 appears as one of the softest sources in its class, with an unfolded $E\,F_{E}$ spectrum peaking at $\approx$1 keV \citep[c.f.~the ULX classification of][]{2013MNRAS.435.1758S}.
The optically-thick thermal continuum component used as seed to the Comptonization models can be equally well modelled with a disk-blackbody or a simple blackbody, given its low temperature at the lowest edge of the ACIS-S sensitivity. Its direct flux contribution to the observed spectrum is small, although difficult to constrain precisely, because of the low number of counts at very soft energies. Individual fits to the three longest observations with a {\it diskir} model suggest a direct disk contribution an order of magnitude lower than the Comptonized component.

The main reason why none of the smooth continuum models described above are really good fits is the presence of strong residuals (F-test level of significance $>$99.99\%) below and around 1 keV. A single-temperature {\it mekal} component is not sufficient to eliminate the residuals. Instead, two {\it mekal} components with temperatures $kT_1 \approx 0.2$ keV and $T_2 \approx 0.9$ keV significantly improve the fits (Figure \ref{b1_model}), providing $\chi^2_{\nu} \approx 1.01$ for the Comptonization models and $\chi^2_{\nu} \approx 1.00$ for the disk models. In the latter case, the temperature profile index is $p \lesssim 0.6$ (``broadened disk"), favouring the slim disk over the standard disk model. 

The presence of soft X-ray residuals and the strong continuum curvature are robust and independent of the choice of cold absorption model. We also tried combinations of neutral and ionized absorbers ({\it tbabs} $\times$ {\it varabs}), but they do not reproduce the strong residual feature at energies $\approx$0.8--1.0 keV. Lower-energy residuals at $\approx$0.5--0.6 keV
are relatively less constrained because of the degraded sensitivity of ACIS-S at low energies, rather than because of intrinsic absorption. For all our Comptonization-type and disk-type models, the intrinsic cold absorption $N_{\rm H}$ is $\lesssim$ a few $10^{20}$ cm$^{-2}$ and in most cases, consistent with 0 within the 90\% confidence level.

Disk models with additional thermal plasma emission produce equally good $\chi^2$ values as Comptonization models with thermal plasma emission (Table \ref{tab:my_label}). Therefore, it is worth examining in more details whether disk models are physically self-consistent, and what their physical interpretation could be. Let us start with a standard disk, to account for the possibility that ULX-1 is a rather massive BH accreting at sub-Eddington rates. We may be tempted to discard this possibility straight away, because the best-fitting temperature profile index $p \lesssim 0.6$ (rather than $p = 0.75$) is generally considered the hallmark of a disk at the Eddington accretion rate, not of a standard disk; however, it was recently shown that broadened disks with $p \approx 0.6$ may also occur in the sub-Eddington regime with accretion rates an order of magnitude lower \citep{2016..Sutton}. Therefore, we will consider that case here. In standard accretion disk models, with the inner disk fixed at the innermost stable circular orbit, there are two approximate relations between the observable quantities $T_{\rm in}$ and $L$, and the non-directly-observable physical properties $\dot{m}$ (Eddington-scaled mass accretion rate) and $M$ (BH mass):
\begin{eqnarray}
L & \approx & 1.3 \times 10^{39} \, \dot{m} M_{10}\ \ {\mathrm {erg~s}}^{-1}\label{SS_Lum}\\
kT_{\rm in} & \approx & 1.3 \, (\dot{m}/M_{10})^{1/4}\ \ {\mathrm{keV}}\label{SS_Temp}
\end{eqnarray}
 \citep{1973A&A....24..337S}, where $M_{10}$ is the BH mass in units of 10 $M_{\odot}$. Equation \eqref{SS_Lum} is simply the luminosity as a fraction of Eddington, in the radiatively efficient regime; Equation \eqref{SS_Temp} follows from the relation $L \approx 4\pi r_{\rm in}^2 \sigma T_{\rm in}^4$. For ULX-1, the best-fitting peak temperature is $kT_{\rm in} \approx (0.7 \pm 0.1)$ keV, and the luminosity $L \approx 2 \times 10^{39}$ erg s$^{-1}$ (as we shall discuss later). From Equations \eqref{SS_Lum} and \eqref{SS_Temp}, this would correspond to a BH mass $M \approx 40$--50 $M_{\odot}$ at accretion rates $\dot{m} \approx 0.3$--0.4. Even if we allow for the possibility of $p < 0.6$ in a sub-Eddington disk, the presence of strong line residuals in the soft X-ray band is another, stronger piece of evidence against the standard disk model; it is instead indicative of Eddington accretion and associated outflows \citep{2015MNRAS.454.3134M,2016Natur.533...64P}. For these reasons, we dis-favour the sub-Eddington standard disk model for ULX-1.
 
 Next, we test the self-consistency of the slim disk model. In the super-Eddington regime, the disk luminosity is modified as
 \begin{equation}
    L \approx 1.3 \times 10^{39} \, (1+0.6 \ln \dot{m})\, M_{10}\ \ {\mathrm {erg~s}}^{-1}\label{slim_Lum} 
 \end{equation}
 \citep{1973A&A....24..337S,2007MNRAS.377.1187P}. A slim disk is no longer truncated exactly at the innermost stable circular orbit: the fitted inner disk radius can be approximated by the empirical scaling $r_{\rm in}(\dot{m}) \approx r_{\rm in}(\dot{m}=1)\,[T_{\rm in}(\dot{m}=1)/T_{\rm in}(\dot{m})]$ \citep{2000PASJ...52..133W}. As a result, the disk luminosity becomes $L \propto M_{10}^{3/2} T_{\rm in}^2$. Substituting $L$ from Equation \eqref{slim_Lum}, and matching the normalization to that of the sub-Eddington case (Equation \eqref{SS_Temp}), we obtain:
 \begin{equation}
     kT_{\rm in} \approx 1.3 \, (1+0.6 \ln \dot{m})^{1/2}\, M_{10}^{-1/4}\ \ {\mathrm{keV}}\label{slim_Temp}.
 \end{equation}
 A proper treatment of the observed properties of a slim disk requires additional parameters such as the viewing angle and the BH spin \citep{2008PASJ...60..653V,2009ApJS..183..171S,2013MNRAS.436...71V}; however, Equations \eqref{slim_Lum} and \eqref{slim_Temp} are already good enough as a first-order approximation to test the consistency of a slim disk model for ULX-1. From Equation \eqref{slim_Temp}, we find that a slim disk peak temperature $kT_{\rm in} \approx 0.7$ keV requires a BH mass $M \ga 100 M_{\odot}$ (confirmed also by the numerical results of \citealt{2008PASJ...60..653V}); but such BH would be far below Eddington at the observed luminosity $L \approx 2 \times 10^{39}$ erg s$^{-1}$, contrary to our initial slim disk assumption. Therefore, the slim disk model cannot be self-consistently applied to the spectrum of ULX-1.
 

Based on those physical arguments, we conclude that the best fits in all three epochs are obtained with a Comptonization model, with the addition of multi-temperature optically-thin thermal-plasma emission. 
From the fits statistics alone, we cannot rule out a broadened disk model, with a heavy stellar-mass BH at sub-Eddington accretion rates; however, the observed presence of strong soft X-ray residuals points to an ultraluminous regime.
We list the best-fitting parameters of two equivalent Comptonization models in Tables \ref{tab_ulx1_comptt}, \ref{tab_ulx1_diskir}; for comparison, we also list the best-fitting parameters of the broadened-disk model (Table \ref{tab_ulx1_diskpbb}).


\begin{table*}[]
    \centering
    \caption{Best-fitting parameters for the spectra of ULX-1 in ObsIDs 13812, 13813 and 13814, modelled with {\it TBabs} $\times$ {\it TBabs} $\times$ ({\it mekal} $+$ {\it mekal} $+$ {\it diskbb} $+$ {\it comptt}). The first {\it TBabs} component (in square brackets) is fixed to the line-of-sight value for all epochs, while the intrinsic absorption is left free. The {\it mekal} components are locked across the three epochs, fitted simultaneously. Errors indicate the 90 per cent confidence interval for each parameter of interest. Fluxes are the observed values; luminosities are corrected for absorption and assume an inclination angle $\theta =80^{\circ}$. Goodness-of-fit $\chi^2_{\nu} = 1.01 (234.4/233)$.}
    \begin{scriptsize}
    {\begin{tabular}{llrrr}
        \hline\hline
        Component & Parameter & \multicolumn{3}{c}{Epoch}\Tstrut\\
        &&\multicolumn{1}{c}{13812} & \multicolumn{1}{c}{13813} & \multicolumn{1}{c}{13814}\Bstrut\\[2pt]
        \hline\hline
        TBabs & $N_{\rm H,0}$ $(10^{22} \centi\meter^{-2})$ & & [0.02] &\Tstrut\Bstrut\\[2pt]
        \hline
        TBabs & $N_{\rm H,int}$ $(10^{22} \centi\meter^{-2})$ &$<0.02$ & $<0.02$  & $<0.02$\Tstrut\Bstrut\\[2pt]
        \hline
        mekal &$kT_1$ $(\kev)$ && \U{0.17}{0.05}{0.04} &\Tstrut\\[2pt]
        &$N_1$\footnote{The {\it mekal} normalizations ($N_1$ and $N_2$) are in units of $10^{-14}/(4\pi d^2)\int n_e\,n_{\rm H}\,{\rm d}V$.} && \U{4.1}{2.1}{1.9}$\times10^{-6}$  &\Bstrut\\[2pt]
        \hline
        mekal&$kT_2$ $(\kev)$ && \U{0.87}{0.09}{0.08} &\Tstrut\\[2pt]
        &$N_2^{\rm ~a}$ && \U{4.6}{1.0}{0.9}$\times10^{-6}$  &\Bstrut\\[2pt]
        \hline
        diskbb&$kT_{\rm in}$ $(\kev)$& \U{0.32}{0.23}{0.08} & \U{0.19}{0.02}{0.03}& \U{0.18}{0.02}{0.01}\Tstrut\\[2pt]
        &$K$\footnote{The {\it diskbb} normalization is in units of $(r_{\rm in}/{\rm km})^2\cos{\theta}\,(d/10\,{\rm kpc})^{-2}$, where $r_{\rm in}$ is the apparent inner-disk radius.} & $<0.6$& $<0.5$& $<2.5$\Bstrut\\[2pt]
        \hline
        comptt & $kT_{0}$ $(\kev)$\footnote{The seed photon temperature for the Comptonizing medium, $kT_{0}$, is locked to peak color temperature of the disk, $kT_{\rm in}$.} & \U{0.32}{0.23}{0.05} & \U{0.19}{0.02}{0.03}& \U{0.18}{0.02}{0.01}\Tstrut\\[2pt]
        &$kT_{\rm e}$ $(\kev)$ & \U{1.1}{*}{0.3} & \U{0.9}{2.0}{0.2}& \U{0.9}{0.9}{0.2}\\[2pt]
       &$\tau$ & \U{7.6}{0.4}{0.4} & \U{7.4}{0.4}{0.4}& \U{8.3}{0.5}{0.4}\\[2pt]    
        &$N_{\rm c}$ & \U{1.4}{0.3}{0.3}$\times10^{-5}$ & \U{8.0}{0.5}{0.6}$\times10^{-5}$ &\U{7.0}{0.7}{0.8}$\times10^{-5}$\Bstrut\\[2pt] 
        \hline
        & $f_{0.3-8.0}$ $(10^{-14}\,{\rm erg}$ ${\rm cm}^{-2}\,{\rm s}^{-1})$ & \U{7.32}{0.19}{0.41}& \U{8.59}{0.27}{0.39} & \U{9.26}{0.39}{0.35} \Tstrut\\[2pt]
        & $L_{0.3-8.0}$ $(10^{39}\,\ergs)$ & \U{1.5}{0.3}{0.3} &\U{1.8}{0.3}{0.3}& \U{2.0}{0.4}{0.4} \\[2pt]        
        & $L_{\rm bol}$ $(10^{39}\,\ergs)$ & \U{2.1}{0.4}{0.4} & \U{2.2}{0.5}{0.5} & \U{2.6}{0.5}{0.5}\Bstrut\\[2pt]
        \hline
    \end{tabular}}
    \end{scriptsize}
    \label{tab_ulx1_comptt}
    \vspace{0.3cm}
\end{table*}

\begin{table*}[]
    \centering
    \caption{As in Table \ref{tab_ulx1_comptt}, for a {\it TBabs} $\times$ {\it TBabs} $\times$ ({\it mekal} $+$ {\it mekal} $+$ {\it diskir}) model. Goodness-of-fit $\chi^2_{\nu} = 1.01 (235.6/233)$.}
    \begin{scriptsize}
    {\begin{tabular}{llrrr}
        \hline\hline
        Component & Parameter & \multicolumn{3}{c}{Epoch}\Tstrut\\
        &&\multicolumn{1}{c}{13812} & \multicolumn{1}{c}{13813} & \multicolumn{1}{c}{13814}\Bstrut\\[2pt]
        \hline\hline
        TBabs & $N_{\rm H,0}$ $(10^{22} \centi\meter^{-2})$ & & [0.02] &\Tstrut\Bstrut\\[2pt]
        \hline
        TBabs & $N_{\rm H,int}$ $(10^{22} \centi\meter^{-2})$ & $<0.06$ & $<0.08$& $<0.12$\Tstrut\Bstrut\\[2pt]
        \hline
        mekal &$kT_1$ $(\kev)$ && \U{0.17}{0.05}{0.03} &\Tstrut\\[2pt]
        &$N_1$ && \U{3.8}{3.9}{1.4}$\times10^{-6}$  &\Bstrut\\[2pt]
        \hline
        mekal&$kT_2$ $(\kev)$ && \U{0.87}{0.08}{0.10} &\Tstrut\\[2pt]
        &$N_2$ && \U{4.5}{1.1}{0.8}$\times10^{-6}$  &\Bstrut\\[2pt]
        \hline
        diskir&$kT_{\rm in}$ $(\kev)$& \U{0.13}{0.40}{0.05} & \U{0.13}{0.09}{0.01}& \U{0.11}{0.21}{0.01}\Tstrut\\[2pt]
        & $\Gamma$ & \U{2.85}{0.16}{*}&\U{2.49}{0.18}{0.15} & \U{2.43}{0.13}{0.13}\\[2pt]
        &$kT_{\rm e}$ $(\kev)$& \U{1.2}{0.5}{0.5} & \U{0.71}{0.11}{0.17}& \U{0.87}{0.17}{0.12}\\[2pt]
        & L$_{\rm c}$/L$_{\rm d}$ & \U{7.2}{*}{0.4}& $>9.5$ & $>4.4$ \\[2pt]
        & $f_{\rm in}$ & [0.1] & [0.1] & [0.1] \\[2pt]
        & $r_{\rm irr}$ & [1.2] & [1.2] & [1.2] \\[2pt]
        & $f_{\rm out}$ & $[1\E{-3}]$ & $[1\E{-3}]$ & $[1\E{-3}]$ \\[2pt]
        & log($r_{\rm out}$) & [4] & [4] & [4] \\[2pt]
        &$K$\footnote{Disk normalization in units of $(r_{\rm in}/{\rm km})^2\cos{\theta}\,(d/10\,{\rm kpc})^{-2}$.} & \U{1.05}{0.05}{0.06}& \U{1.23}{0.06}{0.05}& \U{3.03}{0.15}{0.16}\Bstrut\\[2pt]
        \hline
        & $f_{0.3-8.0}$ $(10^{-14}\,{\rm erg}$ ${\rm cm}^{-2}\,{\rm s}^{-1})$ & \U{7.13}{0.28}{0.29}& \U{8.46}{0.28}{0.36} & \U{9.05}{0.18}{0.60} \Tstrut\\[2pt]
        & $L_{0.3-8.0}$ $(10^{39}\,\ergs)$ & \U{1.5}{0.3}{0.3} &\U{1.8}{0.3}{0.4}& \U{2.0}{0.3}{0.3} \\[2pt]        
        & $L_{\rm bol}$ $(10^{39}\,\ergs)$ & \U{2.0}{0.5}{0.5} & \U{2.3}{0.5}{0.6} & \U{2.6}{0.5}{0.6}\Bstrut\\[2pt]
        \hline
    \end{tabular}}
    \end{scriptsize}
    \vspace{0.3cm}
    \label{tab_ulx1_diskir}
\end{table*}

\subsubsection{Continuum and line luminosity of ULX-1}

The unabsorbed X-ray luminosity $L_{\rm X}$ is related to the absorption-corrected flux $f_{\rm X}$ by the relation $L_{\rm X} = 2\pi d^2 f_{\rm X}/\cos \theta$, where $d$ is the distance to the source and $\theta$ is the viewing angle, when the emission is from a (sub-Eddington) standard disk surface, and $L_{\rm X} = 4\pi d^2 f_{\rm X}$ for a spherical or point-like emitter. We do not have direct information on the geometry of the emitting region in ULX-1; however, analytical models and numerical simulations of near-Eddington and super-Eddington sources predict mild geometrical beaming, that is, most of the X-ray flux is emitted along the direction perpendicular to the disk plane \citep{2012ApJ...752...18K,2014ApJ...796..106J,2016MNRAS.456.3929S}, and the emission should appear fainter and down-scattered in a disk wind when a source is observed at high inclination (as in our case, given the presence of eclipses). Therefore, we choose to use the simplest angle-dependent expression for the luminosity $L_{\rm X} = 2\pi d^2 f_{\rm X}/\cos \theta$. We also identify for simplicity (and in the absence of conflicting evidence) the viewing angle $\theta$ to the plane of the inner disk with the inclination angle of the binary system, which we have constrained to be high from the presence of eclipses; that is, we neglect the possibility of a warped, precessing disk.  
Instead, we estimate the unabsorbed X-ray luminosity of the thermal plasma components as $L_{\rm X} = 4\pi d^2 f_{\rm X}$, because we assume that the distribution of hot plasma is quasi-spherical above and beyond the disk plane, and its emission is approximately isotropic. 
With those caveats in mind, we estimate an emitted 0.3--8.0 keV luminosity of the two-temperature thermal-plasma component $L_{\rm X,mekal} \approx 1.3 \times 10^{38}$ erg s$^{-1}$ (Tables \ref{tab_ulx1_comptt}, \ref{tab_ulx1_diskir}, \ref{tab_ulx1_diskpbb}), essentially all below 2 keV.
Assuming $\theta \approx 80^{\circ}$, we then estimate the total ({\it i.e.}, thermal plasma plus continuum) 0.3--8.0 keV luminosity as $L_{\rm X} \approx (1.5$--$2.0) \times 10^{39}$ erg s$^{-1}$, during the three longest {\it Chandra} observations, regardless of whether the continuum is fitted with a Comptonization model or with a slim disk.

\begin{figure}
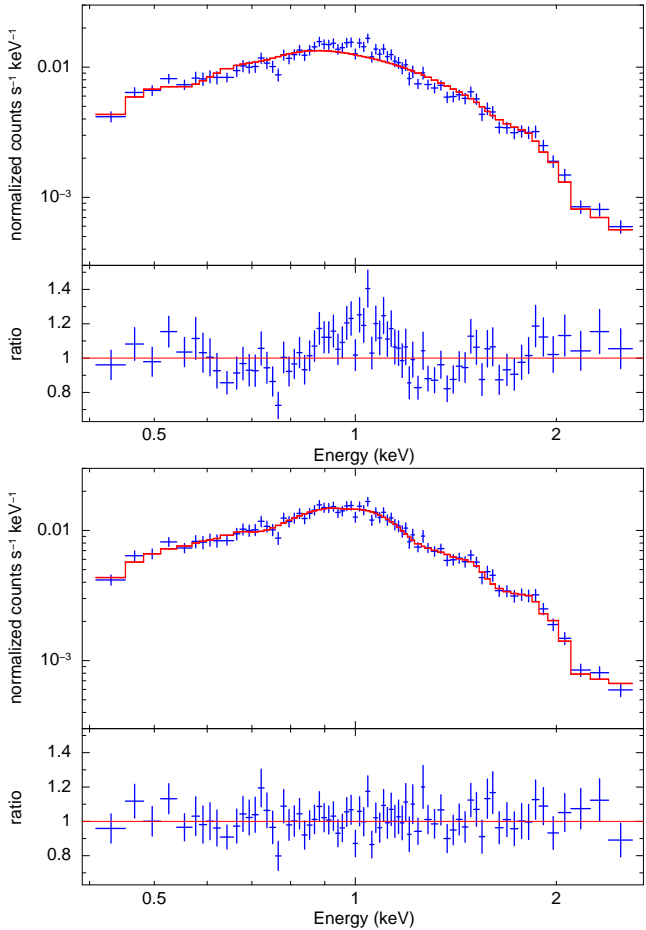

\centering
\hspace{-0.6cm}
\includegraphics[width=0.34\textwidth,angle=270]{ulx1_nomekal.eps}\\
\hspace{-0.6cm}
\includegraphics[width=0.34\textwidth,angle=270]{ulx1_mekal.eps}\\
 \caption{Top panel: datapoints, best-fitting continuum model and spectral residuals for the combined spectrum of all ten {\it Chandra}/ACIS-S observations of ULX-1, selecting only non-eclipse time intervals. The model fitted to the combined spectrum is {\it TBabs}$\times${\it TBabs}$\times$({\it diskbb}+{\it comptt}). Significant residuals are seen at photon energies $\sim$1 keV. The datapoints have been binned to a signal-to-noise ratio $\ge9$. Bottom panel: the same spectrum and residuals after the addition of three thermal-plasma emission components ({\it mekal} model) at $kT_1 \approx 0.13$ keV, $kT_2 \approx 0.73$ keV and $kT_3 \approx 1.74$ keV, which account well for the residuals.} 
  \label{ulx1_residuals}
  \vspace{0.3cm}
\end{figure}

 To increase the signal-to-noise ratio of the thermal-plasma emission component, we extracted and combined (using {\small CIAO}'s {\it specextract} tool) the spectra and responses of all ten {\it Chandra} observations (Table 1), for a grand total of $\approx$700 ks out of eclipse. We fitted the resulting spectrum with the same smooth continuum models (Comptonization and slim-disk models) used for the three long spectra: regardless of the choice of continuum, strong systematic residuals appear at energies around 1 keV. As a representative case, we show the residuals corresponding to a {\it comptt} fit (Figure \ref{ulx1_residuals}, top panel); for this continuum-only model, $\chi^2_{\nu} = 1.54 (217.0/141)$. When two {\it mekal} components are added to the continuum (as we did for the spectra of ObsIDs 13812, 13813 and 13814), the goodness-of-fit improves to $\chi^2_{\nu} = 1.10 (150.7/137)$. We tried introducing a third {\it mekal} component, and obtained a further improvement to the fit (significant to the 99.5\% level), down to $\chi^2_{\nu} = 1.03 (139.5/135)$. We show the 3-{\it mekal} model fit and its residuals in Figure \ref{ulx1_residuals} (bottom panel), and list the best-fitting parameters in Table \ref{tab_ulx1_stack}. The best-fitting {\it mekal} temperatures are $kT_1 \approx 0.13$ keV, $kT_2 \approx 0.7$ keV, and $kT_3 \approx 1.7$ keV. (Instead, adding a third {\it mekal} component to the best-fitting model for ObsIDs 13812, 13813 and 13814 does not significantly improve that fit.) We estimate an unabsorbed 0.3--8.0 keV luminosity of the three-temperature thermal-plasma component $L_{\rm X,mekal} \approx 2.4 \times 10^{38}$ erg s$^{-1}$. This is moderately higher than the value we estimated for a two-temperature model, because now part of the emission at energies $\gtrsim$2 keV is also attributed to optically-thin thermal plasma. The total (continuum plus thermal plasma) unabsorbed luminosity in the 0.3--8.0 keV band is $L_{\rm X} \approx 1.5 \times 10^{39}$ erg s$^{-1}$, consistent with the luminosity estimated in ObsIDs 13812, 13813 and 13814. Alternatively, we replaced the three {\it mekal} components with a {\it cemekl}, which is a multi-temperature thermal-plasma model with a power-law distribution of temperatures. The best-fitting {\it cemekl} $+$ {\it diskbb} $+$ {\it comptt} model has $\chi^2_{\nu} = 1.11 (154.1/139)$, maximum temperature $kT_{\rm max} \approx 2.2$ keV, thermal-plasma luminosity $L_{\rm X,cemekl} \approx 1.4 \times 10^{38}$ erg s$^{-1}$, and total luminosity $L_{\rm X} \approx 2 \times 10^{39}$ erg s$^{-1}$.

 As noted earlier, the role of the disk component in this class of models is to provide the seed photons for the Comptonization component, as well as a soft excess due to the fraction of disk photons that reach us directly. A best-fitting seed temperature $kT_{\rm in} \approx 0.17$ keV and inner-disk size $r_{\rm in} (\cos \theta)^{1/2} \approx 700$ km are consistent with the characteristic temperatures and sizes of the soft thermal components seen in other ULXs ({\it e.g.}, \citealt{2004ApJ...614L.117M, 2006MNRAS.368..397S, 2009MNRAS.398.1450K}). The direct luminosity contribution of the disk in the 0.3--8 keV band is $\approx (4\pm 1) \times 10^{38}$ erg s$^{-1}$.

\begin{table*}
    \centering
    \caption{As in Table \ref{tab_ulx1_comptt}, for a {\it TBabs} $\times$ {\it TBabs} $\times$ ({\it mekal} $+$ {\it mekal} $+$ {\it diskpbb}) model. The intrinsic absorption is fixed at $N_{\rm H,int} = 0$ (a local minimum) for all epochs. Goodness-of-fit $\chi^2_{\nu}$ = 1.00 (238.3/239).}
    \begin{scriptsize}
    {\begin{tabular}{llrrr}
        \hline\hline
        Component & Parameter & \multicolumn{3}{c}{Epoch}\Tstrut\\
        &&\multicolumn{1}{c}{13812} & \multicolumn{1}{c}{13813} & \multicolumn{1}{c}{13814}\Bstrut\\[2pt]
        \hline\hline
        TBabs & $N_{\rm H,0}$ $(10^{22} \centi\meter^{-2})$ & & [0.02] &\Tstrut\Bstrut\\[2pt]
        \hline
        TBabs & $N_{\rm H,int}$ $(10^{22} \centi\meter^{-2})$ & \U{0.05}{0.02}{0.05} & \U{0.06}{0.03}{0.04} & \U{0.04}{0.04}{0.04} \Tstrut\Bstrut\\[2pt]
        \hline
        mekal &$kT_1$ $(\kev)$ && \U{0.18}{0.04}{0.03} &\Tstrut\\[2pt]
        &$N_1$ && \U{8.1}{7.0}{4.5}$\times10^{-6}$  &\Bstrut\\[2pt]
        \hline
        mekal&$kT_2$ $(\kev)$ && \U{0.84}{0.10}{0.08} &\Tstrut\\[2pt]
        &$N_2$ && \U{6.7}{1.1}{1.0}$\times10^{-6}$  &\Bstrut\\[2pt]
        \hline
        diskpbb&$kT_{\rm in}$ $(\kev)$& \U{0.69}{0.11}{0.09} & \U{0.63}{0.08}{0.07}& \U{0.72}{0.12}{0.09}\Tstrut\\[2pt]
        &$p$& $<0.60$ & $<0.57$ & $<0.58$\\[2pt]
        &$K$\footnote{Disk normalization in units of $(r_{\rm in}/{\rm km})^2\cos{\theta}\,(d/10\,{\rm kpc})^{-2}$.} & \U{3.5}{7.6}{1.7}$\times10^{-3}$& \U{7.7}{8.8}{3.2}$\times10^{-2}$& \U{3.8}{5.8}{1.7}$\times10^{-3}$\Bstrut\\[2pt]
        \hline
       & $f_{0.3-8.0}$ $(10^{-14}\,{\rm erg}$ ${\rm cm}^{-2}\,{\rm s}^{-1})$ & \U{7.00}{0.36}{0.38}& \U{8.16}{0.40}{0.42} & \U{9.00}{0.48}{0.50} \Tstrut\\[2pt]
        & $L_{0.3-8.0}$ $(10^{39}\,\ergs)$ & \U{1.8}{0.4}{0.4} & \U{2.4}{0.4}{0.4} & \U{1.5}{0.3}{0.3} \\[2pt]        
        & $L_{\rm bol}$ $(10^{39}\,\ergs)$ & \U{2.8}{0.6}{0.6} & \U{3.2}{0.7}{0.7} & \U{2.8}{0.6}{0.6}\Bstrut\\[2pt]
        \hline
    \end{tabular}    }
    \end{scriptsize}
    \vspace{0.3cm}
    \label{tab_ulx1_diskpbb}
\end{table*}

\begin{table*}[]
    \centering
    \caption{Best-fitting parameters for the combined spectrum of ULX-1 from all 10 {\it Chandra} observations, modelled with {\it TBabs} $\times$ {\it TBabs} $\times$ ({\it mekal} $+$ {\it mekal} $+$ {\it mekal} $+$ {\it diskbb} $+$ {\it comptt}). The first {\it TBabs} component (in square brackets) is fixed to the line-of-sight value for all epochs, while the intrinsic absorption is left free. Errors indicate the 90 per cent confidence interval for each parameter of interest. Fluxes are the observed values; luminosities are corrected for absorption and assume an inclination angle $\theta =80^{\circ}$. Goodness-of-fit $\chi^2_{\nu} = 1.00 (137.0/137)$.}
    \begin{scriptsize}
    {\begin{tabular}{llr}
        \hline\hline
        Component & Parameter & Value\Tstrut\Bstrut\\[2pt]
        \hline\hline
        TBabs & $N_{\rm H,0}$ $(10^{22} \centi\meter^{-2})$ & [0.02] \Tstrut\Bstrut\\[2pt]
        \hline
        TBabs & $N_{\rm H,int}$ $(10^{22} \centi\meter^{-2})$ & \U{0.001}{0.007}{0.001}\Tstrut\Bstrut\\[2pt]
        \hline
        mekal &$kT_1$ $(\kev)$ & \U{0.13}{0.01}{0.01} \Tstrut\\[2pt]
        &$N_1$\footnote{The {\it mekal} normalizations ($N_1$, $N_2$ and $N_3$) are in units of $10^{-14}/(4\pi d^2)\int n_e\,n_{\rm H}\,{\rm d}V$.} & \U{5.0}{1.7}{1.6}$\times10^{-6}$ \Bstrut\\[2pt]
        \hline
        mekal&$kT_2$ $(\kev)$ & \U{0.73}{0.05}{0.05} \Tstrut\\[2pt]
        &$N_2^{\rm ~a}$ & \U{3.2}{0.5}{0.5}$\times10^{-6}$  \Bstrut\\[2pt]
        \hline
        mekal&$kT_3$ $(\kev)$ & \U{1.74}{0.16}{0.12} \Tstrut\\[2pt]
        &$N_3^{\rm ~a}$ & \U{1.10}{0.18}{0.18}$\times10^{-5}$  \Bstrut\\[2pt]
        \hline
        diskbb&$kT_{\rm in}$ $(\kev)$& \U{0.17}{0.02}{0.01}\Tstrut\\[2pt]
        &$K$\footnote{The {\it diskbb} normalization is in units of $(r_{\rm in}/{\rm km})^2\cos{\theta}\,(d/10\,{\rm kpc})^{-2}$, where $r_{\rm in}$ is the apparent inner-disk radius.} & \U{0.72}{0.25}{0.25}\Bstrut\\[2pt]
        \hline
        comptt & $kT_{0}$ $(\kev)$\footnote{The seed photon temperature for the Comptonizing medium, $kT_{0}$, is locked to peak color temperature of the disk, $kT_{\rm in}$.}& \U{0.17}{0.02}{0.01}\Tstrut\\[2pt]
        &$kT_{\rm e}$ $(\kev)$ & \U{0.64}{0.27}{0.18}\\[2pt]
       &$\tau$ & \U{9.7}{0.3}{0.3}\\[2pt]    
        &$N_{\rm c}$ &\U{7.8}{0.7}{0.7}$\times10^{-5}$\Bstrut\\[2pt] 
        \hline
        & $f_{0.3-8.0}$ $(10^{-14}\,{\rm erg}$ ${\rm cm}^{-2}\,{\rm s}^{-1})$ & \U{7.75}{0.22}{0.22} \Tstrut\\[2pt]
        & $L_{0.3-8.0}$ $(10^{39}\,\ergs)$&\U{1.4}{0.3}{0.3} \\[2pt]        
        & $L_{\rm bol}$ $(10^{39}\,\ergs)$&\U{2.2}{0.6}{0.6}\Bstrut\\[2pt]
        \hline
    \end{tabular}}
    \end{scriptsize}
    \label{tab_ulx1_stack}
    \vspace{0.3cm}
\end{table*}











\begin{figure}
\centering
\hspace{-0.4cm}
\includegraphics[width=0.34\textwidth,angle=270]{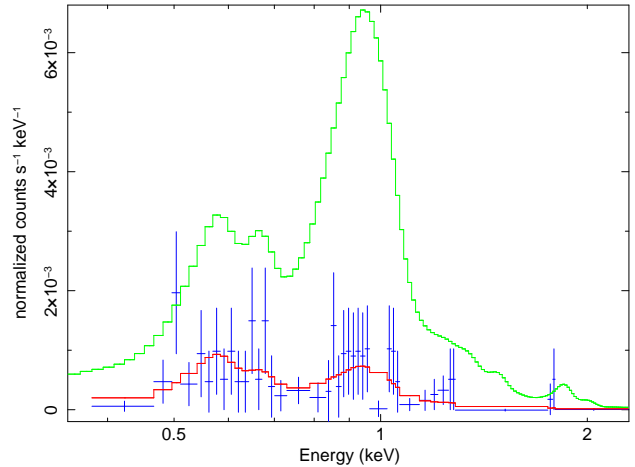}
 \caption{Combined {\it Chandra}/ACIS-S spectrum of ULX-1 during the three eclipses in ObsIDs 1622, 13813 and 13814. The datapoints have been grouped to 1 count per bin. The green curve illustrates the contribution from the best-fitting {\it mekal} components (at $T_1\approx 0.18$ keV and $T_2\approx0.86$ keV) during the non-eclipse intervals of the three longest {\it Chandra} observations (Table \ref{tab_ulx1_comptt}). The red curve is the contribution from two {\it mekal} components at the same fixed temperatures but with free normalizations, fitted to the eclipse data with the Cash statistics. This plot supports our suggestion that the residual emission during eclipses is due to thermal plasma.}
  \label{ULX1_resid}
  \vspace{0.3cm}
\end{figure}

Finally, we inspected the spectral emission in eclipse. We extracted a combined spectrum of the eclipse intervals in ObsIDs 1622, 13813 and 13814. Although we only have $\approx$60 counts, the energy distribution of the counts is similar (Figure \ref{ULX1_resid}) to the thermal-plasma emission component out of eclipse, rather than to the continuum emission. After rebinning the eclipse spectrum to 1 count per bin, we applied the Cash statistics (Cash 1979) to fit the normalization of the same two {\it mekal} components previously found in the out-of-eclipse spectra of ObsIDs 13812, 13813 and 13814 (temperatures fixed at $kT_1 \approx 0.2$ keV and $T_2 \approx 0.9$ keV). We find a C-stat value of 49.5 over 54 degrees of freedom for the best-fitting model. The emitted luminosity $\approx 2.4 \times 10^{37}$ erg s$^{-1}$, consistent with our previous simpler estimate based on count rates (Section \ref{ulx1_bp_sec}). We then let the temperatures free, but did not obtain any significant improvement to the quality of the fit (C-stat value of 49.2 over 52 degrees of freedom). Nor do we improve the fit by adding a third {\it mekal} component.

\subsubsection{Spectral models and luminosity of ULX-2}

As we did for ULX-1, we started by fitting the spectra of ULX-2 
during {\it Chandra} ObsIDs 13812, 13813 and 13814 with a simple power-law model (Table \ref{tab:my_label}). The fit is good ($\chi^2_{\nu} \approx 0.99$), but there are residuals consistent with a high-energy downturn. The best-fitting power-law index is $\Gamma = 2.1 \pm 0.1$; however, this value may be an over-estimate if the high-energy steepening is not properly accounted for. Hence, we re-fitted the spectrum with a cutoff power-law ({\it TBabs $\times$ TBabs $\times$ cutoffpl)}, and found that the fit is significantly improved: $\chi^2_{\nu} \approx 0.93$, with an F-test significance $\approx$ 99.99\% with respect to the unbroken power-law. The power-law index below the cutoff is $\Gamma = 1.1 \pm 0.2$ and the characteristic energy of the cutoff is ($3.0 \pm 0.6$) keV. This is evidence that the spectrum of ULX-2 is significantly curved. Therefore, as we did for ULX-1, we tried a series of models suitable to curved spectra: disk models and Comptonization model.

\begin{figure*}
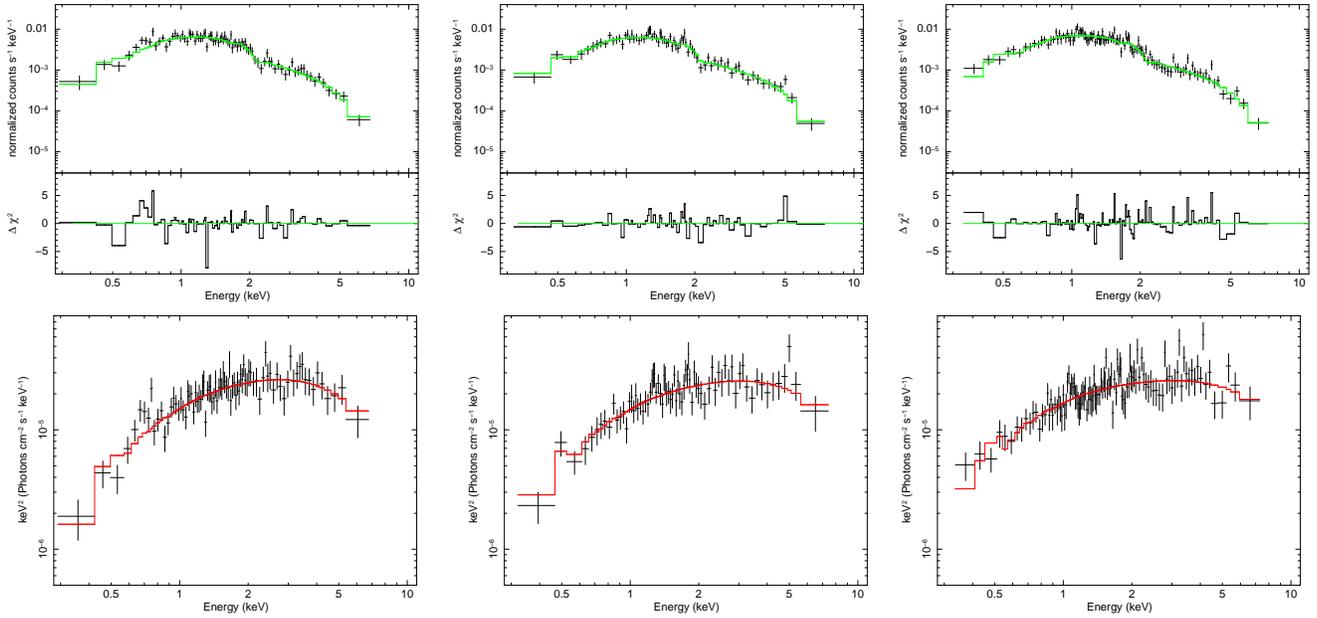

\centering
\includegraphics[width=0.22\textwidth,angle=270]{dat_ULX2_12.eps}
\includegraphics[width=0.22\textwidth,angle=270]{dat_ULX2_13.eps}
\includegraphics[width=0.22\textwidth,angle=270]{dat_ULX2_14.eps}\\[4pt]
\hspace{-0.07cm}
\includegraphics[width=0.22\textwidth,angle=270]{mod_ULX2_12.eps}
\hspace{0.13cm}
\includegraphics[width=0.22\textwidth,angle=270]{mod_ULX2_13.eps}
\includegraphics[width=0.22\textwidth,angle=270]{mod_ULX2_14.eps}
 \caption{Upper panels: {\it Chandra}/ACIS-S spectra of ULX-2 in ObsIDs 13812,13813 and 13814, with model fits and $\chi^2$ residuals. The model is {\it TBabs} $\times$ {\it TBabs} $\times$ {\it diskpbb}; see Table \ref{tab_ulx2_comptt} for the best-fitting parameters. Lower panels: unfolded spectra from the same epochs. The red curve represents the {\it diskpbb} component. This plot confirms that the spectrum of ULX-2 is harder than that of ULX-1, and does not have significant contributions from thermal plasma.}
  \label{b2_model}
  \vspace{0.3cm}
\end{figure*}

Among disk models, we find that a broadened disk is a significantly better fit (F-test significance $>$99.99\%) than a standard disk; a {\it TBabs $\times$ TBabs $\times$ diskpbb} model provides $\chi^2_{\nu} \approx 0.93$ (Table \ref{tab:my_label}). The peak disk temperature $kT_{\rm in} \approx$ 1.4--2.0 keV and $p \approx 0.6$, perfectly in line with the expected values for a mildly super-Eddington slim-disk model around a stellar-mass BH. The best-fit parameters can be found in Table \ref{tab_ulx2_diskpbb}; the model is illustrated in Figure \ref{b2_model}. The {\it diskpbb} normalization, $K$, translates into a characteristic inner disk radius 
\begin{equation}
R_{\rm in} \approx 3.18 K^{1/2} \, d_{10{\rm kpc}} \, (\cos \theta)^{-1/2} \ \ {\rm km},
\end{equation}
using the conversion factors suitable for slim-disk models \citep{2008PASJ...60..653V}; 
$d_{10{\rm kpc}}$ is the distance in units of 10  kpc. A feature of super-critical slim disks is that $R_{\rm in}$ is located slightly inside the innermost stable circular orbit \citep{2003PASJ...55..959W, 2008PASJ...60..653V}. When this correction is taken into account, the mass $M_{\bullet}$ of a non-rotating BH can be estimated as $M_{\bullet} \approx 1.2 \times R_{\rm in}c^2/(6G) \approx 1.2 R_{\rm in}/(8.9\,{\rm km}) M_{\odot}$. Characteristic radii $R_{\rm in} (\cos \theta)^{1/2} \approx$29--56 km are consistent with all the three long {\it Chandra} observations considered here (Table \ref{tab_ulx2_diskpbb}). For $\theta \approx 80^{\circ}$, this corresponds to characteristic masses $M_{\bullet} \approx 9$--18$M_{\odot}$, consistent with the observed mass distribution of Galactic BHs \citep{2012ApJ...757...36K}. For a range of viewing angles $70^{\circ} \lesssim \theta \lesssim 85^{\circ}$, the corresponding BH mass range becomes $M_{\bullet} \approx 7$--25$M_{\odot}$.
The emitted luminosity in the 0.3--8.0 keV band is $\approx 2 \times 10^{39}$ erg s$^{-1}$ (assuming again a viewing angle $\theta = 80^{\circ}$) and the bolometric disk luminosity is $\approx 3 \times 10^{39}$ erg s$^{-1}$ $\approx$1--3$L_{\rm Edd}$ for the range of BH masses estimated earlier. In this model, ULX-2 would be classified as a broadened-disk ULX in the scheme of \citet{2013MNRAS.435.1758S}.

Although we favour the slim disk model because of its self-consistency, we cannot rule out the possibility that ULX-2 is fitted by a Comptonization model (Table \ref{tab_ulx2_comptt}): for example, {\it TBabs $\times$ TBabs $\times$ (diskbb $+$ comptt)} yields $\chi^2_{\nu} \approx 0.93$, statistically equivalent to the slim disk model (Table \ref{tab:my_label}), with electron temperatures $kT_e \approx 1$--1.5 keV and optical depth $\tau \approx 9$--13 (slightly hotter and more optically thick than the best-fitting {\it comptt} models in ULX-1). Similar values of $\chi^2_{\nu}$ and $kT_e$ are also obtained from other Comptonization models such as {\it diskir}. 

Regardless of the model, the unfolded $E\,F_{E}$ spectrum peaks at $\approx$5 keV, similar to the sources classified as hard ultraluminous by \citet{2013MNRAS.435.1758S}. The original definition of the hard ultraluminous regime requires also the presence of a soft excess. In our spectra, it is difficult to constrain the significance of a direct soft emission component (in addition to the Comptonized component or the cutoff power-law) because of the low sensitivity of ACIS-S below 0.5 keV. When we fit the spectrum with a {\it diskbb} $+$ {\it comptt} model, we find that no more than $\sim$50\% of the flux in the 0.3--1.0 keV band is in the direct {\it diskbb} component (90\% upper limit), but the {\it diskbb} normalization is also consistent with 0 within the 90\% confidence limit. Regardless of classification semantics, it is clear that ULX-2 has a hard spectrum in the {\it Chandra} band, with a high-energy curvature.

\begin{figure}
\centering
\hspace{-0.6cm}
\includegraphics[width=0.34\textwidth,angle=270]{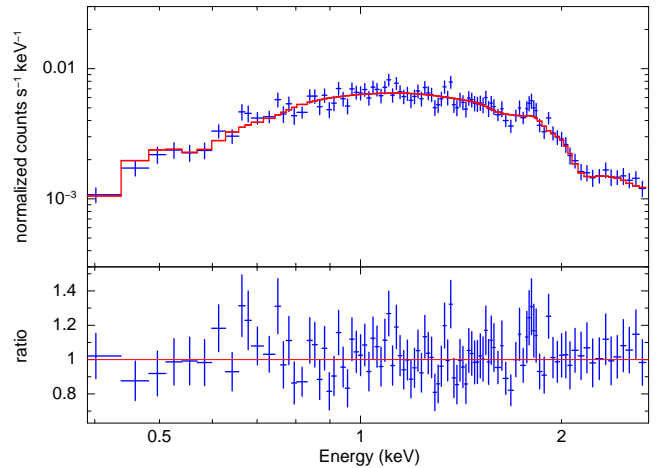}
 \caption{Datapoints, best-fitting continuum model and spectral residuals for the combined spectrum of all ten {\it Chandra}/ACIS-S observations of ULX-2, selecting only non-eclipse time intervals. The model fitted to the combined spectrum is {\it TBabs}$\times${\it TBabs}$\times${\it diskpbb}. The datapoints have been binned to a signal-to-noise ratio $\ge7$. The combined spectrum of ULX-2 does not show any significant systematic residuals around 1 keV (unlike the spectrum of ULX-1).} 
  \label{ulx2_residuals}
  \vspace{0.3cm}
\end{figure}

No significant residuals are found at $\approx$0.8--1 keV in the individual spectra from ObsIDs 13812, 13813 and 13814; however, in at least one observation (ObsID 13812), the spectrum shows two emission features with $>$90\% significance at $E \approx 1.3$ keV and $E \approx 1.8$ keV. Similar lines are typically found in thermal plasma emission. They are usually interpreted as emission from a blend of Mg XI lines at 1.33--1.35 keV, and from a Si XIII line at 1.84 keV (with the likely additional contribution of slightly weaker Mg XII lines at 1.75 keV and 1.84 keV). To investigate these and possible other emission features, we extracted a combined {\it Chandra} spectrum of ULX-2 from all ten observations, as we did for ULX-1. We fitted the combined spectrum with a {\it diskpbb} model, and obtain an excellent fit, $\chi^2_{\nu} = 0.86$ (Table \ref{tab_ulx2_stacked}). The significance of the two candidate emission features seen in ObsID 13812 fades to $<$90\% in the combined spectrum (Figure \ref{ulx2_residuals}). Adding thermal-plasma components to the combined spectrum does not produce any significant improvement. The characteristic disk temperature $kT_{\rm in} \approx 1.6$ keV, radial temperature index $p \approx 0.57$ and inner-disk radius $R_{\rm in} (\cos \theta)^{1/2} \approx 40$ km (Table \ref{tab_ulx2_stacked}) are consistent with those expected for a super-critical disk, and with the values obtained from the individual fits to ObsIDs 13812, 13813 and 13814. The corresponding range of BH masses is $M_{\bullet} \approx 8$--20$M_{\odot}$, for a non-rotating BH and a viewing angle $\theta = 80^{\circ}$.

\begin{figure}
\centering
\includegraphics[width=0.34\textwidth,angle=270]{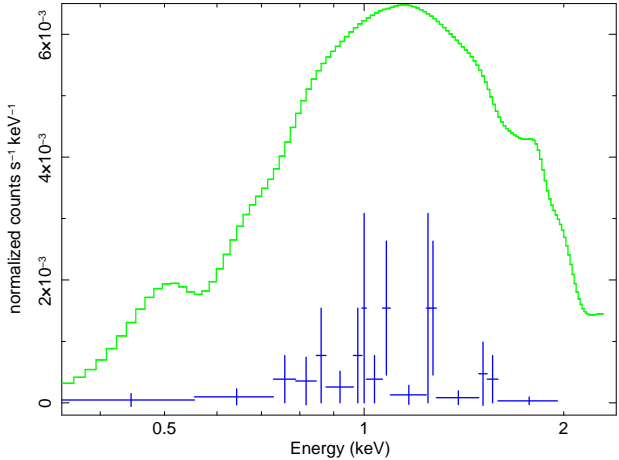}
 \caption{{\it Chandra}/ACIS-S spectrum of ULX-2 during the eclipse in ObsID 13813. The green curve illustrates the contribution from the best-fitting {\it diskpbb} component during the combined non-eclipse observations. The datapoints have been grouped to 1 count per bin. }
  \label{ULX2_resid}
  \vspace{0.3cm}
\end{figure}

Finally, we examined the spectrum of ULX-2 in eclipse (Figure \ref{ULX2_resid}). It appears different from what is seen in ULX-1: there is no evidence of a bimodal distribution of counts and it is not possible (from the few counts available) to determine whether the eclipse emission has the same origin as the out of eclipse continuum ({\it e.g.}, a small fraction of the direct emission scattered into our line-of-sight by an extended corona), or comes from thermal-plasma at higher temperatures or from bremsstrahlung emission.


\section{Discussion} \label{disc_sec}

\subsection{Two eclipsing ULXs in one field: too unlikely?}

Luminous stellar-mass BH X-ray binaries or ULXs with X-ray eclipses are very rare sources. SS\,433 in the Milky Way shows eclipses of its X-ray emission caused by the donor star on a 13.1-d binary period \citep[\textit{e.g., }][]{1987MNRAS.228..293S, 2004ASPRv..12....1F, 2005A&A...431..575B, 2010PASJ...62..323K, 2013MNRAS.436.2004C, 2013ApJ...775...75M}. Unlike M\,51 ULX-1 and ULX-2, SS\,433 does not appear as luminous as a ULX because the direct X-ray emission from the inner disk/corona region is already occulted from us. Its donor star periodically eclipses the thermal bremsstrahlung radiation ($L_{\rm X} \sim 10^{36}$ erg s$^{-1}$) from the base of the jet. The first unambiguous eclipsing behaviour in a candidate BH X-ray binary outside the Milky Way was found in IC\,10 X-1, located in a Local Group dwarf galaxy, with a Wolf-Rayet donor star, a binary period of 1.45 days, and an X-ray luminosity $L_{\rm X} \approx 10^{38}$ erg s$^{-1}$ \citep{2007ApJ...669L..21P, 2015MNRAS.446.1399L, 2016ApJ...817..154S}. For IC\,10 X-1, it is still disputed whether the accreting compact object is a BH or a neutron star \citep{2015MNRAS.452L..31L}. Outside the Local Group, NGC\,300 X-1 ($L_{\rm X} \approx 5 \times 10^{38}$ erg s$^{-1}$; binary period $\approx$33 hr) shows X-ray dips, consistent with occultation from geometrically thick structures in the outer disk, or absorption in the wind of the donor star, but not with true eclipses \citep{2015MNRAS.451.4471B}. A strong candidate for a true eclipse is the sharp dip in the {\it Swift}/X-Ray Telescope flux recorded once from the ULX P13 in NGC\,7793, at an orbital phase consistent with the inferior conjunction of its supergiant donor star \citep{2014Natur.514..198M}; however, there is no further confirmation of that single monitoring datapoint at subsequent epochs. Thus, we argue that the two M\,51 ULXs discussed in this paper are the first unambiguous eclipsing sources observed at or near the Eddington regime. 

It is rare enough to find two such bright sources projected close to each other in what is not a particularly active starburst region: it is obviously even stranger that both of them show eclipses. Therefore, we tried to assess the statistical significance of this finding. Firstly, we assume that any distance between ULX-1 and ULX-2 not in the plane of M\,51 is negligible and thus only consider the $\approx350\,\parsec$ separation. We want to discover the chances of finding two randomly distributed, luminous X-ray binaries in the same galaxy within $350\,\parsec$ of each other, both having inclination angles $>$80$^{\circ}$. Assuming for example 10 ULXs with $L_{\rm X} \ga 10^{39}$ erg s$^{-1}$ in the same spiral galaxy within a radius of $8\,\kilo\parsec$ (an over-estimate of the real number of ULXs detected in local-universe galaxies), we used a Monte-Carlo simulation, placing ULXs at random and recording the number of occurrences in which two ULXs were found within a radius of $350\,\parsec$; for 10 million trials, we find $P_1 \approx 6.6\%$. The probability of finding two nearby ULXs then has to be multiplied by the probability $(P_2)^2$ that they both show eclipses.
Assuming no preferential orientation angle, the likelihood of finding a ULX with a viewing angle, for example, $\theta > \theta_{\rm min} = 70^{\circ}$ is $P_2 = \cos 70$. However, if the orientation is too close to $90^{\circ}$, the direct X-ray emission is likely blocked by the outer disk and the source would not appear as a ULX. The thickness of the disk in ULXs is unknown (likely a few degrees), and the minimum angle $\theta_{\rm min}$ that produces eclipses is model-dependent, as a function of the ratio between stellar radius $R_{\ast}$ and binary separation $a$, namely $\cos \theta_{\rm min} \approx R_{\ast}/a$. For plausible distributions of such quantities, $P_2 \la 0.3$ \citep{2005ApJ...634L..85P}.
The final probability becomes $P_1(P_2)^2 \la$ a few $10^{-3}$: we only expect this to happen once every few hundred major galaxies. 

\subsection{Spectral properties: broadened disks, Comptonization and thermal plasma}

Luminosity (or more precisely, mass accretion rate) and viewing angle are thought to be the main parameters that determine the observational appearance of ULXs in the X-ray band \citep[{\it e.g.}, ][]{2013MNRAS.435.1758S,2015MNRAS.447.3243M,2016MNRAS.456.1859U}; disentangling and quantifying their roles is still an unsolved problem. M\,51 ULX-1 and ULX-2 have approximately the same luminosity and inclination angle (as they both show eclipses); however, they spectral appearance is substantially different. ULX-1 has soft colors in the {\it Chandra} band and is well modelled by a soft thermal component (blackbody or disk-blackbody) plus Comptonization; ULX-2 has hard colors and is well modelled by a slim disk with $kT_{\rm in} \approx  1.5$--2.0 keV (hotter than a standard disk). Also, ULX-1 has significant line residuals around 1 keV (consistent with thermal-plasma emission), which are not seen in ULX-2. Thus, we propose that there are other physical parameters that determine the spectral appearance of a ULX in addition to Eddington ratio and viewing angle. We also discovered that ULX-1 has strong radio and optical evidence of a jet (as we will discuss in a separate paper; Soria et al., in prep.) while ULX-2 does not; understanding the relation between outflow structure and spectral appearance remains a key unsolved problem.

\begin{table*}[]
    \centering
    \caption{Best-fitting parameters for the spectrum of ULX-2 in ObsIDs 13812, 13813 and 13814, modelled with {\it TBabs} $\times$ {\it TBabs} $\times$ {\it diskpbb}. The first {\it TBabs} component (in square brackets) is fixed to the line-of-sight value for all epochs, while the intrinsic absorption is left free but locked across all epochs. Errors indicate the 90 per cent confidence interval for each parameter of interest. Unabsorbed luminosities assume an inclination angle $\theta =80^{\circ}$. Goodness-of-fit $\chi^2_{\nu} = 0.93 (238.9/258)$.}
    \begin{scriptsize}
    {\begin{tabular}{llrrr}
        \hline\hline
        Component & Parameter & \multicolumn{3}{c}{Epoch}\Tstrut\\[2pt]
        &&\multicolumn{1}{c}{13812} & \multicolumn{1}{c}{13813} & \multicolumn{1}{c}{13814}\Bstrut\\[2pt]
        \hline\hline
        TBabs & $N_{\rm H,0}$ $(10^{22} \centi\meter^{-2})$ & & [0.02] &\Tstrut\Bstrut\\[2pt]
        \hline
        TBabs & $N_{\rm H,int}$ $(10^{22} \centi\meter^{-2})$ & & \U{0.09}{0.04}{0.04}  &\Tstrut\Bstrut\\[2pt]
        \hline
        diskpbb&$kT_{\rm in}$ $(\kev)$& \U{1.5}{0.4}{0.2} & \U{1.7}{0.8}{0.4}& \U{1.8}{0.7}{0.4}\Tstrut\\[2pt]
        &$p$& \U{0.58}{0.07}{0.04} & \U{0.56}{0.06}{0.04} & \U{0.54}{0.04}{0.03}\\[2pt]
        &$K$\footnote{Disk normalization in units of $(r_{\rm in}/{\rm km})^2\cos{\theta}\,(d/10\,{\rm kpc})^{-2}$.} & \U{4.5}{7.9}{3.2}$\times10^{-4}$& \U{2.1}{5.6}{1.7}$\times10^{-4}$& \U{1.5}{3.3}{1.1}$\times10^{-4}$\Bstrut\\[2pt]
        \hline
        & $f_{0.3-8.0}$ $(10^{-14}\,{\rm erg}$ ${\rm cm}^{-2}\,{\rm s}^{-1})$ & \U{8.18}{0.49}{0.50} & \U{8.23}{0.60}{0.59} & \U{8.67}{0.50}{0.53}\Tstrut\\[2pt]
        & $L_{0.3-8.0}$ $(10^{39}\,\ergs)$ & \U{2.2}{0.4}{0.4} & \U{2.3}{0.4}{0.4}&\U{2.5}{0.4}{0.4}\\[2pt] 
        & $L_{\rm bol}$ $(10^{39}\,\ergs)$ & \U{2.7}{0.4}{0.4} & \U{3.0}{0.5}{0.5} & \U{3.9}{0.6}{0.6}\Bstrut\\[2pt]
        \hline
    \end{tabular}}
    \end{scriptsize}
    \label{tab_ulx2_diskpbb}
    \vspace{0.3cm}
\end{table*}

\begin{table*}[]
    \centering
    \caption{As in Table \ref{tab_ulx2_diskpbb}, for a {\it TBabs} $\times$ {\it TBabs} $\times$ ({\it diskbb} $+$ {\it comptt}) model. Goodness-of-fit $\chi^2_{\nu} = 0.93 (234.1/252)$.}
    \begin{scriptsize}
    {\begin{tabular}{llrrr}
        \hline\hline
        Component & Parameter & \multicolumn{3}{c}{Epoch}\Tstrut\\[2pt]
        &&\multicolumn{1}{c}{13812} & \multicolumn{1}{c}{13813} & \multicolumn{1}{c}{13814}\Bstrut\\[2pt]
        \hline\hline
        TBabs & $N_{\rm H,0}$ $(10^{22} \centi\meter^{-2})$ & & [0.02] &\Tstrut\Bstrut\\[2pt]
        \hline
        TBabs & $N_{\rm H,int}$ $(10^{22} \centi\meter^{-2})$ & & $<0.15$  &\Tstrut\Bstrut\\[2pt]
        \hline
        diskbb&$kT_{\rm in}$ $(\kev)$& \U{0.16}{0.76}{0.06} & \U{0.26}{0.66}{0.11}& \U{0.23}{0.60}{0.15}\Tstrut\\[2pt]
        &$K$\footnote{The {\it diskbb} normalization is in units of $(r_{\rm in}/{\rm km})^2\cos{\theta}\,(d/10\,{\rm kpc})^{-2}$, where $r_{\rm in}$ is the apparent inner-disk radius.} & $<0.25$& $<0.12$& $<0.40$\Bstrut\\[2pt]
        \hline
        comptt & $kT_{0}$ $(\kev)$\footnote{The seed photon temperature $kT_{0}$ is locked to the peak temperature of the disk, $kT_{\rm in}$.} & \U{0.16}{0.76}{0.06} & \U{0.26}{0.66}{0.11}& \U{0.23}{0.60}{0.15}\Tstrut\\[2pt]
        &$kT_{\rm e}$ $(\kev)$ & \U{1.0}{0.2}{0.1} & \U{1.5}{*}{0.4}& \U{1.1}{1.0}{0.3}\\[2pt]
       &$\tau$ & \U{13.3}{2.6}{2.7} & \U{9.2}{3.3}{7.4}& \U{11.7}{*}{3.9}\\[2pt]    
        &$K_{\rm c}$ & \U{4.5}{0.8}{0.9}$\times10^{-5}$ & \U{2.5}{1.3}{2.2}$\times10^{-5}$ &\U{3.4}{2.0}{1.7}$\times10^{-5}$\Bstrut\\[2pt] 
        \hline
        & $f_{0.3-8.0}$ $(10^{-14}\,{\rm erg}$ ${\rm cm}^{-2}\,{\rm s}^{-1})$ & \U{8.12}{0.47}{0.48} & \U{8.28}{0.62}{0.63} & \U{8.65}{0.53}{0.52}\Tstrut\\[2pt]
        & $L_{0.3-8.0}$ $(10^{39}\,\ergs)$ & \U{1.9}{0.3}{0.3} & \U{1.9}{0.3}{0.3}&\U{2.1}{0.3}{0.3}\\[2pt] 
        & $L_{\rm bol}$ $(10^{39}\,\ergs)$ & \U{1.9}{0.3}{0.3} & \U{2.1}{0.4}{0.4} & \U{2.3}{0.4}{0.4}\Bstrut\\[2pt]
        \hline
    \end{tabular}}
    \end{scriptsize}
    \label{tab_ulx2_comptt}
    \vspace{0.3cm}
\end{table*}

\begin{table*}[]
    \centering
    \caption{Best-fitting parameters for the combined {\it Chandra} spectrum of ULX-2, modelled with {\it TBabs} $\times$ {\it TBabs} $\times$ {\it diskpbb}. The first {\it TBabs} component (in square brackets) is fixed to the line-of-sight value, while the intrinsic absorption is left free. Errors indicate the 90 per cent confidence interval for each parameter of interest. Unabsorbed luminosities assume an inclination angle $\theta =80^{\circ}$. Goodness-of-fit $\chi^2_{\nu} = 0.87 (190.4/219)$.}
    \begin{scriptsize}
    {\begin{tabular}{llr}
        \hline\hline
        Component & Parameter & Value\Tstrut\Bstrut\\[2pt]
        \hline\hline
        TBabs & $N_{\rm H,0}$ $(10^{22} \centi\meter^{-2})$ & [0.02] \Tstrut\Bstrut\\[2pt]
        \hline
        TBabs & $N_{\rm H,int}$ $(10^{22} \centi\meter^{-2})$& \U{0.08}{0.03}{0.03} \Tstrut\Bstrut\\[2pt]
        \hline
        diskpbb&$kT_{\rm in}$ $(\kev)$ & \U{1.5}{0.2}{0.2}\Tstrut\\[2pt]
        &$p$&  \U{0.58}{0.04}{0.03}\\[2pt]
        &$K$\footnote{Disk normalization in units of $(r_{\rm in}/{\rm km})^2\cos{\theta}\,(d/10\,{\rm kpc})^{-2}$.} & \U{3.9}{3.9}{2.1}$\times10^{-4}$\Bstrut\\[2pt]
        \hline
        & $f_{0.3-8.0}$ $(10^{-14}\,{\rm erg}$ ${\rm cm}^{-2}\,{\rm s}^{-1})$ & \U{8.12}{0.22}{0.23} \Tstrut\\[2pt]
        & $L_{0.3-8.0}$ $(10^{39}\,\ergs)$ & \U{2.2}{0.4}{0.4} \\[2pt] 
        & $L_{\rm bol}$ $(10^{39}\,\ergs)$ & \U{2.9}{0.5}{0.5}\Bstrut\\[2pt]
        \hline
    \end{tabular}}
    \end{scriptsize}
    \label{tab_ulx2_stacked}
    \vspace{0.3cm}
\end{table*}


The different role played by an optically-thick thermal component in the modelling of ULX-1 and ULX-2 exemplifies the confusion sometimes found in the literature about the properties of ULX disks. In ULX-1, the ``disk" emission is much cooler ($kT \sim 0.1$--0.2 keV) and comes from a large area, with characteristic size $\approx$2000--3000 km (as inferred from the normalization of the {\it diskir} component in Table \ref{tab_ulx1_diskir} and/or the normalization of the {\it diskbb} component in Table \ref{tab_ulx1_stack}). This is much further out than the innermost stable circular orbit around a BH; it is probably located at, or just outside, the spherization radius, where massive radiation-driven outflows are predicted to be launched. This thermal component represents what is sometimes referred to as the ``soft excess" in ULXs \citep[{\it e.g.}, ][]{2003ApJ...585L..37M,2005MNRAS.357.1363R,2006MNRAS.368..397S,2009MNRAS.397.1836G}. In M\,51 ULX-1 and in many other similar ULXs, it contributes $\la 10\%$ of the continuum flux in the {\it Chandra} band. Despite being often modelled with a {\it diskbb} component for practical purposes, it is by no means clear whether or not it originates from the disk; it could come instead from the more optically thick parts of the outflow \citep{2015MNRAS.447.3243M, 2016MNRAS.456.1859U}. On the other hand, the ``disk" in ULX-2 is the dominant continuum emission component. It is probably emitted by a non-standard, geometrically thicker disk with advection, photon trapping and outflows (slim disk model), extending all the way down to the innermost stable circular orbit and possibly even a little further inside it \citep{2008PASJ...60..653V}. This state is the natural progression from the high/soft state of stellar-mass BHs \citep[$L \la 0.3 L_{\rm Edd}$; ][]{2006ARA&A..44...49R}, to the apparently standard regime \citep{2004ApJ...601..428K} and the super-Eddington regime. It is also sometimes referred to as the ``broadened disk" ultraluminous regime \citep{2013MNRAS.435.1758S}.

A temperature $kT_e \approx 0.8$ keV for the Comptonizing region in ULX-1 is certainly unusually low for a ULX, but not unique. The ULX NGC\,55 X-1 has a similar Comptonizaton temperature, similar seed photon temperature $kT_0 \approx 0.2$ keV, similar optical depth $\tau \approx 10$ and similar luminosity $L_{\rm X} \approx 2 \times 10^{39}$ erg s$^{-1}$ \citep{2009MNRAS.397.1836G}. It is a classic example of a ULX in the soft ultraluminous regime \citep{2013MNRAS.435.1758S}. NGC\,55 X-1 is also viewed at high inclination, as proved by X-ray dips attributed to clumps of obscuring material in the outer disk \citep{2004MNRAS.351.1063S}. As for M\,51 ULX-1, NGC\,55 X-1 gets softer during the dips: this is consistent with the obscuration of the harder emission from the inner disk region, while a more extended source of soft X-ray photons remains partially unocculted \citep{2004MNRAS.351.1063S}.



It is important to underline the detection of the thermal-plasma emission in ULX-1 out of eclipse, with a luminosity $L_{\rm X,mekal} \approx 2 \times 10^{38}$ erg s$^{-1}$ and an emission measure $\sim n_e^2 V$ $\approx 10^{61}$ cm$^{-3}$ (as fitted to the spectra of ObsIDs 13812, 13813 and 13814). The detection of residual soft emission in eclipse, with a luminosity $L_{\rm X} \approx 2 \times 10^{37}$ erg s$^{-1}$, is consistent with a fraction of the emitting hot gas (perhaps the outer part of the same outflow responsible for the Comptonized component) extending on a scale similar to, or larger than, the size of the companion star; namely, a radius $\ga$ a few $\times 10^{12}$ cm (as we shall discus in Section \ref{donor_sec}). Conversely, the fact that $\approx$90\% of the thermal-plasma emission seen out of eclipse also disappears in eclipse is evidence that the emission comes directly from a region of comparable size to the binary system, and not for example from the hot spots of a compact jet on a scale of a few pc (which would still be unresolved by {\it Chandra} but unaffected by eclipses). An extended hot halo is a characteristic feature of the best-studied eclipsing X-ray binary, the low-mass Galactic system EXO\,0748$-$676 \citep{2001A&A...365L.282B}. Soft X-ray residuals consistent with thermal plasma emission (and/or absorption) have been reported in several other (non-eclipsing) ULXs such as NGC\,5408 X-1 \citep{2014MNRAS.438L..51M,2015MNRAS.454.3134M,2015ApJ...814...73S,2016arXiv160408593P}, NGC\,6946 X-1 \citep{2014MNRAS.438L..51M}, Ho\,II X-1 \citep{2001AJ....121.3041M, 2004ApJ...608L..57D}, NGC\,4395 X-1 \citep{2006MNRAS.368..397S}, NGC\,4559 X-1 \citep{2004MNRAS.349.1193R}, NGC\,7424 ULX2 \citep{2006MNRAS.370.1666S}, NGC\,1313 X-1 \citep{2013ApJ...778..163B, 2016arXiv160408593P} and Ho\,IX X-1 \citep{2014ApJ...793...21W}; the last two of those ULXs are hard ultraluminous sources, while all the others are classified as soft ultraluminous.



\begin{figure}
    \centering
    \includegraphics[width=0.45\textwidth]{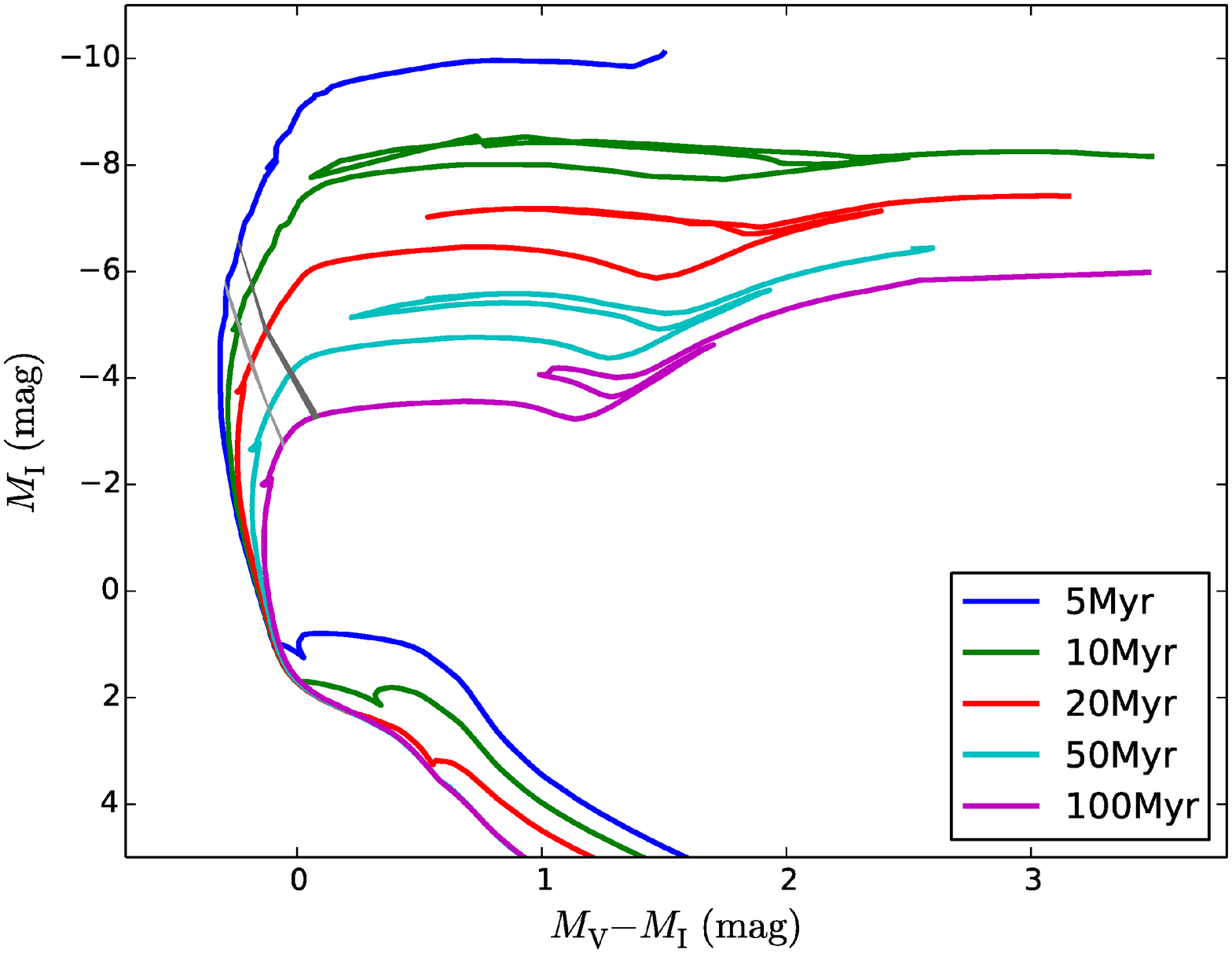}\\
    \includegraphics[width=0.45\textwidth]{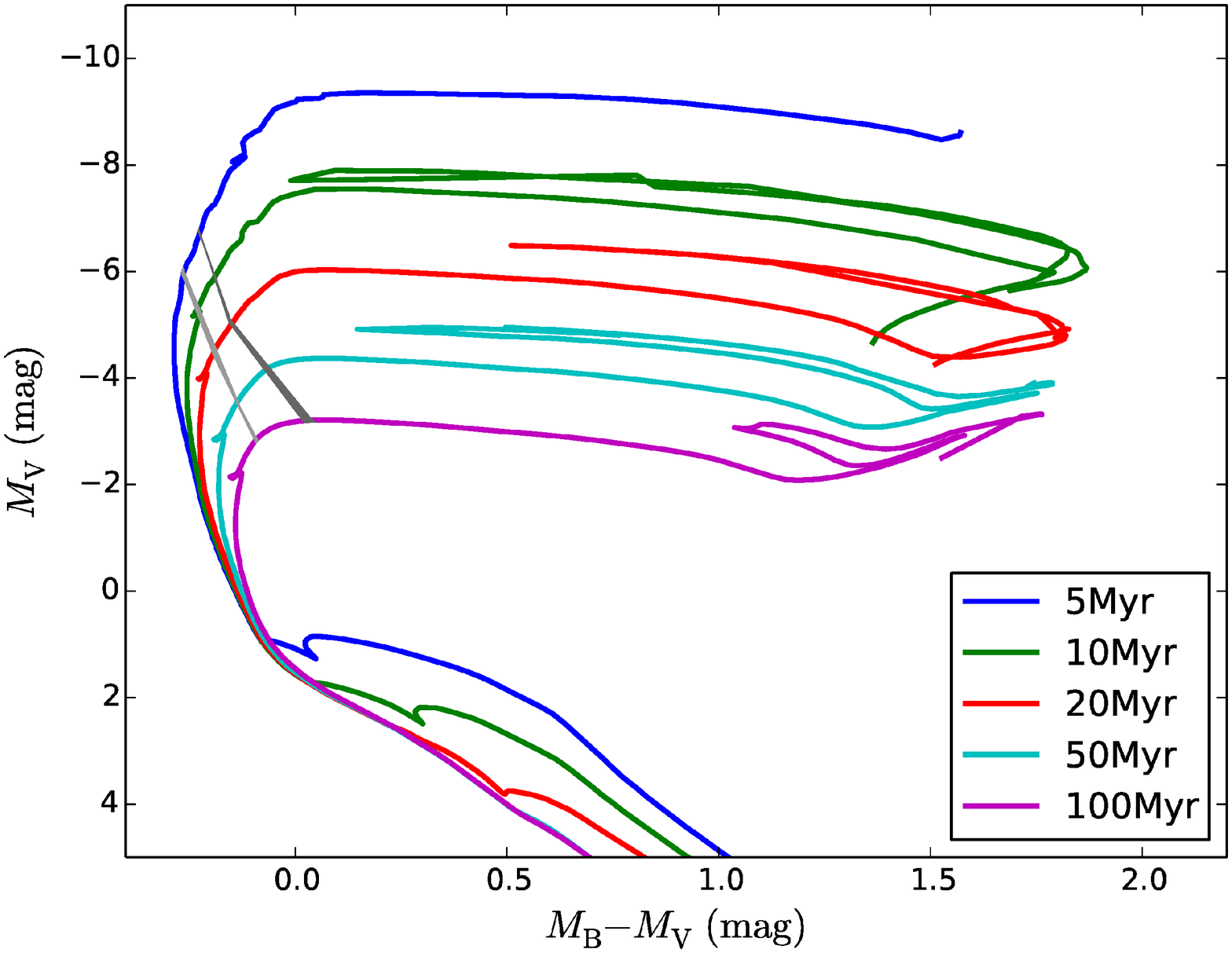}
    \caption{Top panel: theoretical stellar-population ($I$, $V-I$) isochrones, with the location of the potential donor stars of ULX-1 consistent with the permitted range of binary periods. The dark shaded grey band represents (young) stars with a mean density consistent with a period $12.2\,\days \leq P \leq 13.1\,\days$. The light shaded grey band represents stars with a mean density consistent with a period $6.1\,\days \leq P \leq 6.4\,\days$. Bottom panel: as in the the top panel, for the ($V$, $B-V$) isochrones.}
    \label{col_mag_fig}
    \vspace{0.3cm}
\end{figure}

\subsection{Constraints on the donor star of ULX-1}\label{donor_sec}
Wind accretion is not an effective mechanism to produce X-ray luminosities $\ga 10^{39}$ erg s$^{-1}$; at such luminosities, stellar-mass BHs require feeding via Roche-lobe overflow, or at the very least, via a focused wind from a donor star that is almost filling its Roche lobe. For the following discussion we will assume that the donor star in ULX-1 (and in ULX-2, although not discussed here for a lack of constraints) is at least close to filling its Roche lobe. Therefore, we will express the radius of the donor star $R_{\ast}$ as a function of binary separation $a$ as,
\begin{equation}
R_{\ast}/a \approx \frac{0.49 q^{2/3}}{0.6 q^{2/3} + \ln\left(1+q^{1/3}\right)},
\end{equation}
valid to better than 1\% for any $q$ \citep{1983MNRAS.204..449E}.
We have already shown (Figure \ref{ec_test}) that there are only selected pairs of values for the binary period $P$ and the eclipse duration $\tau_{\rm ecl}$ consistent with the empirical data. Each value of $\phi \equiv \pi \tau_{\rm ecl}/P$ corresponds to one particular solution \citep[Equations 4 and 5 in][]{1976ApJ...208..512C} for the pair of $(\theta, q)$ where $\theta$ is, as usual, the viewing angle, and $q \equiv M_{\ast}/M_{\bullet}$ is the ratio of donor star mass over compact object mass. Analytic solutions of $q(\phi)$ can be obtained \citep{2005ApJ...634L..85P} in the limiting case of $\theta = 90^{\circ}$: 
\begin{eqnarray}
\phi &=& \arcsin(R_{\ast}/a) \nonumber \\
& \approx &
\arcsin \left[ \frac{0.49 q^{2/3}}{0.6 q^{2/3} + \ln\left(1+q^{1/3}\right)} \right].
\end{eqnarray}
As an example, in Figure \ref{ec_test} we labelled 4 representative values of $q(\theta=90^{\circ})$ corresponding to 4 permitted values of $\phi$ (marked as A,B,C,D). For a fixed value of $\phi$, $q$ increases going to lower (less edge-on) values of $\theta$ \citep[Table 1 in][]{1976ApJ...208..512C}. For example, for $\phi = \tau_{\rm ecl}/P = 0.17$, $q(\theta = 80^{\circ}) \approx 1.3 q(\theta = 90^{\circ})$, and $q(\theta = 70^{\circ}) \approx 2.8 q(\theta = 90^{\circ})$; for $\tau_{\rm ecl}/P = 0.08$, $q(\theta = 80^{\circ}) \approx 2.0 q(\theta = 90^{\circ})$, and $q(\theta = 70^{\circ}) \approx 7.4 q(\theta = 90^{\circ})$. Regardless of the uncertainty in the true value of $\theta$ for ULX-1, the robust result is that permitted periods of $\approx$6 days always correspond to $q(\theta) \geq q(90^{\circ})\ga 4$ (with a more likely range $q \sim 5$--10), while permitted periods of $\approx$12--13 days correspond to $q(\theta) \geq q(90^{\circ}) \approx 0.25$--1.2. In the young stellar environment in which ULX-1 is located, with a likely OB donor star, the higher range of mass ratios (longer eclipse fraction) is indicative of a neutron star accretor, or a low-mass stellar BH seen almost edge-on; instead, the lower range of mass ratios (shorter eclipse fraction) is consistent with a larger range of BH masses, or with a neutron star seen at intermediate angles $\theta \sim 60^{\circ}$--$70^{\circ}$.

If $q$ is known or well constrained, we can then derive a period-density relation for the donor star, and constrain its mass and evolutionary stage. In the limiting case of $q \la 0.5$, such a relation reduces to $\bar{\rho} \approx 110 P_{\mathrm hr}^{-2}$ g cm$^{-3}$; however, in the more general case \citep{1983MNRAS.204..449E}, 
\begin{equation}
\bar{\rho} \approx \frac{10.89}{P_{\rm hr}^2}\left(\frac{q}{1+q}\right)\left[\frac{0.49 q^{2/3}}{0.6 q^{2/3} + \ln\left(1+q^{1/3}\right)}\right]^{-3}\text{g cm}^{-3}.
\end{equation}
For example, for $q=1$, $\bar{\rho} \approx 99 P_{\mathrm hr}^{-2}$ g cm$^{-3}$; for $q=5$, $\bar{\rho} \approx 65 P_{\mathrm hr}^{-2}$ g cm$^{-3}$.


We chose two representative values of $q$ consistent with the 6-day range of period solutions ($q=4$ and $q=10$), and two values of $q$ consistent with periods in the 12-day range ($q=0.5$ and $q=1$). For those four values of $q$, we calculated the average density of the Roche-lobe-filling donor star (Table \ref{tab_ulx1_donor}). Typical values are $\bar{\rho} \approx 3 \times 10^{-3}$ g cm$^{-3}$ for the shorter period solution, and $\bar{\rho} \approx 10^{-3}$ g cm$^{-3}$ for the longer one.
Finally, we used the latest set of Padova isochrones \footnote{Available at http://stev.oapd.inaf.it/cgi-bin/cmd} \citep{2012MNRAS.427..127B,2015MNRAS.452.1068C} with metallicity $Z=0.019$, to estimate what types of stars have such densities, for a series of stellar population ages. In practice, we know that both ULXs reside in a region of the M\,51 disk with recent star formation (Soria et al., in prep.). Therefore, we only focused on population ages $\le$100 Myr as the most likely candidates for the ULX donor stars. We find (Figure \ref{col_mag_fig}) that both ranges of permitted periods correspond to blue supergiants ($B-V$ color index $\approx$ $-$0.2--0 mag) with absolute brightness $M_V$ spanning the range between $M_V \approx -3$ mag and $M_V \approx -6$ mag, depending on their age; stars corresponding to the longer period approximately half a magnitude brighter than those associated with the shorter period. For the youngest ages ($\approx$5 Myr), the characteristic periods allowed for ULX-1 are consistent with donor stars of mass $\approx$29--31\,$M_{\odot}$, and radii $\approx$24--35\,$R_{\odot}$; for an age of $\approx$20 Myr, the predicted mass is $\approx$11\,$M_{\odot}$, with radii $\approx$17--25\,$R_{\odot}$; for an age of $\approx$50 Myr, the predicted mass is $\approx$7\,$M_{\odot}$, with radii $\approx$14--20\,$R_{\odot}$ (Table \ref{tab_ulx1_donor}).
In follow-up work, we will discuss how the observed optical brightness of the ULX-1 counterpart and of the neighbouring stars overlaps with these predictions.

\begin{table*}[]
    \centering
    \caption{Main properties of Roche-lobe-filling donor stars for a representative sample of acceptable binary periods for ULX-1.}
    \begin{scriptsize}
	{\begin{tabular}{c|ccccc|ccccc}
        \hline\hline
        Age & $M_\ast$ & $R_\ast$ & $M_V$ & $T_{\rm eff}$ & $M_\bullet$ & $M_\ast$ & $R_\ast$ & $M_V$ & $T_{\rm eff}$ & $M_\bullet$\Tstrut\\
        (Myr) & $\left(M_{\odot}\right)$ & $\left(R_{\odot}\right)$ & (mag) & (K) & $\left(M_{\odot}\right)$ & $\left(M_{\odot}\right)$ & $\left(R_{\odot}\right)$ & (mag) & (K) & $\left(M_{\odot}\right)$ \Bstrut\\[2pt]
        \hline\hline\\[-3pt]
         &  \multicolumn{5}{c|}{$P = 6.2$d, $q = 4.0$} &  \multicolumn{5}{c}{$P = 6.3$d, $q=10.0$}\Tstrut\\[3pt]
        5 & 29.7 & 23.8 & $-6.2$ & 26,800 & 7.4 & 30.3 & 26.9 & $-6.4$& 25,600 & 3.0\Tstrut\\[2pt]
        10 & 17.3 & 19.8 & $-5.4$ & 22,300 & 4.3 & 17.3 & 22.3 & $-5.6$& 21,000 & 1.7 \\[2pt]
        15 & 13.1 & 18.1 & $-4.9$ & 19,000 & 3.3 & 13.1 & 20.3 & $-5.1$& 18,000 & 1.3 \\[2pt]
        20 & 11.0 & 17.1 & $-4.6$ & 17,100 & 2.8 & 11.0 & 19.2 & $-4.8$& 16,200 & 1.1\\[2pt]
        30 & 8.8 & 15.8 & $-4.2$ & 14,900 & 2.2 & 8.8 & 17.8 & $-4.3$& 14,100 & 0.9\\[2pt]
        40 & 7.7 & 15.1 & $-3.9$ & 13,600 & 1.9 & 7.7 & 17.0 & $-4.1$& 12,800 & 0.8 \\[2pt]
        50 & 6.9 & 14.6 & $-3.7$ & 12,600 & 1.7 & 6.9 & 16.4 & $-3.8$& 11,900 & 0.7 \\[2pt]
        70 & 5.9 & 13.9 & $-3.4$ & 11,300 & 1.5 & 5.9 & 15.6 & $-3.5$& 10,700 & 0.6\\[2pt]
        100 & 5.1 & 13.2 & $-3.0$ & 10,100 & 1.3 & 5.1 & 14.8 & $-3.1$& 9,400 & 0.5\Bstrut\\[2pt]
        \hline\\[-3pt]
&  \multicolumn{5}{c|}{$P = 12.8$d, $q=0.5$} &  \multicolumn{5}{c}{$P = 13$d, $q=1.0$}\Tstrut\\[3pt]
        5 & 30.8 & 33.2 & $-6.7$ & 23,600 & 61.7 & 30.9 & 35.1 & $-6.8$& 23,000 & 30.9\Tstrut\\[2pt]
        10 & 17.3 & 27.4 & $-5.8$ & 19,100 & 34.6 & 17.3 & 28.9 & $-5.9$& 18,600 & 17.3\\[2pt]
        15 & 13.1 & 25.0 & $-5.3$ & 16,200 & 26.2 & 13.1 & 26.2 & $-5.4$& 15,800 & 13.1\\[2pt]
        20 & 11.0 & 23.6 & $-5.0$ & 14,600 & 22.1 & 11.0 & 24.8 & $-5.1$& 14,200 & 11.0\\[2pt]
        30 & 8.8 & 21.9 & $-4.6$ & 12,700 & 17.7 & 8.8 & 23.0 & $-4.6$& 12,400 & 8.8\\[2pt]
        40 & 7.7 & 20.9 & $-4.3$ & 11,500 & 15.3 & 7.7 & 22.0 & $-4.3$& 11,200 & 7.7\\[2pt]
        50 & 6.9 & 20.1 & $-4.0$ & 10,700 & 13.8 & 6.9 & 21.2 & $-4.1$& 10,400 & 6.9\\[2pt]
        70 & 5.9 & 19.1 & $-3.7$ & 9,500 & 11.9 & 5.9 & 20.2 & $-3.7$& 9,300 & 5.9\\[2pt]
        100 & 5.1 & 18.2 & $-3.2$ & 8,400 & 10.2 & 5.1 & 19.2 & $-3.2$& 8,200 & 5.1\Bstrut\\[2pt]
        \hline
    \end{tabular}}
    \end{scriptsize}
    \label{tab_ulx1_donor}
    \vspace{0.3cm}
\end{table*}

The mass $M_\bullet$ of the compact object is still unknown, but from the analysis outlined above we can see how observational constraints on $q$ and $M_\ast$ lead to constraints on the nature of the accretor. For example, for a period in the 6-day range, there are intermediate-age, evolved donor stars that have a mean density consistent with the period-density relation, but would imply (Table \ref{tab_ulx1_donor}) a mass of the accreting object $\la 2M_{\odot}$, consistent only with a neutron star accretor. On the other hand, an $\approx$6-day period is consistent with a stellar-mass BH accretor only for a narrow range of massive, young ($<$10 Myr) donor stars. Conversely, mass ratios $\la1$ (corresponding to a period in the 12-day range) are consistent only with a BH accretor. Independent observational constraints on the mass and age of the donor star in ULX-1 from the brightness of its optical counterpart will be presented and discussed in follow-up work currently in preparation.


Mass transfer from a donor star more massive than the accretor shrinks the binary separation and therefore causes higher, sustained mass transfer rates; this happens for $q > 5/6$ for the conservative mass transfer case, but we must account for possible additional shrinking of the system due to angular momentum losses in a wind \citep{2002apa..book.....F}. Blue supergiants have radiative envelopes; hence, mass transfer for $q \ga 1$ should proceed on a thermal (Kelvin-Helmholtz) timescale of the envelope, $\sim 10^4$ yr. For $q \la 5/6$ (permitted only for periods in the 12-day range), mass transfer would proceed instead on the nuclear timescale of the donor as it expands to the supergiant state. Therefore, determining the binary period of ULX-1 with future observations may reveal whether thermal-timescale or nuclear-timescale mass transfer is associated with strong ULX outflows.

Semi-detached, eclipsing system such as ULX-1 and ULX-2 offer also the best chance to determine the accretor mass from optical spectroscopic observations.
Let us assume for example that with future observations we will measure the binary period and strongly constrain the mass $M_\ast$ and radius $R_\ast$ of the donor star, and that we take spectra of the optical counterpart. If the donor star has absorption lines, phase-resolved optical spectroscopy might reveal its radial velocity curve, and hence the mass function $f(M_\bullet)$ of the compact object,
\begin{equation}
f(M_\bullet) = \frac{M_\bullet^3 \sin^3 \theta}{\left(M_\bullet + M_\ast \right)^2} 
\approx \frac{M_\bullet^3}{\left(M_\bullet + M_\ast \right)^2}, 
\end{equation}
from which $M_\bullet$ can be determined. Even without phase-resolved spectroscopy (hard to schedule on an 8-m telescope), one can still constrain the accretor mass if double-peaked (disk) emission lines are detected in the optical spectrum (typically, H$\alpha$, H$\beta$ and He\,{\footnotesize{II}} $\lambda4686$). Such lines are usually emitted from the outer rings of the accretion disk, and their full-width at half-maximum $V_{\rm fwhm}$ depends on the projected velocity of rotation of the gas at the outer disk radius $R_{\rm d}$: 
\begin{equation}
V^2_{\rm fwhm} \approx \frac{4GM_\bullet}{R_{\rm d}}\,\sin^2 \theta 
\approx  \frac{4GM_\bullet}{0.7 R_{\rm RL}}, \label{eq_V}
\end{equation}
where we have used the empirical and theoretical constraint \citep{1988MNRAS.232...35W} that the accretion disks extends to an outer radius of $\approx$70\% of the primary Roche lobe radius $R_{\rm RL}$. 
The Roche lobe radius is also a function of $q$; for example a useful approximation is 
\begin{equation}
R_{\rm RL} \approx R_{\ast} (M_\bullet/M_\ast)^{0.45},\label{eq_RL}
\end{equation}
\citep{2002apa..book.....F}. From Equations \eqref{eq_V} and \eqref{eq_RL}, $M_\bullet$ can be obtained without the need for phase-resolved spectroscopy.

\section{Conclusions}

Using archival {\it Chandra} and {\it XMM-Newton} observations, we found X-ray eclipses in two ULXs in the same region of M\,51. 
Eclipsing systems among the ULX and luminous BH X-ray binary populations are very rare: finding two of them not only in the same galaxy but a few arcsec from each other is a surprising result. Our serendipitous discovery in the archival data suggests that perhaps other eclipsing sources may have been missed, or mis-classified as variable/transient in previous X-ray source catalogs. If persistent ULXs are stellar-mass BHs fed by Roche-lobe-filling B-type supergiants, with a mass ratio $q \sim 1$, systems seen at inclination angles $\ga 75^{\circ}$ are expected to spend up to $\approx$15\% of their time in eclipse, over characteristic binary periods $\sim$10 days. Neutron star accretors are expected to have even longer eclipse fractions when seen edge-on ($\approx$20\%), and to have eclipses for viewing angles as low as $55^{\circ}$. Thus, a statistical study of the observed eclipse fractions in ULXs is a possible way to determine whether the ULX population is dominated by BHs or neutron stars.

We analyzed the presence and duration of the eclipses in ULX-1 and ULX-2 using a sequence of archival {\it Chandra} and {\it XMM-Newton} observations. For ULX-1, we argued that the most likely binary period is either $\approx$6.3 days, or $\approx$12.5--13 days. Assuming that the donor star fills its Roche lobe (a plausible assumption in ULXs, given the accretion rate needed to power them), we used the period-density relation to constrain the mass and evolutionary state of the donor star corresponding to those periods. For example, we showed that for a characteristic age $\approx$10 Myr, the donor-star mass is $\approx$17$M_{\odot}$, while for a characteristic age $\approx$20 Myr, $M_{\ast} \approx 11M_{\odot}$.

We compared and discussed the X-ray spectral and timing properties of the two eclipsing ULXs. ULX-1 is softer, and has a spectrum well-fitted by Comptonization models in a cool, dense medium. ULX-2 is harder, consistent with either a slim disk or Comptonization in a hotter medium. Both sources are clearly seen at high inclination, given the presence of eclipses; however, neither of them is an ultraluminous supersoft source (ULS). This supports our earlier suggestion \citep{2016MNRAS.456.1859U} that ULSs require not only a high viewing angle, but also an accretion rate high enough to produce effectively optically thick outflows. 

ULX-1 has strong spectral residuals around 0.8--1.0 keV: a spectral feature seen in other ULXs (usually those with a softer spectrum, thought to be viewed at higher inclination) but not well understood yet. Its most likely interpretation is a combination of thermal-plasma emission and absorption lines from a dense outflow. In ULX-1, a residual thermal-plasma emission ($\sim$10\% of the thermal-plasma emission out of eclipse) is still seen in eclipse, while the continuum component completely disappears. This suggests that the thermal-plasma emission originates from a region slightly larger than the size of the eclipsing star (that is, from a characteristic size of a few $10^{12}$ cm), rather than from the inner disk (which would be completely eclipsed) or from pc-scale shock-ionized gas (which would not be eclipsed at all). Instead, ULX-2 does not show significant thermal-plasma emission, although it does show residual emission in eclipse.
In conclusion, ULX-1 and ULX-2 are important sources to help us disentangle the effects of inflow/outflow structure versus viewing angle, and deserve further follow-up multiband studies.

\section*{Acknowledgements}

We thank the anonymous referee for several useful suggestions. We thank Manfred Pakull, Christian Motch and Fabien Gris\'{e} for their contribution to the multiband analysis and interpretation, Kiki Vierdayanti for clarifications on slim disk models, Megumi Shidatsu, Yoshihiro Ueda, Andrew Sutton and Matt Middleton for useful discussions on Comptonization models in stellar-mass BHs and ULXs, Douglas Swartz and Allyn Tennant for consultations on the {\it Chandra} data analysis, and Duncan Galloway for his suggestions on X-ray dips. We also thank our Curtin University colleagues Sam McSweeney, Vlad Tudor, Bradley Meyers, Steven Tremblay, Peter Curran, James Miller-Jones, Gemma Anderson, Richard Plotkin and Thomas Russell for their useful suggestions and comments. This paper benefited from discussions at the 2015 and 2016 International Space Science Institute workshops `The extreme physics of Eddington and super-Eddington accretion onto black holes' in Bern, Switzerland (team PIs: Diego Altamirano \& Omer Blaes). RU acknowledges support from Curtin University's 2015 Department of Applied Physics Scholarship and the Australian Postgraduate Award. RS acknowledges hospitality and scientific discussions at Kyoto University and at the Strasbourg Observatory during part of this work.

\label{lastpage}

\end{document}